\documentclass[prb,twocolumn,superscriptaddress,amsmath,amssymb,aps,nofootinbib]{revtex4-2}
\usepackage[utf8]{inputenc}
\usepackage{amsmath,amsthm,amssymb,amsfonts,braket}
\usepackage{placeins}[verbose]
\usepackage{tabularx, longtable, multirow}
\usepackage{graphicx}
\usepackage{dcolumn}
\usepackage{bm,bbm,dsfont}
\usepackage{musicography}
\usepackage[english]{babel}
\usepackage{comment}
\usepackage[dvipsnames]{xcolor}
\usepackage{siunitx}
\usepackage[normalem]{ulem}
\usepackage[T1]{fontenc}
\usepackage{mdframed}
\usepackage{cancel}

\providecommand{\ignore}[1]{}

\def\i{{\sf i}}
\def\c{\bar c}
\def\a{\bar a}
\def\b{\bar b}

\begin{document}
	
\title{Fusion mechanism for quasiparticles and \\ topological quantum order in the lowest Landau level}

\author{Arkadiusz Bochniak}
\email{arkadiusz.bochniak@mpq.mpg.de}
\affiliation{Max-Planck-Institut f{\"u}r Quantenoptik, Hans-Kopfermann-Str. 1, 85748 Garching, Germany}
\affiliation{Munich Center for Quantum Science and Technology, Schellingstraße 4, 80799 M{\"u}nchen, Germany}
\author{Gerardo Ortiz}
\email{ortizg@indiana.edu}
\affiliation{Department of Physics, Indiana University, Bloomington IN 47405, USA}
\affiliation{Institute for Quantum Computing, University of Waterloo, Waterloo, N2L 3G1, ON, Canada}
\date{\today}
	
\begin{abstract}
Starting from Halperin multilayer systems we develop a hierarchical scheme, {\it dubbed symmetrized multicluster construction}, that generates bosonic and fermionic single-layer quantum Hall states (or vacua) of arbitrary filling factor. Our scheme allows for the insertion of quasiparticle excitations with either Abelian or non-Abelian statistics and quantum numbers that depend on the nature of the original vacuum. Most importantly, it reveals a fusion mechanism for quasielectrons and magnetoexcitons that generalizes ideas about particle fractionalization introduced in A. Bochniak, Z. Nussinov, A. Seidel, and G. Ortiz, Commun. Phys. {\bf 5}, 171 (2022) for the case of Laughlin fluids. In addition, in the second quantization representation, we uncover the inherent topological quantum order (or the off-diagonal long-range order) characterizing these vacua. In particular, we illustrate the methodology by constructing generalized composite (generalized Read) operators for the non-Abelian Pfaffian and Hafnian quantum fluid states.  
\end{abstract}
	
\maketitle
	
\section{Introduction}

Fractional quantum Hall (FQH) fluids have been intensely studied for decades \cite{Jainbook}. These topological fluids portray quantum phases of interacting electronic matter, in the presence of strong magnetic fields, with emergent Abelian and non-Abelian excitations, anomalous electromagnetic response, and exquisitely quantized transport properties. At a fundamental mathematical level these fluid states are described by multivariate homogeneous polynomials of complex particle coordinates with zeros, when particles coalesce, that effectively encode the topological characteristics of their excitations.  
These polynomials are, generically, non-holomorphic if they are constructed from higher Landau level orbitals, and may display different types of zeros, i.e., different clustering properties. These, on the other hand, may relate to the type of particle interaction of some local parent Hamiltonian stabilizing such a fluid state as its ground state. For instance, parton states, the Laughlin sequence being a representative, satisfy a uniform clustering condition and are stabilized by two-body interactions \cite{Ahari22}. In contrast, states such as Pfaffians and Hafnians do not satisfy a uniform clustering condition and are stabilized by ${\sf k}$-body interactions with ${\sf k}>2$. 

In this paper, we show how to systematically generate translationally and rotationally (i.e., homogeneous) invariant multivariate polynomials,  of arbitrary filling factors (fractions), representing fermionic and bosonic FQH fluid states. For the sake of simplicity, we focus on the case of holomorphic polynomials on a disk geometry, that is, states  defined within the lowest-Landau-level (LLL) subspace which includes the Pfaffian and Hafnian states among its members (Gaussian factors are absorbed in the measure \cite{QP22}). Our symmetrized multicluster construction (SMC) starts from a Halperin bilayer (two clusters) system which, through an appropriate antisymmetrization (or symmetrization) procedure, turns this system into a single-layer fermionic (or bosonic)  FQH state.  Most importantly, we show how to generate quasihole, quasielectron, and magnetoexcitonic excitations in a natural and straightforward manner without invoking conformal field theory constructions. The basic idea for quasiholes is simple and consists of a complete antisymmetrization (symmetrization) procedure over the Halperin bilayer coordinate systems with (or without) magnetic fluxes attached in distinct layers, thus explicitly breaking translational invariance. This methodology, by construction, takes advantage of the diversity of possible clustering conditions among particle layer species. Exploiting the fusion mechanism for quasielectrons advanced for Laughlin's fluids in Ref. [\onlinecite{QP22}] we present quasiparticle excitations with proper particle fractionalization and quantum numbers, in arbitrary filling factor LLL FQH fluids. Our construction extends naturally to multilayer (multicluster) systems, thus providing a hierarchical scheme in search of incompressible FQH fluids and their quasiparticle excitations. In a related vein, we would also like to unveil the intricate many-body correlations present in non-Abelian topological fluids at the most fundamental level. In particular, we want to establish the topological (non-local) quantum order characterizing FQH fluid ground states. Working in a second quantization representation we express those ground states in a manner that easily helps establish their corresponding generalized composite (generalized Read) operators. We illustrate our many-body technique  in the Pfaffian and Hafnian state cases.

We start Sec.~\ref{sec:Multilayer-to-single-layer} presenting the main idea behind our symmetrized multicluster construction. The simplest example of a LLL FQH state beyond Laughlin's sequence is the fermionic (respectively bosonic) Pfaffian state $\mathrm{\Psi}_{\mathrm{Pf}}^q$ corresponding to the filling factor $\nu=\frac{1}{q}$ with $q\in 2\mathbb{N}$ (respectively $q\in 2\mathbb{N}+1$). This generic Pfaffian state is closely related to Halperin bilayer system \cite{Halperin83}, with the correspondence mimicking the construction of the Moore-Read (MR) state (corresponding to the choice $q=2$) out of Halperin $331$ bilayer system \cite{Greiter92a}. More precisely, the Pfaffian state $\mathrm{\Psi}_{\mathrm{Pf}}^q$ with  parameter $q$ can be thought of as the total antisymmetrization (respectively symmetrization) of the fermionic (respectively bosonic) Halperin $\Psi_{q+1\, q+1\, q-1}$ state. In the fermionic case, the proof can be found in \cite[Section~13.2]{Jainbook} and references therein. For completeness, here we present a combinatorial proof for fermions that we extend to the bosonic case. In addition, we construct the family of Hafnian states, both bosonic and fermionic, from the bilayer system $\Psi_{q+2\, q+2\,q-2}$ (the proof is in Appendix \ref{app:proof1}), generalizing previous observations for $\Psi_{551}$ and $\Psi_{440}$ \cite{Jeong}. Based on number theoretic results in \cite{Kuo}, our scheme can generate states within the LLL with arbitrary filling factors. 
We remark that a related idea of total antisymmetrization of a bilayer system was also discussed in the context of composite fermions, Halperin homogeneous multilayer systems \cite{Cappelli01,Regnault08}, and shares some common features with models based on conformal field theories \cite{Hermanns10, Hansson07, Hansson09, Hansson17}. We close Sec.~\ref{sec:Multilayer-to-single-layer} with a brief description of  multilayer and hierarchical generalizations that we plan to expand on in a future publication.  Here, we only present a derivation of a candidate $\nu=3/5$ incompressible fluid state.

Remarkably, our hierarchical construction scheme allows for the study of quasiparticle excitations in a quite natural way. In Sec.~\ref{sec:Quasiparticles}, we explain the methodology to insert (or remove) magnetic fluxes to generate quasiparticles with Abelian and non-Abelian exchange statistics. In this way, we generate quasihole excitations in arbitrary Pfaffian states and compare the resulting wave functions with the previous proposal \cite{MooreRead91, Greiter91}. It was observed in \cite{Cappelli01} that the procedure of placing quasiholes on particular layers and subsequent antisymmetrization leads to a wave function identical to that proposed by Moore and Read (and, more generally, in the case of multilayered systems this leads to quasiholes in Read-Rezayi states \cite{RR99}). We present explicit mathematical proof of this fact in Appendix \ref{app:profPfQh}. Armed with our quasihole construction we apply it to the family of Hafnian states and present a combinatorial proof in Appendix \ref{app:proof2}. 

Quasihole excitations are physically linked to magnetic flux insertion. On the other hand, their quasiparticle (quasielectron) excitations, physically associated with flux removal, were until recently \cite{QP22} a matter of debate. Advancing a second-quantization formalism we established a particle fractionalization principle that allowed derivation of quasielectron wave functions in Laughlin fluids \cite{QP22}. Furthermore, the obtained quasielectron turned out to be a composite object, made out of a certain number of quasiholes and a bare electron, satisfying a fusion rule leading to consistent quasiparticle quantum numbers. Here, we show that the same fusion mechanism \cite{QP22} is operative in non-Abelian FQH fluids allowing us to study quasielectrons and magnetoexcitons in arbitrary FQH fluids. In particular, we have checked numerically (by quantum Monte Carlo simulations) that the fusion mechanism works seamlessly in the case of Pfaffian fluids. In Appendix \ref{app:Pfaffiansimul} we explain how to efficiently simulate the Pfaffian of a skew-symmetric matrix. 

In the present paper, we are extensively using both first and second quantization representations. It turns out that each representation manifestly uncovers different aspects of the intricate many-body correlations present in those FQH states. For instance, while the nodal structure and root pattern of FQH states are revealed in the first quantization, fermionic pairing and Bose-Einstein condensation of bosons become manifest in the second quantization representation. In Sec.~\ref{sec:TQO}, we advance a derivation of the Pfaffian and Hafnian classes of states in second-quantization that manifestly uncovers their intrinsic topological order, or a special type of off-diagonal long-range order (ODLRO) \cite{Girvin87}. It shows, in particular, the reasons behind the lack of local particle condensation and, most critically, reveals the corresponding composite (generalized Read) operators whose expectation values signal the ODLRO present in those non-Abelian topological fluids. Proofs by recursion are presented in Appendix~\ref{app:proofPf}.

Finally, in Sec.~\ref{sec:Conclusions}, we summarize our main findings and reflect on remaining open questions and current directions we are pursuing to rigorously prove conjectures formulated in the current manuscript. 

\section{From Halperin multilayer to single-layer Quantum Hall states}
\label{sec:Multilayer-to-single-layer}

We start considering systems of two independent layers each comprising $n_1$ and $n_2$ particles. The positions of particles from the first layer are described by a tuple of complex numbers $Z_{n_1}=(z_1,\ldots,z_{n_1})$, while for the second layer, we use $\,\overline{\! Z}_{n_2}=(z_{n_1+1},\ldots,z_{n_1+n_2})$. The ordered union $(z_1,\ldots, z_{n_1+n_2})$ of these two tuples is indicated by $Z_{N}$ with $N=n_1+n_2$. We will also use the notation $\hat{z}_i$ to express that the variable $z_i=x_i+ \i y_i$ is not present in the set $Z_{N}$, while $(Z_{N})_{\hat{i}\hat{j}}$ stands for the set $(z_1,\ldots,\widehat{z_i},\ldots,\widehat{z_j},\ldots, z_{N})$, for $i<j$. The Vandermonde determinant in $Z_n$ variables is denoted by
\begin{equation}
 \Psi_L(Z_{n})=\prod_{i<j}^{n}(z_i-z_j).
 \end{equation}

The most general Halperin inhomogeneous bilayer state \cite{Halperin83} is of the form
\begin{equation}
\begin{split}
      \Psi_{a_1a_2b}(Z_{N})&=\Psi_L(Z_{n_1})^{a_1}\,\Psi_L(\,\overline{\! Z}_{n_2})^{a_2} \, {\cal P}_b(Z_{N}),
\end{split}
\end{equation}
with {\it pairing} between layers,
\begin{eqnarray}
{\cal P}_b(Z_{N})=\prod\limits_{\substack{i \in Z_{n_1}\\ j \in \,\overline{\! Z}_{n_2}}}\!\!\! (z_i-z_{j})^{b},
\end{eqnarray}
and has (degree equal to the) total angular momentum
\begin{eqnarray}
    J=n_1\left[ \left ( \frac{a_1+a_2}{2}+b\mu\right)n_1-\left ( \frac{a_1+\mu a_2}{2}\right)\right] \, \hbar,
\end{eqnarray}
with $\mu=\frac{n_2}{n_1}$, $\hbar$ the reduced Planck constant, and filling factor
\begin{equation}
       \nu=\frac{N-1}{\max\{n_1 a_1 +n_2 b, n_2a_2+ n_1 b\}}.
\end{equation}
In the original Halperin's construction \cite{Halperin83}, it was assumed that the two components covered the same area, thus, leading to an effective constraint where the two components' maximal Landau orbital angular momenta become equal, $n_1a_1+n_2 b=n_2a_2+n_1 b$, or equivalently, $\mu=\frac{a_1-b}{a_2-b}$, which leads to a 
total angular momentum
\begin{equation}
    J=\frac{N}{2} \left[  \frac{N}{2}\left(\frac{a_1+a_2}{2}+b\right)-\left ( \frac{a_1+ a_2}{2}\right)\right] \, \hbar,
\end{equation}
and filling factor \cite{Tong}
\begin{eqnarray} \nu=\frac{a_1+a_2-2b}{a_1a_2-b^2} .
\end{eqnarray} 
We refer to this assumption as Halperin's constraint. Such states can produce any filling factor $\nu\in \mathbb{Q}_+$ with $a_1,a_2>0$, $b\geq 0$ and $a_1a_2>b^2$ \cite{Kuo}. Despite being physically motivated this condition may be ignored if one is only interested in the ring of polynomials with different patterns of zeros. However, had we imposed that constraint it implies $n_1=n_2$ if and only if $a_1=a_2$.

One of the most broadly studied examples is Halperin $331$ ($n_1=n_2=n$) state \cite{Halperin83},  introduced as a generalization of Laughlin's proposal for bilayer or spinful systems. Such systems allow for a FQH fluid with $\nu=\frac{1}{2}$. It is known that the antisymmetrization among all $2n$ variables in the Halperin $331$ state produces the MR state \cite{Greiter92a, Ho95, Barkeshli10}, and more generally, starting from Halperin bilayer states of the form $\Psi_{q+1\, q+1\, q-1}$ with $q$ even one gets the Pfaffian state $\Psi_\mathrm{Pf}^q$. (We will later prove that this is true also for $q$ odd.) The (polynomial part of the) Pfaffian wave function is given by
\begin{equation}
    \Psi_{\mathrm{Pf}}^q(Z_{2n}) =\mathrm{Pf}_{2n}\left(\frac{1}{z_i-z_j}\right)\Psi_L(Z_{2n})^q,
\end{equation}
where $q\in \mathbb{N}$ is fixed, and the Pfaffian of an $2n\times 2n$ skew-symmetric matrix $A$ is defined as
\begin{equation}
{\rm Pf}_{2n}(A)\!=\!\frac{1}{2^n n!}\!\!\sum_{\sigma \in S_{2n}}\!\!{\rm sgn}(\sigma) \prod_{i=1}^n A_{\sigma(2i-1),\sigma(2i)},
\end{equation}
where $\mathrm{sgn}(\sigma)=\pm 1$ is the signature of the permutation $\sigma$ of $2n$ variables. Here $S_{2n}$ stands for the group of all such permutations. For $q$ even we have a fermionic wave function, while $q$ odd corresponds to a bosonic one. The filling factor of this state is $\nu=\frac{1}{q}$.
The special case $q=2$, the MR state, was advanced \cite{MooreRead91} as a possible incompressible wave function describing the observed  $\nu=\frac{5}{2}$ FQH fluid (considered effectively in the LLL it is associated to $\nu=\frac{1}{2}$). It was shown that this state is a ground (zero energy) state of a three-body Hamiltonian \cite{Greiter92, Wan08} with the smallest total angular momentum, $J=\frac{N(2N-3)}{2} \hbar$ \cite{Wan08}.

Here we generalize this idea, expand it to multilayer systems, and propose a hierarchical scheme, the SMC. For the sake of clarity, we next describe the construction for bilayer systems. The symmetric group has several irreducible representations; the totally symmetric or antisymmetric ones define the fully polarized fluids considered in our SMC, while others may require a more involved procedure.
In the situation where $n_1=n_2=n$, not necessarily satisfying Halperin's constraint, we will refer to as fermionic (bosonic) Halperin state whenever both $a_1, a_2$, as well as the pairing term $b$, are odd (even) integers. Fermionic (bosonic) Halperin states are obtained after the application of a total antisymmetrization $\mathcal{A}_{2n}$ (symmetrization $\mathcal{S}_{2n}$), with respect to all $Z_{2n}$ coordinates, to Halperin bilayer systems
\begin{eqnarray}
\label{conv}
    \mathcal{A}_{2n}\Psi_{a_1a_2b}(Z_{2n})&=&\frac{1}{N!}\sum\limits_{\sigma\in S_{2n}}\mathrm{sgn}(\sigma)\Psi_{a_1a_2b}(\sigma (Z_{2n})), \nonumber\\
    \mathcal{S}_{2n}\Psi_{a_1a_2b}(Z_{2n})&=&\frac{1}{N!}\sum\limits_{\sigma\in S_{2n}}\Psi_{a_1a_2b}(\sigma (Z_{2n})),
\end{eqnarray}
where $\sigma(Z_{2n})=(z_{\sigma(1)},\ldots,z_{\sigma(2n)})$. Note that the total antisymmetrization or symmetrization procedures preserve the value of the original Halperin state filling factor. For $n_1=n_2$ and $a_1=a_2$ the nonzero states are obtained from antisymmetrization (respectively symmetrization) if and only if $b$ is odd (respectively even). For $a_1\neq a_2$ (and also for $n_1\neq n_2$), there is no such parity constraint on $b$, and we can in certain cases still produce nonzero states in the LLL by applying the above procedure. 

Since in the bosonic case, $a_1=2k_1$, $a_2=2k_2$, and $b=2l$, the filling factor (under the condition that $a_1n_1+bn_2=a_2n_2+bn_1$) is $\nu=\frac{k_1+k_2-2l}{2(k_1k_2-l^2)}$, the symmetrization procedure allows generation of bosonic fluids of arbitrary $\nu\in \mathbb{Q}_+\cap (0,1)$ due to the number-theoretical results from \cite{Kuo}. The fermionic case is, however, more subtle and requires some analysis beyond \cite{Kuo}. We leave the general question for future research, however, we mention an illustrative example. Relaxing Halperin's constraint, it is possible to obtain a fermionic state with $\nu=\frac{3}{4}$ by taking any odd $a_1>a_2>0$ and $b=1$ such that $\mu=3a_1-4>1$. (There exists also an analogous family with $\mu<1$.) In this case, the total angular momentum reads $J=n_1\left(\frac{7a_1+a_2-8}{2}n_1 -\frac{a_1+a_2}{2}\right)\, \hbar$, which is minimized if one chooses $a_1=3$, $a_2=1$ and $\mu=5$, in which case $J=n_1(7n_1-2)\, \hbar$. We remark that after relaxing Halperin's constraint, it is not in principle guaranteed that the resulting state corresponds to a liquid. However, in the present work, we will be mostly interested in the situation where $a_1=a_2=a$ and $n_1=n_2=n$, the homogeneous bilayer case, with corresponding filling factor $\nu=\frac{2}{a+b}$. We parametrize these (bosonic or fermionic) states as $\Psi_{q+s\, q+s\, q-s}$ with $q\ge s$. There are essentially two subclasses here. For $s$ odd, $q$ even corresponds to fermionic Halperin's states, while $q$ odd to bosonic ones. For $s$ even, however, we have the opposite: $q$ even leads to bosonic states, while $q$ odd corresponds to fermionic ones.

We conclude this brief description of the general idea with the following remarks. As mentioned in the Introduction, the zeros and clustering properties of FQH states (without and with translational-symmetry-breaking magnetic fluxes) encode their topological characteristics. The class of polynomials obtained from Halperin bilayer systems display a $M$-clustering property with $M=\min\{a_1,a_2,b\}$. Moreover, using the fact that the ring of multivariate polynomials over the complex field is a unique factorization domain (UFD), or factorial, it was shown in  Ref. \cite{Ahari22} that the set of parton-like states spans the ring of (anti-)symmetric holomorphic polynomials with the $M$-clustering property. This means that except for the case $a_1=a_2=b$, which represents a single parton state, all other single-layer states derived from Halperin bilayer systems can be efficiently expanded in terms of parton-like states with identical $M$-clustering properties \cite{Ahari22}. In addition, the root pattern of the FQH fluid uniquely characterizes it, leading to a root state or DNA that encodes all topological properties of the fluid \cite{Ahari22}. In the LLL, the root pattern (or state) is represented as a string of positive integers defining the occupation numbers of Landau angular momenta orbitals in the nonexpandable Slater determinant (or permanent) component of the fluid state \cite{Ortiz2013}. For instance, if particles occupy the orbital angular momenta $(j_1,j_2)=(0,1) \ \mbox{mod(4)}$, the (bulk) string becomes $\{1100\} = 11001100110011001100\ldots$. Root patterns for fermionic systems can only include $1$s and $0$s, because of Pauli exclusion, while bosonic systems may in principle include arbitrary positive integers. For the class of holomorphic polynomials considered in this work, the corresponding (bulk) root pattern can be determined by using the algorithm proposed in Ref.~\cite{Seidel08} for Halperin bilayer states, which amounts to maximization of the quantity 
\begin{eqnarray}
\Delta_{J}= \sum_{i=1}^Nj_i^2 .
 \label{algorithm-root}
\end{eqnarray}
We would also like to emphasize that establishing which of the multiple FQH homogeneous fluid states of a given filling factor $\nu$ corresponds to an incompressible one is a nontrivial mathematical task. Later on, we will formulate a conjecture. 

We next present a few paradigmatic examples resulting from our construction, including  derivations of Pfaffian and Hafnian families of states, and conclude the section with a generalization to multilayer systems and a proposal for hierarchical construction. 

\subsection{The case $(q+1,q+1,q-1)$} 

Let us consider first the case with $s=1$. We reproduce here the (adjusted version of the) argument from \cite{Jainbook} used for the fermionic case ($q$ even), extending it then to bosonic systems ($q$ odd).

First, using the Cauchy determinant formula \cite{Cauchy41},
    \begin{equation}
        {\rm det}_{n}\!\left(\frac{1}{z_i-z_{n+k}}\right) \!=\!\frac{\prod\limits_{i<k}^{n}(z_k-z_i)(z_{n+i}-z_{n+k})}{\prod\limits_{i,k=1}^{n}(z_i-z_{n+k})},
    \end{equation}
   it follows that
\begin{equation}
\begin{split}
        \Psi_{q+1\, q+1\, q-1}(Z_{2n})\!=\!\varepsilon_n\!\left[ {\rm det}_{n}\!\left(\!\frac{1}{z_i-z_{n+k}}\!\right)\right]\! \Psi_L(Z_{2n})^q,
\end{split}        
    \end{equation}
where
\begin{eqnarray}
\label{eq:varepsilon}
\varepsilon_n=(-1)^{\frac{n(n-1)}{2}}. \end{eqnarray}

For $q$ even (i.e., in the fermionic case), we have
\begin{equation}
\label{eq:anti_L}
\begin{split}
    &\mathcal{A}_{2n}\left({\rm det}_{n}\!\left(\frac{1}{z_i-z_{n+k}}\right)\Psi_L(Z_{2n})^q\right)\\
    &=\mathcal{A}_{2n}\left({\rm det}_{n}\!\left(\frac{1}{z_i-z_{n+k}}\right)\right)\Psi_L(Z_{2n})^q,
\end{split}    
\end{equation}
and also
\begin{equation}
\begin{split}
    &\mathcal{A}_{2n}{\rm det}_{n}\!\left(\!\frac{1}{z_i-z_{n+j}}\!\right)\!=\!\!\sum\limits_{\sigma\in S_n}\mathrm{sgn}(\sigma)\mathcal{A}_{2n}\prod\limits_{j=1}^n \frac{1}{z_j-z_{n+\sigma(j)}}\\
    &=n! \, \mathcal{A}_{2n}\prod\limits_{j=1}^n\frac{1}{z_j-z_{n+j}}\!=\!-\frac{2^n(n!)^2}{(2n)!}\,\mathrm{Pf}_{2n}\!\left(\frac{1}{z_i-z_j}\right),
\end{split}    
\end{equation}
where in the last step we have used the fact that $(1,n+1,2,n+2,\ldots, n,2n)$ is an odd permutation of $(1,\ldots,2n)$, which produces an additional minus sign. This leads to the conclusion that total antisymmetrization (i.e., with respect to all coordinates) of  Halperin bilayer state $\Psi_{q+1\, q+1\, q-1}(Z_{2n})$ leads to the fermionic Pfaffian state $\Psi_{\mathrm{Pf}}^q$, up to an overall constant.

We show next how the above proof can be modified to cover also the bosonic case, i.e., situations with $q$ odd. First, we notice that
\begin{equation}
\begin{split}
        &\Psi_{q+1\, q+1\, q-1}(Z_{2n})\\ &\!=\!\varepsilon_n \!\left[{\rm det}_{n}\!\left(\frac{1}{z_i-z_{n+k}}\right)\!\Psi_L(Z_{2n}) \right]\!\Psi_L(Z_{2n})^{q-1}.
\end{split}        
\end{equation}
In contrast to the antisymmetrization performed for fermions, in the case of bosonic systems, we  symmetrize the Halperin bilayer state with respect to all variables. Since $q-1$ is even, the problem reduces to the total symmetrization of ${\rm det}_{n}\!\left(\frac{1}{z_i-z_{n+k}}\right)\Psi_L(Z_{2n})$. Notice that
\begin{equation}
\begin{split}
    &\mathcal{S}_{2n}\left({\rm det}_{n}\!\left(\frac{1}{z_i-z_{n+k}}\right)\Psi_L(Z_{2n})\right)\\ 
    &=\mathcal{A}_{2n}\left({\rm det}_{n}\left(\frac{1}{z_i-z_{n+k}}\right)\right)\Psi_L(Z_{2n}). 
\end{split}    
\end{equation}

Indeed, this claim is equivalent to 
\begin{equation}
\begin{split}
    &\sum\limits_{\sigma\in S_{2n}}\!\!\!{\rm det}_{n} \left(\frac{1}{z_{\sigma(i)}-z_{\sigma(n+k)}}\right)\Psi_L(\sigma(Z_{2n}))\\
    &=\!\!\sum\limits_{\sigma\in S_{2n}}\!\!\! \mathrm{sgn}(\sigma){\rm det}_{n}\!\left(\frac{1}{z_{\sigma(i)}-z_{\sigma(n+k)}}\right)\Psi_L(Z_{2n}).
\end{split}    
\end{equation}
It is sufficient to show that for any $\sigma\in S_{2n}$, we have
\begin{equation}
\begin{split}  
&{\rm det}_{n}\!\left(\frac{1}{z_{\sigma(i)}-z_{\sigma(n+k)}}\right)\Psi_L(\sigma(Z_{2n}))\\ 
&=\mathrm{sgn}(\sigma){\rm det}_{n}\!\left(\frac{1}{z_{\sigma(i)}-z_{\sigma(n+k)}}\right)\Psi_L(Z_{2n}),
\end{split}
\end{equation}
but this simply follows from the total antisymmetry of Laughlin's factor. This shows that in the bosonic case, the total symmetrization also leads to the bosonic Pfaffian state $\Psi_{\mathrm{Pf}}^q$, up to an overall constant.

We can establish the bulk root patterns corresponding to this family of Pfaffian states. For example, the bulk root pattern originating from the state $\Psi_{220}$ (describing a $\nu=1$ bosonic state) is simply $\{20\}$, while the one for $\Psi_{331}$ (describing fermions with $\nu=\frac{1}{2}$) is $\{1100\}$. Yet another example is the state describing $\nu=\frac{1}{3}$ bosons and obtained from $\Psi_{442}$. In this case, the corresponding pattern is $\{101000\}$. Similarly to homogeneous vacuum fluids, one may consider quasihole states of the form $(\prod_{i=1}^n z_i) \, \Psi_{q+1\, q+1\, q-1}(Z_{2n})$, properly (anti-)symmetrized, and study the corresponding bulk root patterns.  For the aforementioned examples they are $\{1\}$, $\{10\}$ and $\{100100\}$, respectively. 

\subsection{The case $(q+2,q+2,q-2)$} 

Let us now consider the case with $s=2$. It is known (see \cite{Jeong} and references therein) that the so-called Hafnian state, ${\rm det}_{2n}\!\left(\frac{1}{z_i-z_j}\right)\Psi_{L}(Z_{2n})^{3}$, can be obtained from the antisymmetrization of the fermionic Halperin bilayer state $\Psi_{551}$. A similar statement can be shown to be true for its bosonic counterpart - symmetrization of $\Psi_{440}$ leads to ${\rm det}_{2n}\!\left(\frac{1}{z_i-z_j}\right)\Psi_{L}(Z_{2n})^{2}$. Notice that these states correspond to $q=3$ and $q=2$, respectively. However, what is known in the literature \cite{Jeong} as a `Hafnian state' can be really understood as a special case of the whole family of Hafnians in the same way as the MR state is a special case, that with $q=2$, of the Pfaffian family.

The family of Hafnian states parameterized by $q\in \mathbb{N}$ is defined as 
\begin{equation}
\label{eq:haf_def}
\begin{split}
    \Psi_{\mathrm{Hf}}^q(Z_{2n})=&\mathrm{Hf}_{2n}\left(\frac{1}{(z_i-z_j)^2}\right)\Psi_L(Z_{2n})^{q}\\=&{\rm det}_{2n}\!\left(\frac{1}{z_i-z_j}\right)\Psi_{L}(Z_{2n})^{q},
\end{split}    
\end{equation}
where the Hafnian of an $2n\times 2n$ symmetric matrix $B$ is expressed as
\begin{equation}
{\rm Hf}_{2n}(B)=\frac{1}{2^n n!}\sum_{\sigma \in S_{2n}} \prod_{i=1}^n B_{\sigma(2i-1),\sigma(2i)} .
\end{equation}
Notice that now the case with $q$ odd corresponds to fermions, while $q$ even to bosons.

Below we show that the Hafnian state $\Psi_{\mathrm{Hf}}^q$ is associated with an antisymmetrization (respectively symmetrization) of the Halperin bilayer state $\Psi_{q+2\, q+2\, q-2}$ for $q$ odd (respectively even), with $q\ge 2$. We start with the fermionic case and notice that for any odd $q\ge 3$, we have
\begin{equation}
\label{eq:hf01}
\begin{split}
        &\mathcal{A}_{2n}\Psi_{q+2\, q+2\, q-2}(Z_{2n})\\
        &=\!\mathcal{S}_{2n}\!\left(\!\prod\limits_{i<j}^n(z_i-z_j)^4(z_{n+i}-z_{n+j})^4\!\right)\! \Psi_L(Z_{2n})^{q-2}.
\end{split}        
\end{equation}
In Appendix \ref{app:proof1} we prove that
\begin{equation}
\label{eq:HafPROOF}
    \begin{split}
        &\mathcal{S}_{2n}\left(\prod\limits_{i<j}^n(z_i-z_j)^4(z_{n+i}-z_{n+j})^4\right)\\
        &= \frac{2^n (n!)^2}{(2n)!}\mathrm{Hf}_{2n}\!\left(\frac{1}{(z_i-z_j)^2}\right)\Psi_{L}(Z_{2n})^{2},
    \end{split}
\end{equation}
and this leads to the conclusion that
\begin{equation}
\label{eq:Hafnianfamily}
    \begin{split}
        &\mathcal{A}_{2n}\Psi_{q+2\, q+2\, q-2}(Z_{2n})\\
        &=\frac{2^{n}(n!)^2}{(2n)!}\mathrm{Hf}_{2n}\!\left(\frac{1}{(z_i-z_j)^2}\right) \Psi_L(Z_{2n})^{q}.
    \end{split}
\end{equation}
Notice that from the above relation, we also obtain the identity
\begin{equation}
\begin{split}
     \mathcal{A}_{2n}\!\Biggl(\Biggl[&{\rm det}_{n}\!\left(\frac{1}{z_i-z_{n+k}}\right)\Biggr]^{2}\Biggr)\\ &=\frac{2^n (n!)^2}{(2n)!}\mathrm{Hf}_{2n}\!\left(\frac{1}{(z_i-z_j)^2}\right).  
\end{split}     
\end{equation}

For the bosonic case ($q\ge 2$ even) we immediately get
\begin{equation}
\begin{split}
    \!&\mathcal{S}_{2n}\Psi_{q+2\, q+2\, q-2}(Z_{2n})\\
    &\!=\!\mathcal{S}_{2n}\left(\prod\limits_{i<j}^n(z_i-z_j)^4(z_{n+i}-z_{n+j})^4\right)\Psi_L(Z_{2n})^{q-2}\\
    &\!=\!\frac{2^{n}(n!)^2}{(2n)!}\mathrm{Hf}_{2n}\!\left(\frac{1}{(z_i-z_j)^2}\right)\Psi_L(Z_{2n})^{q}.
\end{split}    
\end{equation}
This finishes the proof of our claim. 

We close the discussion of Hafnian states by presenting explicit bulk root patterns for the states originating from $\Psi_{440}$ (bosonic $\nu=\frac{1}{2}$) and $\Psi_{551}$ (fermionic $\nu=\frac{1}{3}$) for both homogeneous fluids (vacua) as well as the ones with quasiholes $(\prod_{i=1}^n z_i) \, \Psi_{q+2\, q+2\, q-2}(Z_{2n})$. For $\Psi_{440}$ we obtain $\{2000\}$ for the vacuum and $\{1100\}$ for the state with quasiholes, while the corresponding bulk root patterns originating from the $\Psi_{551}$ state are $\{110000\}$ and $\{101000\}$, respectively.

\subsection{The case $(q+s,q+s,q-s)$} 

For the general odd $s$ and even $q>s$ case, the resulting fermionic state is of the form
\begin{equation}
\begin{split}
    \Psi_{\mathrm{Pf}}^{q;s}(Z_{2n})\!=\!\varepsilon_n^s\mathcal{A}_{2n}\!\left(\left[{\rm det}_{n}\!\left(\frac{1}{z_i-z_{n+k}}\right)\right]^{s}\right)\Psi_L(Z_{2n})^q,
\end{split}    
\end{equation}
with $\varepsilon_n$ defined in Eq. \eqref{eq:varepsilon}. We refer to this class of states as fermionic Pfaffian-like states. We remark that, for $s\geq 3$,
\begin{equation}
    \Psi_{\mathrm{Pf}}^{q;s}(Z_{2n})\neq  \left[\mathrm{Pf}_{2n}\left(\frac{1}{z_i-z_j}\right)\right]^s\Psi_L(Z_{2n})^q,
\end{equation}
because in the antisymmetrization $\mathcal{A}_{2n}$ there is no totally symmetric function involved. In contrast, the total symmetry of Laughlin's factor in Eq. \eqref{eq:anti_L}  allowed for factorization. Moreover, notice that 
\begin{equation}
    \left[{\rm det}_{n}\!\left(\frac{1}{z_i-z_{n+k}}\right)\right]^s\neq {\rm det}_{n}\!\left(\frac{1}{(z_i-z_{n+k})^s}\right),
\end{equation}
since on the left-hand side we have a determinant of the $s$th power of a matrix with entries $A_{ik}=\frac{1}{z_i-z_{n+k}}$ but, in particular, $(A^2)_{ik}=\sum_{j}\frac{1}{(z_i-z_{n+j})(z_j-z_{n+k})}\neq \frac{1}{(z_i-z_{n+k})^2}$. We use the notion of bosonic Pfaffian-like states for those generated by a symmetrization process applied to Halperin bilayer states $\Psi_{q+s\, q+s\, q-s}$ with $q$ odd. 

Similarly, for $s$ even, we will 
call Hafnian-like states to those 
obtained from Halperin bilayer states $\Psi_{q+s\, q+s\, q-s}$ after application of a 
total antisymmetrization or symmetrization process. We also remark that the special case with $s=0$ corresponds to states from Laughlin's sequence.

\subsection{Generalization to multilayer systems}
\label{sec:multi}

The above construction can be generalized in a straightforward manner to multilayer systems. In particular, the most general Halperin inhomogeneous trilayer system is of the form
\begin{equation}
\label{eq:trilayer}
    \begin{split}
        \Psi_{a_1a_2a_3b_1b_2b_3}(Z_{N})\!=&\Psi_L(Z_{n_1})^{a_1}\,\Psi_L(\,\overline{\! Z}_{n_2})^{a_2}\,\Psi_L(\,\widetilde{\! Z}_{n_3})^{a_3}  \\
    \times&{\cal P}_{b_1}(Z_{1,2}){\cal P}_{b_2}( Z_{1,3}){\cal P}_{b_3}(Z_{2,3}),
    \end{split}
\end{equation}
where the third cluster of particle coordinates is indicated as $\widetilde{Z}_{n_3}=(z_{n_1+n_2+1},\ldots,z_{N})$ with $N=n_1+n_2+n_3$, and, to simplify the notation, we denote $Z_{1,2}=Z_{n_1}\sqcup \overline{Z}_{n_2}$, $Z_{1,3}=Z_{n_1}\sqcup \widetilde{Z}_{n_3}$ and $Z_{2,3}=\overline{Z}_{n_2}\sqcup \widetilde{Z}_{n_3}$. The generalized angular momentum condition (i.e., the trilayer analog of Halperin's constraint) takes the form
\begin{eqnarray}
n_1 a_1+ n_2 b_1+n_3 b_2&=& n_2 a_2+ n_2 b_1+n_3 b_2 \nonumber \\
&=&n_3 a_3+ n_1 b_2+n_2 b_3,
\end{eqnarray}
where all $a_i$ and $b_i$, $i=1,2,3$, share the same parity. The case $a_1=a_2=a_3=a$ and $b_1=b_2=b_3=b$ (with $n_1=n_2=n_3=n$, the homogeneous case) leads to a filling factor $\nu=\frac{3}{a + 2 b}$.  The generalization of the latter homogeneous case for $n_l$ layers gives $\nu=\frac{n_l}{a + (n_l-1) b}$.

In this way, for instance, one can generate from $\Psi_{333111}(Z_{N})$ (after antisymmetrization) a $\nu=3/5$ fermionic fluid state with bulk root pattern $\{11100\}$. The resulting wave function $\Psi_{3/5}(Z_N)$ is then of the form
\begin{equation}
   \mathcal{A}_{3n}\Biggl\{\frac{\Psi_{L}(Z_{n_1})\Psi_{L}(\overline{Z}_{n_2})\Psi_{L}(\widetilde{Z}_{n_3})}{{\cal P}_1(Z_{1,2}){\cal P}_1( Z_{1,3}){\cal P}_1(Z_{2,3})}\Biggr\}\Psi_{L}(Z_N)^2.
\end{equation}
The most general (homogeneous, trilayer) case is associated to
\begin{equation}
   \mathcal{O}_{3n}\Biggl\{\frac{\left(\Psi_{L}(Z_{n_1})\Psi_{L}(\overline{Z}_{n_2})\Psi_{L}(\widetilde{Z}_{n_3})\right)^{s}}{{\cal P}_{s}(Z_{1,2}){\cal P}_s( Z_{1,3}){\cal P}_s(Z_{2,3})}\Biggr\}\Psi_{L}(Z_N)^{q} ,
\end{equation}
with parametrization $a=q+s$, $b=q-s$, whose filling factor is $\nu=\frac{3}{3 q - s}$. 

\begin{figure}[htb] 
\centering
\includegraphics[width=0.7\columnwidth]{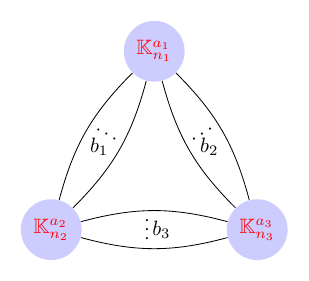}
\caption{Complete multigraph representing the Halperin trilayer state $\Psi_{a_1a_2a_3b_1b_2b_3}$.  $\mathbb{K}_{n_l}^{a_l}$ stands for a complete multigraph on $n_l$ vertices with edges of multiplicities $a_l$, $l=1,2,3$.
} 
\label{fig:00}
\end{figure}

Our SMC can be represented schematically in terms of (undirected) graphs with vertices corresponding to particle labels $i,j=1,\ldots, N$, and edges $\langle i j\rangle$ which are associated to the polynomial $z_i-z_j$. Figure ~\ref{fig:00} illustrates the correspondence for a trilayer system. Laughlin factor $\Psi_L^{a_l}$ in layer $l=1,2,3$ is represented by a complete multigraph $\mathbb{K}_{n_l}^{a_l}$ on $n_l$ vertices with all edges of multiplicities $a_l$. Pairing among  layers is pictured as edges connecting those three complete multigraphs, with multiplicities identified with pairing exponents. The resulting Halperin multilayer state is mapped onto a complete multigraph. In particular, for $a_1=a_2=a_3=b_1=b_2=b_3=a$, we obtain a complete multigraph of $N=n_1+n_2+n_3$ vertices with all edges of multiplicity $a$ which realizes a Laughlin state. In general, every distinct complete multigraph of $N$ vertices represents a  pattern of zeros of the corresponding polynomial. 

\subsection{A hierarchical construction} 
\label{sec:hierarchy}

Starting with a family of Halperin multilayer systems we obtained single-layer LLL states  with particular filling factors $\nu$. Let us now generalize this construction and propose a new hierarchical scheme. 

We begin writing a Halperin bilayer state ($n_1=n_2=n$) in the following manner, where notation will become clear in a moment,
\begin{equation}
\begin{split}
        \Psi^{(0)}_{a_1^{(0)}a_2^{(0)}b^{(1)}}\!(Z_{2n})\!=\!\Psi_{L}(Z_n)^{a_1^{(0)}}\Psi_{L}(\,\overline{\! Z}_{n})^{a_2^{(0)}}\mathcal{P}_{b^{(1)}}(Z_{2n}),
\end{split}
\end{equation}
where $\Psi_{L}(Z_n)^{a_1^{(0)}}=\Psi^{(0)}_{a_1^{(0)}00}(Z_n)$ and $\Psi_{L}(\,\overline{\! Z}_{n})^{a_2^{(0)}}=\Psi^{(0)}_{0a_2^{(0)}0}(\,\overline{\! Z}_{n})$ are the (polynomial parts of) Laughlin's wave functions with powers $a_1^{(0)}$ and $a_2^{(0)}$, respectively.  The pairing between layers is indicated by $\mathcal{P}_{b^{(1)}}(Z_{2n})=\prod\limits_{i,k}(z_i-z_{n+k})^{b^{(1)}}$. We stress that all parameters, $a_1^{(0)},a_2^{(0)}$, and $b^{(1)}$, used in this construction are chosen with the same parity. Let $\mathcal{O}\in\{\mathcal{A},\mathcal{S}\}$ be either an antisymmetrization or a symmetrization operation, depending on the nature of the state. Defining
\begin{equation}
    \Psi^{(1)}_{a_1^{(0)} a_2^{(0)} b^{(1)}}(Z_{2n})=\mathcal{O}_{2n}\Psi^{(0)}_{a_1^{(0)}a_2^{(0)}b^{(1)}}(Z_{2n})
\end{equation}
corresponds to the families of wave functions described in previous sections. 

Assume now that two sets of numbers, $\{a_{1,1}^{(0)},a_{1,2}^{(0)},b_1^{(1)}\}$ and $\{a_{2,1}^{(0)},a_{2,2}^{(0)},b_2^{(1)}\}$, are given. Following the above prescription we can generate two new wave functions
\begin{equation}
    \Psi^{(1)}_{a_{1,1}^{(0)} a_{1,2}^{(0)} b_1^{(1)}}(Z_{2n}) \quad \mbox{and} \quad \Psi^{(1)}_{a_{2,1}^{(0)} a_{2,2}^{(0)} b_2^{(1)}}(Z_{2n}).
\end{equation}
Imagine a situation where we put these two new states on two layers and introduce pairing between them,
\begin{equation}
\begin{split}
       \Psi^{(1)}_{a_1^{(1)}a_2^{(1)}b^{(2)}}(Z_{4n})=&\Psi^{(1)}_{a_{1,1}^{(0)} a_{1,2}^{(0)} b_1^{(1)}}(Z_{2n})\Psi^{(1)}_{a_{2,1}^{(0)} a_{2,2}^{(0)} b_2^{(1)}}(\overline{Z}_{2n})\\
       &\times \mathcal{P}_{b^{(2)}}(Z_{4n}),
\end{split}
 \end{equation}
where $a_1^{(1)}=\{a_{j,1}^{(0)}\}_{j=1}^2$, $a_2^{(1)}=\{a_{j,2}^{(0)}\}_{j=1}^2$ and $b^{(2)}=\{b_j^{(1)}\}_{j=1}^2$. Again, only wave functions being of the same bosonic (respectively fermionic) nature can be paired together, and the pairing term has to have an even (respectively odd) exponent. This new Halperin-like state can be then used to produce a new state within the LLL:
\begin{equation}
    \Psi^{(2)}_{a_1^{(1)}a_2^{(1)}b^{(2)}}(Z_{4n})=\mathcal{O}_{4n}\Psi^{(1)}_{a_1^{(1)}a_2^{(1)}b^{(2)}}(Z_{4n}).
\end{equation}
This hierarchical procedure can be continued. Starting from a set $\{\{a^{(0)}_{j,1}, a^{(0)}_{j,2},b_j^{(1)}\}\}_{j=1}^{2^\ell}$ we can construct the $\ell$-th level hierarchy wave function recursively:
\begin{equation}
\begin{split}
    &\Psi^{(\ell)}_{a_1^{(\ell-1)} a_2^{(\ell-1)}b^{(\ell)}}\!\left(Z_{2^\ell n}\right)\!=\!\mathcal{O}_{2^\ell n}\Psi^{(\ell -1)}_{a_1^{(\ell-1)} a_2^{(\ell-1)}b^{(\ell)}}\!\left(Z_{2^{\ell}n}\right)
\end{split}    
\end{equation}
with parameters $a_i^{(\ell-1)}=\{a_{j,i}^{(0)}\}_{j=1}^{2^\ell}$ for $i=1,2$, and $b^{(\ell)}=\{b_j^{(1)}\}_{j=1}^{2^\ell}$.

There is a series of natural questions that arises. First of all, suppose that in the SMC, we produce a state of the filling factor $\nu$. By the discussion in the previous sections, we already know that any such a filling factor can be obtained already after one step of the hierarchy scheme, possibly with different Laughlin's powers $a_1,a_2$ and the pairing parameter $b$. Therefore, from this perspective, one can then think that a one-step hierarchy is enough to generate any possible states. Is it indeed the case? In other words, what is the complete set of characteristics that fully distinguishes states with the same filling factor but differing in the level in the hierarchy? 

Secondly, the above hierarchical scheme is not the only possible one that one can deduce from generalizing the construction performed on the first level. For a reason that will be clear in a moment, we refer to this specific hierarchical construction as a {\it symmetric} scheme. Starting from $2^\ell$-layered system we can choose any order of interlayer antisymmetrizations or symmetrizations. More generally, let us have $p$ layers of Laughlin's states characterized by $a_1,\ldots,a_p$, respectively. We can perform $\mathcal{O}$'s operations in several orders and for different subsets of layers. For example, for a trilayer system, we can choose, e.g., the following: 
\begin{equation}
\label{eq:hier1}
\begin{split}
    &\Psi_{a_1}\Psi_{a_2}\Psi_{a_3}\xrightarrow{\mathcal{O}_{12}}{\mathcal{O}_{12}}\left(\Psi_{a_1}\Psi_{a_2}\right)\Psi_{a_3}\xrightarrow{\mathcal{O}_{12,3}}\\
    &\xrightarrow{\mathcal{O}_{12,3}}\mathcal{O}_{12,3}\left({\mathcal{O}_{12}}\left(\Psi_{a_1}\Psi_{a_2}\right)\Psi_{a_3}\right)
\end{split}    
\end{equation}
or
\begin{equation}
\begin{split}
    &\Psi_{a_1}\Psi_{a_2}\Psi_{a_3}\xrightarrow{\mathcal{O}_{23}}\Psi_{a_1}{\mathcal{O}_{23}}\left(\Psi_{a_2}\Psi_{a_3}\right)\xrightarrow{\mathcal{O}_{12,3}}\\
    &\xrightarrow{\mathcal{O}_{12,3}}\mathcal{O}_{1,23}\left(\Psi_{a_1}{\mathcal{O}_{23}}\left(\Psi_{a_2}\Psi_{a_3}\right)\right),
\end{split}    
\end{equation}
etc. The natural question of equivalence between different schemes arises. One would like to rigorously classify all such schemes and group them into suitable equivalence classes. Last but not least, one can ask about potential generalizations of the above hierarchical construction beyond the $n_1=n_2=n$ case. We leave these problems for future research.

The hierarchical scheme includes, as a particular case, the multilayer construction of the last section. Consider the $\nu=3/5$ fluid state discussed in Sec.~\ref{sec:multi}. We next show that this state can also be obtained from our hierarchical construction, e.g., by using the scheme from Eq.~\eqref{eq:hier1} with $a_1=a_2=a_3=3$ and the interlayer pairing given by ${\cal P}_1$ for all pairs of layers. If $b_1=b_2=b_3=b$, the wave function $\Psi_L(\widetilde{Z}_{n_3})^{a_3}{\cal P}_{b_2}(Z_{1,3}){\cal P}_{b_3}(Z_{2,3})$ is symmetric with respect to the $Z_{1,2}$ variables. Since ${\cal A}_{3n}=\mathcal{A}_{3n}\circ\mathcal{A}_{2n}$, this means that the total antisymmetrization of the wave function $\Psi_{a_1a_2a_3bbb}$ of Eq. \eqref{eq:trilayer} can be written as 
\begin{equation}
\begin{split}
    \mathcal{A}_{3n}\Bigl[&\mathcal{A}_{2n}\Bigl(\Psi_L(Z_{n_1})^{a_1}\Psi_L(\overline{Z}_{n_2})^{a_2}{\cal P}_b(Z_{1,2})\Bigr)\\
    &\times \Psi_L(\widetilde{Z}_{n_3})^{a_3}{\cal P}_b(Z_{1,3}){\cal P}_b(Z_{2,3})\Bigr].
\end{split}    
\end{equation}
Notice that ${\cal P}_b(Z_{1,3}){\cal P}_b(Z_{2,3})$ represents the pairing between $Z_{1,2}$ and $\widetilde{Z}_{n_3}$. For $a_1=a_2=a_3=q+1$ and $b=q-1$, this can be further rewritten as
\begin{equation}\hspace{-0.2cm}
    {\cal A}_{3n}\Bigl(\Psi_{\mathrm{Pf}}^q(Z_{1,2}) \Psi_L(\widetilde{Z}_{n_3})^{q+1}{\cal P}_{q-1}(Z_{1,3}){\cal P}_{q-1}(Z_{2,3})\Bigr).
\end{equation}
In particular, for the $333111$ state we have $q=2$ and the resulting wave function is
\begin{equation}
    {\cal A}_{3n}\Bigl(\Psi_{\mathrm{MR}}(Z_{1,2}) \Psi_L(\widetilde{Z}_{n_3})^{3}{\cal P}_{1}(Z_{1,3}){\cal P}_{1}(Z_{2,3})\Bigr).
\end{equation}

A natural question arises: which state among the ones with the same filling fraction corresponds to an incompressible fluid? Motivated by the analysis of Pfaffian and Hafnian families and other multilayer families of states we postulate the following conjecture.

{\it Conjecture:} ($K$-incompressibility) Given the filling factor $\nu$, the bulk root pattern corresponding to an infinite-size (translationally and rotationally invariant) interacting particle Hall system having the smallest Kolmogorov complexity and minimal total angular momentum $J$ is associated with an incompressible FQH fluid. 

According to this conjecture, the aforementioned $\nu=3/5$ state qualifies as a possible candidate for an incompressible FQHE fluid state. 

\section{Quasiparticle excitations}
\label{sec:Quasiparticles}

\subsection{Quasiholes}
\label{se:exc_qh}

Our next goal is to determine the quasihole and quasiparticle excitations of the states (vacua) obtained by the SMC presented in the previous section. For the MR state, the fundamental quasihole excitations were proposed to consist of two quasiholes \cite{MooreRead91,Greiter91} located at different positions $\eta_1, \eta_2$ in the plane, with corresponding wave functions
\begin{equation}
\label{eq:MReta}
\begin{split}
    &\Psi_{\mathrm{MR};\eta_1,\eta_2}^{2\mathrm{qh}}(Z_{2n})\equiv \Psi_{\mathrm{Pf};\eta_1,\eta_2}^{q=2, 2\mathrm{qh}}(Z_{2n})\\
    &=\!\mathrm{Pf}_{2n}\!\left(\!\frac{(z_i-\eta_1)(z_j-\eta_2)+(i\leftrightarrow j)}{z_i-z_j}\!\right)\! \Psi_L(Z_{2n})^2.
\end{split}    
\end{equation}
This object can be interpreted as a charge-$|e|/2$ vortex, and every single quasihole possesses a charge $-e/4$ ($e$ is the elementary charge of the fluid's particle, typically, the  electron) \cite{Toke}. The generalization of the above wave function to a system of arbitrary even number of quasiholes was also studied \cite{Nayak}, and the braiding properties of quasiholes in the MR state led to the concept of non-Abelian statistics.

Since Halperin's construction is a straightforward generalization of Laughlin's fluid to bilayer systems, it is natural to insert two Laughlin's quasiholes in the Halperin $\Psi_{331}$ state, one per layer, and conjecture that after  antisymmetrization with respect to all particle coordinates, one ends up with a system of two quasiholes in the MR fluid. This was observed in Ref.~\cite{Cappelli01}. Here we present explicit proof of this fact and generalize the procedure for multilayer systems. In particular, we consider both fermionic and bosonic Halperin bilayer states $\Psi_{q+1\, q+1\, q-1 }$ on an equal footing. 

The general quasihole construction proceeds as follows. We start with Halperin bilayer state $\Psi_{a_1a_2b}$ and obtain its (anti-)symmetrized version $\mathcal{O}\Psi_{a_1a_2b}$. Quasiholes for this vacuum can be generated as 
\begin{equation}
   \mathcal{O}\Bigl\{ \Psi^{a_1,m_1}_{L,\eta_1}(Z_n)\Psi_{L,\eta_2}^{a_2, m_2}(\,\overline{\! Z}_{n})\, {\cal P}_b(Z_{2n})\Bigr\},
\end{equation}
where
\begin{equation}
    \Psi^{a_l,m_l}_{L,\eta}(Z_n)=\prod\limits_{i=1}^n(z_i-\eta)^{m_l} \Psi_L(Z_n)^{a_l}
\end{equation}
represents a cluster of $m_l$ quasiholes at $\eta$ in the $\nu=\frac{1}{a_l}$ Laughlin's fluid in layer $l=1,2$. We now demonstrate that this construction indeed reproduces the MR quasiholes from \eqref{eq:MReta} ($m_1=m_2=1$).

Let us start motivating the proof with the simplest case of $N=2$ particles, one per layer and $m_1=m_2=1$. Then, $\Psi_{331}(Z_{2})=z_1-z_2$, and $\Psi_{331;\eta_1,\eta_2}^{2\mathrm{qh}}=(z_1-\eta_1)(z_2-\eta_2)(z_1-z_2)$. Its antisymmetrization leads to
\begin{equation}
\begin{split}
       &\mathcal{A}_{2n}\Psi_{331;\eta_1,\eta_2}^{2\mathrm{qh}}(Z_2)\\
       &=\frac{1}{2} \bigl[(z_1-\eta_1)(z_2-\eta_2)+(z_1-\eta_2)(z_2-\eta_1)\bigr](z_1-z_2).
\end{split}       
 \end{equation}
On the other hand, the wave function for the two-quasihole system in the MR state, as proposed in Ref.~\cite{MooreRead91}, is given by \eqref{eq:MReta} with $n=1$. Since $\mathbb{Z}_2$ has only two elements, the Pfaffian reduces to $\frac{(z_1-\eta_1)(z_2-\eta_2)+(z_2-\eta_1)(z_1-\eta_2)}{z_1-z_2}$, while the Laughlin's factor is simply $(z_1-z_2)^2$. Therefore the resulting wave function for two quasiholes in the MR state coincides (up to an irrelevant constant factor) with the one obtained from the construction above.

Consider now the Pfaffian state with any even number of particles $N=2n$. For $m_1+m_2=m$ Laughlin  quasiholes in Halperin bilayer state $\Psi_{q+1\, q+1 \, q-1}$, with $m_l$ quasiholes inserted in layer $l=1,2$, the wave function is given by
\begin{equation}
\begin{split}
    \Psi_{q+1\, q+1\, q-1;\, \eta_1,\eta_2}^{(m_1,m_2)}(Z_{2n})&\!=\!\varepsilon_n\!\prod\limits_{i=1}^n(z_i-\eta_1)^{m_1}\!\!\prod\limits_{k=1}^n(z_{n+k}-\eta_2)^{m_2}\\
    &\times{\rm det}_{n}\!\left(\frac{1}{z_i-z_{n+k}}\right)\Psi_L(Z_{2n})^q,
\end{split}    
\end{equation}
Its antisymmetrization leads to the Pfaffian state,
\begin{equation}
\label{eq:PfQhP}
\begin{split}
&\mathcal{A}_{2n}\Biggl[\prod\limits_{i=1}^n(z_i-\eta_1)^{m_1}\prod\limits_{k=1}^n(z_{n+k}-\eta_2)^{m_2}{\rm det}_{n}\!\left(\!\frac{1}{z_i-z_{n+k}}\!\right)\!\Biggr]\\
&=\!-\frac{(n!)^2}{(2n)!} \, \mathrm{Pf}_{2n}\!\left(\!\frac{(z_i-\eta_1)^{m_1}(z_j-\eta_2)^{m_2}+(i\leftrightarrow j)}{z_i-z_j}\!\right)\!.
\end{split}    
\end{equation}
For details of the proof see Appendix~\ref{app:profPfQh}. This means that the addition of $m_1=m_2=1$ Laughlin quasiholes in layers of the Halperin bilayer state $\Psi_{q+1\ q+1\, q-1}$ leads to quasiholes in the Pfaffian state consistent with the proposal for the MR fluid.

\begin{figure*}[htb]
\centering
\includegraphics[scale=1.25]{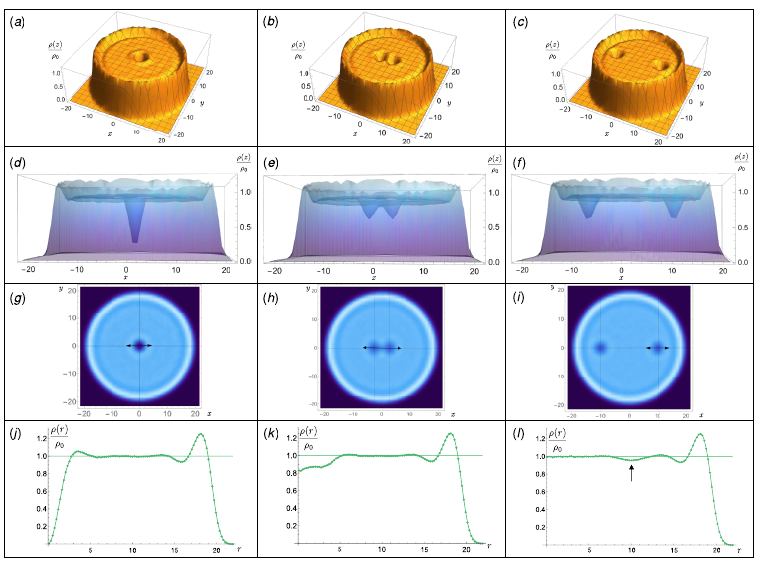}
\caption{{\bf [(a)--(c)]} Density profiles $\rho(z)$ (in units of the uniform density $\rho_0=\frac{\nu}{2\pi}$) for the MR system with two paired quasiholes located at $(\eta_1,\eta_2)=(0,0)$, $(-2.5\ell, 2.5\ell)$ and $(-10\ell,10\ell)$, respectively. {\bf [(d)--(f)]} Density profiles from {\bf (a)--(c)}, respectively, with reduced opacity. {\bf [(g)--(i)]} Contour plots for density profiles from {\bf (a)--(c)}, respectively. The size of the corresponding excitations is marked by arrows. {\bf [(j)-(l)]} Normalized radial densities $\rho(r)$ from the panels above. 
} 
\label{fig:01}
\end{figure*}

We performed variational Monte Carlo simulations to visualize a Pfaffian fluid with quasiholes. In order to implement this efficiently, we used the algorithm discussed in detail in Appendix~\ref{app:Pfaffiansimul}. For a system with $50$ particles per layer, we study three situations that differ by the positions of the two quasiholes. Lengths are measured in units of the magnetic length $\ell=\sqrt{\frac{\hbar c}{\vert e\vert  B}}$, where $B$ is the magnetic field strength, $c$ the speed of light, and $\vert e\vert $ the magnitude of the elementary charge. In each case, we place them symmetrically, at positions $(\eta_1,\eta_2)=(0,0)$, $(\eta_1,\eta_2)=(-2.5\ell,2.5\ell)$ and $(\eta_1,\eta_2)=(-10\ell,10\ell)$, respectively. We will use units s.t. $\ell=1$. The resulting three-dimensional density plots for these systems are shown in Figs.~\ref{fig:01}(a)--\ref{fig:01}(c). To improve visualization of the characteristic features of these systems, we reduce the opacity and present these data in panels (d)--(f), respectively, as well as we provide in panels (g)--(i) the two-dimensional projection of the corresponding density plots. Panels (j)--(l) show the corresponding radial densities. We remark that in those cases where the quasiholes are not located at the origin [cases (k) and (l)], the lack of cylindrical symmetry influences the depth of the valleys present in these plots.
 
To determine the value of the charge of such excitations, we performed Monte Carlo simulations for a system with $100$ particles per layer, located two quasiholes at the origin (to take advantage of the circular symmetry), see Fig. 3, and and made use of the expression
\begin{equation}
\label{eq:charge}
    \delta\rho=2\pi \int_0^{r_{\mathrm{cut-off}}}\left[\rho(r) - \rho_{\mathrm{Pf}}(r) \right]rdr,
\end{equation}
where $\rho(r)$ stands for the radial density with quasiholes and $\rho_{\mathrm{Pf}}(r)$ is the radial density of the homogeneous Pfaffian fluid. The cutoff radius $r_{\mathrm{cut-off}}$ has to be chosen such that the entire quasihole is contained within a disk of this radius and, simultaneously, its value must be sufficiently smaller than the size of the droplet to avoid influence from the boundary of the droplet \cite{Kivelson}. Here we choose $0\leq r_{\mathrm{cut-off}}\leq 20\ell$, and observe the saturation of $\delta\rho$ at $|e|/2$, as expected for the MR state with two quasiholes at the same position. (The mean value computed in the range $10\ell\leq r_{\mathrm{cut-off}}\leq 20\ell$ is $0.496(1)|e|$.) Since the two dips in the density profile present in the state $\Psi_{\mathrm{MR};\eta_1,\eta_2}^{2\mathrm{qh}}(Z_{2n})$ (with positions $\eta_1$ and $\eta_2$ well separated) are identical, each of them possesses the elementary charge $|e|/4$. 

\begin{figure}[htb]
\centering
\includegraphics[width=\columnwidth]{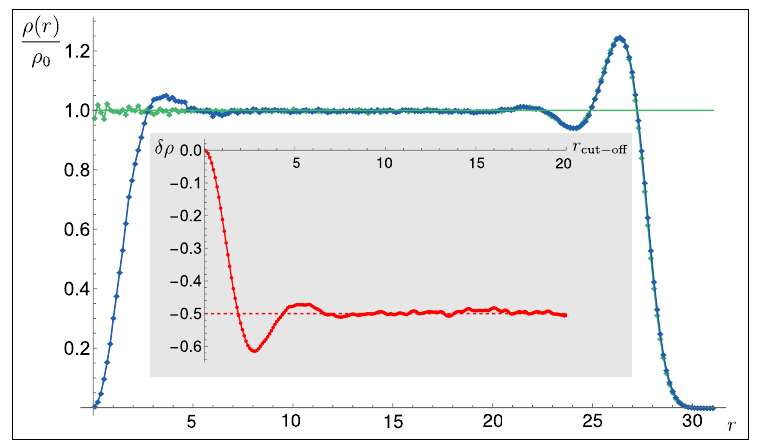}
\caption{Normalized radial density for the MR system with $2n=200$ particles (green) and for the two paired quasiholes located at $\eta_1=\eta_2=0$ (blue). {(\bf Inset)} Computation of the charge $\delta \rho$ as a function of $r_{\mathrm{cut-off}}$, according to \eqref{eq:charge}.} 
\label{fig:02}
\end{figure}

\begin{figure}[htb]
\centering
\includegraphics[width=\columnwidth]{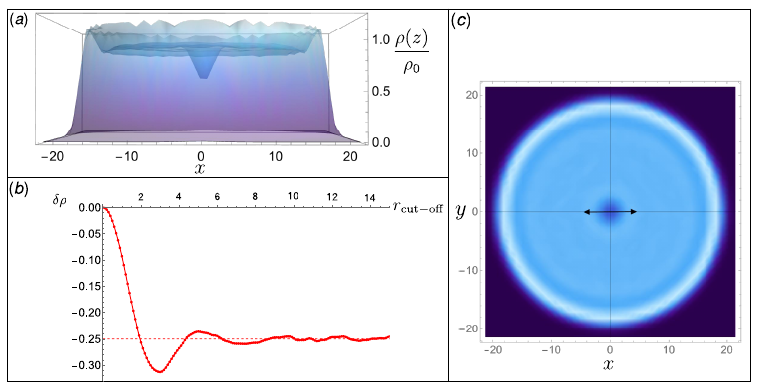}
\caption{Single quasihole \eqref{eq:single} located at $\eta=0$ for the MR fluid with $2n=100$ particles. {\bf(a)} The three-dimensional normalized density profile (as a function of $z=x+iy$), with reduced opacity in the figure. {\bf(b)} The charge $\delta \rho$ as a function of $r_{\mathrm{cut-off}}$, computed according to Eq. \eqref{eq:charge}. {\bf(c)} Contour plot with the size of the excitation indicated.
} 
\label{fig:03}
\end{figure}

\begin{figure}[htb]
\centering
\includegraphics[width=\columnwidth]{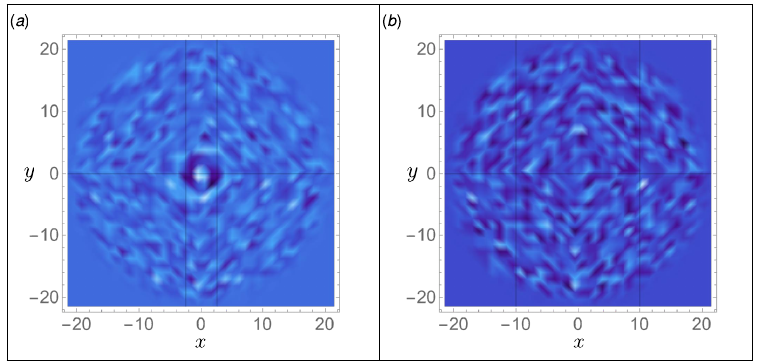}
\caption{Difference in densities (as a function of $z=x+iy$) between the state of two isolated quasiholes, each of them defined by \eqref{eq:single}, and the one with paired quasiholes obtained according to the (anti-)symmetrization procedure for {\bf (a)} quasiholes located at $(\eta_1,\eta_2)=(-2.5\ell,2.5\ell)$ and {\bf (b)} quasiholes located at $(\eta_1,\eta_2)=(-10\ell,10\ell)$.} 
\label{fig:04}
\end{figure}

It has been postulated \cite{Wan08} that a single quasihole excitation in the MR fluid can be effectively described by the wave function
\begin{equation}
\label{eq:single}
    \Psi_{\mathrm{MR};\eta}^{1\mathrm{qh}}=\mathrm{Pf}_{2n}\left(\frac{z_i+z_j-2\eta}{z_i-z_j}\right)\Psi_L(Z_{2n})^2.
\end{equation}
We have performed a Monte Carlo simulation for this system with $2n=100$ particles and $\eta=0$. The resulting density profiles as well as the computation of the quasihole charge are shown in Fig.~\ref{fig:03}. The latter saturates at $\delta \rho=|e|/4$. (The mean value computed in the range $7\ell\leq r_{\mathrm{cut-off}}\leq 14\ell$ is $0.251(1)|e|$.) We also notice that the size of this single quasihole agrees with the size of each of the two quasiholes that form the paired state $\Psi_{\mathrm{MR};\eta_1,\eta_2}^{2\mathrm{qh}}$ when they are well-separated (e.g., in the case with $\eta_1=-\eta_2=10\ell$), as well as with the size of the excitation formed when $\eta_1=\eta_2=0$. To quantitatively check this fact, we performed Monte Carlo simulations with $2n=100$ particles for $\Psi_{\mathrm{MR};\eta_1}^{1\mathrm{qh}}+\Psi_{\mathrm{MR};\eta_2}^{1\mathrm{qh}}-\Psi_{\mathrm{MR}}$, when $(\eta_1,\eta_2)=(-2.5\ell, 2.5\ell)$ and $(\eta_1,\eta_2)=(-10\ell, 10\ell)$, and analyzed the difference from the density corresponding to $\Psi_{\mathrm{MR};\eta_1,\eta_2}^{2\mathrm{qh}}$. The results are shown in Fig.~\ref{fig:04}.

We notice that when $(\eta_1,\eta_2)=(-2.5\ell, 2.5\ell)$ we clearly see a substantial discrepancy near the origin. However, when the two paired quasiholes are well-separated (i.e., for $(\eta_1,\eta_2)=(-10\ell, 10\ell)$), the difference is below $3\%$, which is within statistical uncertainty. 
In other words, from a charge density perspective, the MR state with well-separated quasiholes in the paired state behaves like the one with two independent single quasiholes, each of them described by \eqref{eq:single}. This seemingly surprising observation can be understood from the perspective of Halperin bilayer systems. In the original MR construction quasiholes in Pfaffian fluids were always appearing in pairs. However, allowing for arbitrary combinations of quasiholes in the two layers, we see that the wave function given by \eqref{eq:single} is nothing else than the (anti-) symmetrization of the system with a single Laughlin's quasihole placed in only one of the two layers (arbitrarily chosen). In other words, the system of two paired but well-separated quasiholes is densitywise equivalent to the one with two single quasiholes, both systems obtained from the bilayer construction. Indeed, our quasihole construction is flexible enough to accommodate all possible combinations of Laughlin's quasiholes per layer.  

\begin{figure}[htb]
\centering
\includegraphics[width=\columnwidth]{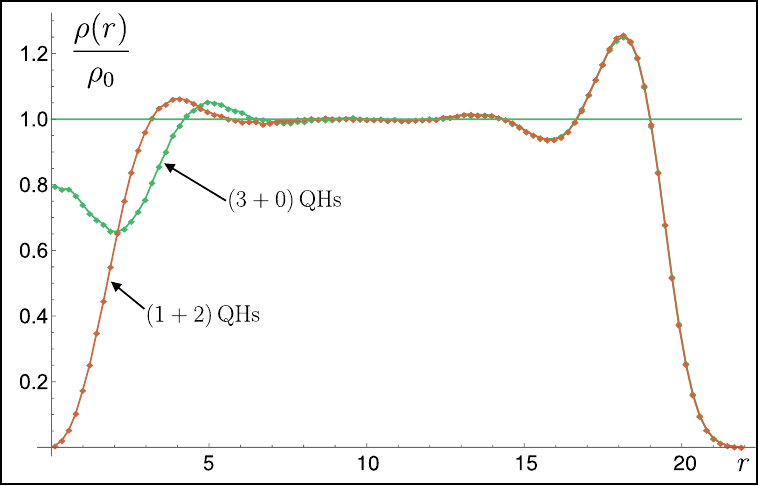}
\caption{Normalized radial density for the MR state with $50$ particles per layer and with different distribution of $3$ quasiholes: all of them in one layer only vs one of them in the first layer and the remaining two in the second one.} 
\label{fig:06}
\end{figure}

\begin{figure}[htb]
\centering
\includegraphics[width=\columnwidth]{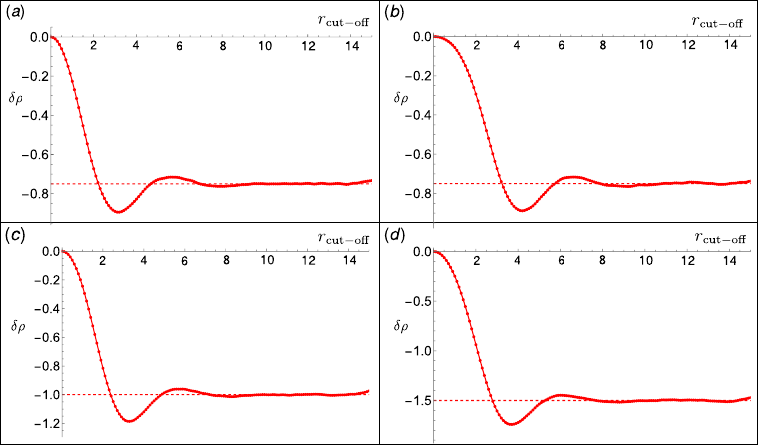}
\caption{Computation of the charge of quasiholes $\delta \rho$ as a function of $r_{\mathrm{cut-off}}$, according to \eqref{eq:charge}, for the MR system with $2n=100$ particles. {\bf (a)} One quasihole in the first layer and two of them in the second one, {\bf (b)} three quasiholes in a single layer and no quasiholes in the other, {\bf (c)} two quasiholes per layer, and {\bf (d)} three quasiholes per layer. All quasiholes are located at the origin of the disk. }
\label{fig:07}
\end{figure}

\begin{figure}[htb]
\centering
\includegraphics[width=\columnwidth]{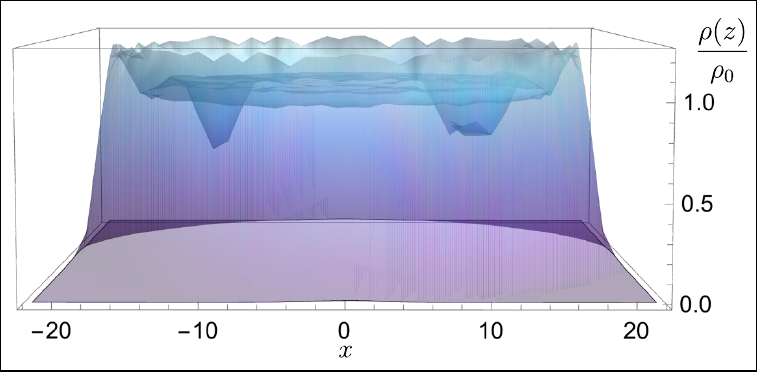}
\caption{Two quasiholes in the MR fluid with $2n=100$ particles, obtained from a Halperin bilayer system with one quasihole in the first layer placed at $\eta_1=-10\ell$ and two in the second one at $\eta_2=10\ell$. The two dips corresponding to different number of quasiholes at a given position differ in size.}
\label{fig:11}
\end{figure}

For different distributions of quasiholes per layer in the MR fluid, we performed Monte Carlo simulations with $2n=100$ particles and $\eta_1=\eta_2=0$, obtaining density profiles and their corresponding charges. In Fig.~\ref{fig:06}, we show that the case of three quasiholes in a single layer differs significantly from the one with two of the quasiholes in the first layer and the remaining one inserted in the second layer. However, both of them lead to a charge $3|e|/4$, as expected for a system of three quasiholes at the same position, each of charge $|e|/4$ -- see Fig.~\ref{fig:07}. We observe the following additive property for the charges of quasiholes: the total charge of an excitation at $\eta$ consisting of a given number of quasiholes (regardless of its distribution between layers) is the multiplicity of the charge of a single excitation. Furthermore, for quasiholes obtained from the (anti-)symmetrization of a Halperin bilayer system with one quasihole in the first layer and two quasiholes in the second one, if the positions of the quasiholes in particular layers are different (and they are well-separated), then the quasihole charge located at $\eta$ is proportional to the exponent associated to that position. We illustrate this point on the MR state with $2n=100$ particles and the aforementioned distribution of quasiholes at $\eta_1=-10\ell$ and $\eta_2=10\ell$ -- see the resulting density plot in Fig.~\ref{fig:11}.

However, agreement in density profiles does not necessarily mean that the system of well-separated paired quasiholes and the ones obtained as a superposition of two single ones are indeed fully equivalent. For that to be true, one has to also examine the Berry connection and compute the braiding statistics, which is beyond the scope of the current paper. In a rigorous proof, one should demonstrate that for $|\eta_1-\eta_2|\gg 1$ the Pfaffian factor in $\Psi_{\mathrm{MR};\eta_1,\eta_2}^{2\mathrm{qh}}$ splits into two pieces, each of them corresponding to a single quasihole. The subleading term for the computation of the charge might become crucial for examining the statistics. We postpone rigorous study of these aspects for future research. Here we concentrate only on the charge density aspects. 

The above quasihole construction can be applied to any member of the Pfaffian and Hafnian families of states as well as any states obtained from Halperin bilayer systems. In particular, notice that
\begin{equation}
\label{eq:hafQH}
\begin{split}
    &\mathcal{S}_{2n}\!\Biggl[\!\prod\limits_{i=1}^n(z_i-\eta_1)(z_{n+i}-\eta_2)\prod\limits_{i<j}^n(z_i-z_j)^4(z_{n+i}-z_{n+j})^4\!\Biggr]\\
    &=\!\frac {(n!)^2}{(2n)!}\mathrm{Hf}_{2n}\!\left[\!\frac{(z_i-\eta_1)(z_j-\eta_2)+(i\leftrightarrow j)}{(z_i-z_j)^2}\!\right]\!\Psi_L(Z_{2n})^2\!,
\end{split}
\end{equation}
and, therefore, the symmetrization or antisymmetrization of the Halperin's state $\Psi_{q+2\, q+2\, q-2}$ with quasiholes inserted one per layer leads to the system of two quasiholes in the Hafnian state:
\begin{equation}
    \Psi_{\mathrm{Hf};\eta_1,\eta_2}^q\!\!=\!\mathrm{Hf}_{2n}\!\left[\!\frac{(z_i-\eta_1)(z_j-\eta_2)+(i\leftrightarrow j)}{(z_i-z_j)^2}\!\right]\!\Psi_L(Z_{2n})^q. 
\end{equation}
In order to prove the above claim \eqref{eq:hafQH} we follow exactly the same steps as for the Hafnian state without quasiholes --- see Appendix~\ref{app:proof2} where we consider the more general case of $m_1+m_2=m$ quasiholes.

\subsection{Fusion mechanism for quasielectrons}

In this section, we show how to generate quasiparticle (quasielectron) excitations in topological fluids derived from our SMC. The idea behind the generation of these quasiparticles is rooted in the fusion mechanism discovered in Ref.~\cite{QP22}, which is based on the algebraic concept of particle fractionalization \cite{QP22}. For a Laughlin fluid with filling factor $\nu=\frac{1}{a}$ the analytical form of the quasielectron wave function was rigorously derived in Ref.~\cite{QP22} from the requirement that placing $a$ of them at the same position is equivalent to localizing a single electron. The resulting quasielectron is effectively a composite object, made out of $a-1$ quasiholes and 1 electron, possessing the right quantum numbers (charge and exchange statistics). Furthermore, it can be shown that it also has the correct value of the {\it topological spin} introduced in Ref.~\cite{Comparin2022} and satisfies the spin-statistics relation in the sense of Ref.~\cite{Nardin2023}.

Here we make use of the latter fusion mechanism and construct quasiparticle excitations in properly \mbox{(anti-)}symmetrized Halperin multilayer systems. In other words, a quasielectron is defined by placing a localized bare electron on top of $m$ elementary quasiholes with the constraint that the quasielectron density satisfies the fusion rule: $\rho_{\mathrm{qp}}(z)=m\rho_{\mathrm{qh}}(z)+\rho_e(z)$, where $\rho_e(z)$ is the density of a bare electron. This leads to an ``effective plasma analogy'' \cite{QP22}: Since the quasielectron emerges out of the fusion of a bare electron with a cluster of quasiholes, and for the latter there exists a (local) plasma analogy, we can infer its universal characteristics.

Based on the properties of quasiholes observed in Sec.~\ref{se:exc_qh}, we consider next a single isolated quasiparticle excitation. To determine the number $m$ of elementary quasiholes required to generate a quasiparticle of a given charge $e_{\mathrm{qp}}$, we notice that the elementary charge of a single quasihole in a topological fluid with $\nu=\frac{2}{a+b}$ is $e^\ast=-\frac{e}{a+b}$. This implies the following relation:  
\begin{equation}
   me^\ast +e=e_{\mathrm{qp}}.
\end{equation}
For quasiparticles having charges $e_{\mathrm{qp}}=-d \, e^\ast$ with some positive integer $d$, this condition leads to $m=a+b-d>0$. 
Since $m$ quasiholes can be distributed among the various layers in many different ways, there are several potentially inequivalent choices leading to the same total charge. Let $m_l$ denote the number of quasiholes placed in the $l$th layer. For the sake of clarity, we next focus on the case of two layers. 

In order to compactly represent our construction, we denote by $\begin{bmatrix} Q_1 \\ Q_2\end{bmatrix}_b$ the bilayer system with quasihole excitation of type $Q_1$ in the first layer, and of type $Q_2$ in the second one. For Laughlin's fluids with filling factor $\frac{1}{a}$ we denote by $QH_{a}(\eta)$ its corresponding quasihole (located at $\eta$) wave function. In particular, we have observed before that in the limit $|\eta_1-\eta_2|\gg 1$ the following holds 
\begin{equation}
    \mathcal{O}\begin{bmatrix}
       m_1 QH_a(\eta_1) \\ m_2 QH_a(\eta_2)
    \end{bmatrix}_b=m_1 QH(\eta_1)\sqcup m_2 QH(\eta_2),
\end{equation}
with $\mathcal{O}$ representing either a symmetrization or antisymmetrization operation depending on the parity of $a$, and $QH(\eta)$ defining a single quasihole (located at $\eta$) wave function of the fluid obtained from the bilayer system following our construction. Here, the disjoint union symbol $\sqcup$ stresses the fact that the two quasiholes are well-separated. Note that, in general, for $m_1+m_2=m_1'+m_2'$
\begin{equation}
    \mathcal{O}\begin{bmatrix}
       m_1 QH_a(\eta) \\ m_2 QH_a(\eta)
    \end{bmatrix}_b\neq \mathcal{O}\begin{bmatrix}
       m_1' QH_a(\eta) \\ m_2' QH_a(\eta)
    \end{bmatrix}_b.
\end{equation}
Furthermore, the fusion mechanism from \cite{QP22} can be symbolically written as 
\begin{equation}
\label{eq:fusionL}
    QE_a(\eta)=(a-1)QH_a(\eta) \oplus e(\eta),
\end{equation}
with $\oplus$ representing the operation of fusion of the corresponding objects.

We now postulate the following fusion mechanism for a single quasiparticle (quasielectron) of charge $e_{\mathrm{qp}}=me^\ast+e$ in arbitrary fluids obtained from bilayer systems
\begin{equation}
    QE(\eta)=\mathcal{O}\begin{bmatrix}
        m_1 QH_a(\eta) \\ m_2 QH_a(\eta)
    \end{bmatrix}_b\oplus e(\eta),
\end{equation}
where $m_1+m_2=m$. Using the parametrization $a=q+s$, for $s=1$, the corresponding wave function is
\begin{equation}
\begin{split}
    &\Psi_{\mathrm{Pf};\eta}^{q;(m_1,m_2)}(Z_{2n+1})=\mathcal{O}_{2n+1}\Bigl\{e^{\frac{z_{2n+1}\eta^\ast}{2}}\\
    &\hspace{0.2cm}\times\!\mathrm{Pf}_{2n}\!\!\left[\frac{(z_i-\eta)^{m_1}(z_j-\eta)^{m_2}+(i\leftrightarrow j)
    }{z_i-z_j}\right]\!\!\Psi_L(Z_{2n})^q\Bigr\},
\end{split}    
\end{equation}
while for $s=2$ we have
\begin{equation}
\begin{split}
    &\Psi_{\mathrm{Hf};\eta}^{q;(m_1,m_2)}(Z_{2n+1})=\mathcal{O}_{2n+1}\Bigl\{e^{\frac{z_{2n+1}\eta^\ast}{2}}\\
    &\hspace{0.2cm}\times\!\mathrm{Hf}_{2n}\!\!\left[\frac{(z_i-\eta)^{m_1}(z_j-\eta)^{m_2}+(i\leftrightarrow j)
    }{(z_i-z_j)^2}\right]\!\!\Psi_L(Z_{2n})^q\Bigr\}.
\end{split}    
\end{equation}

The choice of $m_1$ and $m_2$ whose sum is $m$ is obviously not unique. We illustrate the above construction considering the MR state. Since this state is realized by the antisymmetrization of  Halperin's $\Psi_{331}$ state ($q=2$), the elementary charge is $e/4$, and for $d=1$ we have $m=3$. Hence there are two potential distributions of quasiholes: (a) two of them in the first layer and the remaining one in the second one, or (b) all three quasiholes in a single layer. It turns out that only case (a) leads to the correct fusion of charge densities. To numerically prove this statement, we performed Monte Carlo simulations for a small number of particles, a stringent test.

\begin{figure}[htb]
\centering
\includegraphics[width=\columnwidth]{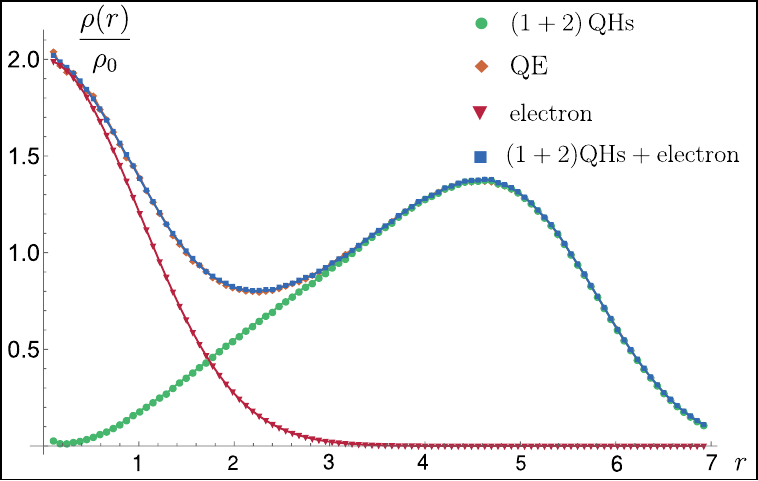}
\caption{Test of the fusion mechanism for the MR state with $n=5$ particles per layer.  The (normalized) radial density of the composite charge-$e/4$ quasielectron satisfies the required fusion rule: it agrees with the sum of the density of a state obtained from the bilayer system with two quasiholes in one layer and the remaining one in the second layer, and the density of a single electron. All quasiholes as well as the electron are placed at the origin of the disk. } 
\label{fig:08}
\end{figure}

In Fig.~\ref{fig:08} we present the corresponding (normalized) density profiles for a system with $n=5$ particles per layer and $(m_1,m_2)=(1,2)$ quasiholes located at $\eta=0$ (a single quasiparticle requires, by construction, one additional particle, i.e., a system with $11$ particles in total). Figure~\ref{fig:08} clearly shows the fusion mechanism at work, i.e., the QE and (1+2)QHs+electron are indistinguishable. 

\begin{figure}[htb]
\centering
\includegraphics[width=\columnwidth]{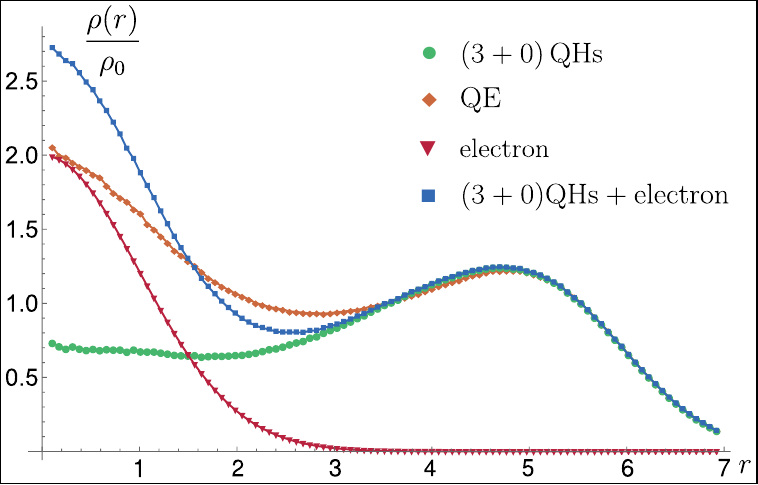}
\caption{Test of the fusion mechanism for the MR state with $n=5$ particles per layer. The potential quasielectron is constructed on top of a Halperin bilayer system with three quasiholes in a single layer. All quasiholes as well as the electron are placed at the origin of the disk. The fusion rule is clearly violated in this case.} 
\label{fig:09}
\end{figure}

Similar computations show also a violation of the fusion rule for case (b), i.e., $(m_1,m_2)=(3,0)$ -- see Fig.~\ref{fig:09}. The significant discrepancy originates from the fact that, in this case, the system of quasiholes has a high density at the origin (see also Fig.~\ref{fig:06} for results obtained for a larger system with no impact from the boundary and a density at the origin of $\sim 0.8$). Adding a Gaussian distribution representing the (properly normalized) radial density of a single electron, $\rho_e(r)=2e^{-r^2/2}$, to the cluster density of quasiholes (three quasiholes in the same layer), does not match the density of this type, $(m_1,m_2)=(3,0)$, of quasielectron excitation.

\begin{figure}[htb]
\centering
\includegraphics[width=\columnwidth]{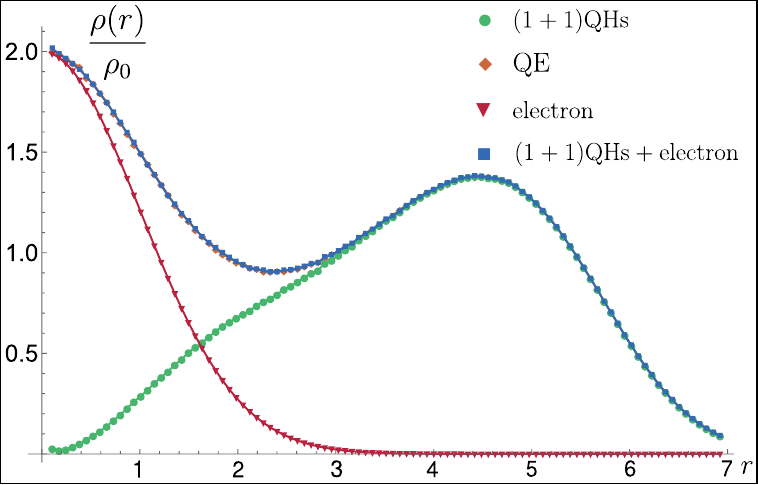}
\caption{Test of the fusion mechanism for the MR state with $n=5$ particles per layer. The (normalized) radial density of the composite charge-$e/2$ quasielectron satisfies the required fusion rule: it agrees with the sum of the density for a state obtained from the bilayer system with one quasihole per layer and the density of a single electron. All quasiholes as well as the electron are placed at the origin of the disk.} 
\label{fig:10}
\end{figure}

\begin{figure}[htb]
\centering
\includegraphics[width=\columnwidth]{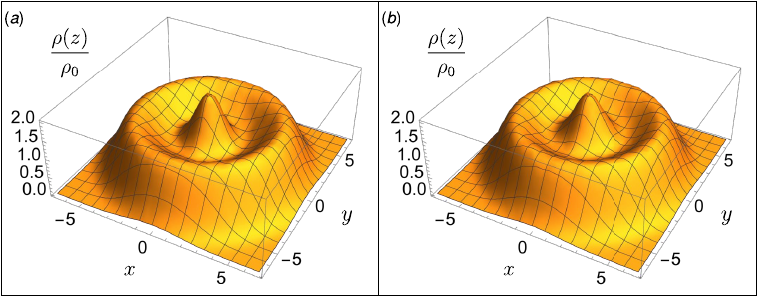}
\caption{Density profiles $\rho(z)$ for {\bf (a)} a charge-$e/4$ quasielectron and {\bf (b)} a charge-$e/2$ quasielectron. Monte Carlo simulations were performed for a system with $n=5$ particles per layer. The two types of excitations have identical amplitudes at the origin and, therefore, 
the charge-$e/2$ quasielectron exhibits a broader density profile. } 
\label{fig:12}
\end{figure}

The charge-$e/4$ quasielectron in the MR state is then constructed in a nontrivial way. We have to first realize, in a very specific way ($(m_1,m_2)=(1,2)$), a cluster of quasiholes located at $\eta$ and then add a single electron at the same position $\eta$. Only particular values of $(m_1,m_2)$ satisfy the fusion mechanism. We claim that these are the proper quasiparticle excitations. Can one anticipate what are those proper $(m_1,m_2)$ values? Are all possible quasiparticle excitations in one-to-one correspondence with those $(m_1,m_2)$ satisfying the fusion mechanism? Consider again the case of the MR fluid. We may wonder whether a charge-$e/2$ quasielectron excitation for the MR state does comply with our fusion mechanism. To satisfy the latter we need in principle two quasiholes. Remarkably, it turns out that the choice $(m_1,m_2)=(1,1)$ does the job. In Fig.~\ref{fig:10} we demonstrate (for a small number of particles) that the resulting quasielectron indeed satisfies the fusion rule. 
Density profiles for both charge-$e/4$ and charge-$e/2$ quasielectrons, located at the origin, are displayed in Fig.~\ref{fig:12}. 
\footnote{One could naively think that  quasielectrons should be obtained by (anti-)symmetrizing Halperin bilayer systems with quasielectrons placed in each Laughlin layer. In the case of the fermionic  Halperin's $\Psi_{q+1\, q+1\, q-1}$ states (i.e., with $q$ even) the corresponding wave function would then be given (up to global normalization and a Gaussian factor) by the total antisymmetrization of
\begin{equation}
\begin{split}
&\mathcal{A}_n\!\!\!\left[e^{\frac{z_n \eta_1^\ast}{2}}\prod\limits_{k=1}^{n-1}(z_k-\eta_1)^q\!\!\!\!\!\!\prod\limits_{1\leq i< j\leq n-1}\!\!\!\!\!\!(z_i-z_j)^{q+1}\right]\\
    \times &\mathcal{A}_n\!\!\!\left[e^{\frac{z_{2n} \eta_2^\ast}{2}}\prod\limits_{k=1}^{n-1}(z_{n+k}-\eta_2)^q\!\!\!\!\!\!\!\prod\limits_{1\leq i< j\leq n-1}\!\!\!\!\!\!(z_{n+i}-z_{n+j})^{q+1}\right]\!\!\\
    &\times\prod\limits_{i,k=1}^n(z_i-z_{n+k})^{q-1}.
\end{split}    
\end{equation}
This construction, however, is not satisfactory. In particular, it is impossible to generate a single charge-$e/2$ quasielectron satisfying the fusion rule since, because of the Pauli exclusion principle, we cannot put two electrons at the same position. In contrast, our above construction provides both charge-$e/4$ and charge-$e/2$ quasielectrons satisfying the fusion rule.} 

\subsection{Fusion Mechanism for Magnetoexcitons}

Having established both quasiholes and quasielectrons in arbitrary LLL topological fluids, we now postulate a particular form of magnetoexciton: objects composed of these two types of excitations. We first generate, within our bilayer construction, clusters of quasiholes necessary to realize a quasielectron at $\eta_1$, and an additional quasihole at $\eta_2$. We then put an extra electron at $\eta_1$. In other words, this construction reads symbolically
   \begin{equation} \hspace*{-0.2cm}
\mbox{\it ME}(\eta_1,\eta_2)=    {\cal O}    \begin{bmatrix}
        m_1 QH_a(\eta_1)\oplus QH_a(\eta_2) \\ m_2 QH_a(\eta_1)
    \end{bmatrix}_b\oplus e(\eta_1).
   \end{equation}
The mathematical structure of this object allows also for efficient Monte Carlo simulations since the first summand is easily implementable due to the Cayley formula \cite{Cayley}. The above equality can be understood as an effective plasma analogy for magnetoexcitons. In particular, our findings provide a concrete proposal to resolve the issue anticipated in \cite{Ma22}: ``{\it Unfortunately, accurate and well-tested wave functions of the quasiparticle-quasihole excitations at $5/2$ are still lacking.}''

We now determine the magnetoexcitons that can exist in a fluid obtained from the SMC. Since for $|\eta_1-\eta_2|\gg 1$ we expect this system to reduce to $QE(\eta_1)\sqcup QH(\eta_2)$, the allowed values $(m_1,m_2)$ should be the ones for the quasielectrons satisfying the fusion mechanism. On the other hand, for $\eta:=\eta_1=\eta_2$, the system reduces to
\begin{equation}
    \mbox{\it ME}(\eta)=    {\cal O}    \begin{bmatrix}
        (m_1+1) QH_a(\eta)\\ m_2 QH_a(\eta)
    \end{bmatrix}_b\oplus e(\eta),
\end{equation}
therefore an additional constraint on $m_1$ arises: $0\leq m_1+1<a$. Moreover, the counting of total charge introduces yet another constraint,
\begin{equation}
    m_1+m_2=a+b-1.
\end{equation}
For example, for the MR state ($a=3$, $b=1$) the latter implies that charge-$e/2$ quasielectrons cannot form magnetoexcitons since they originate from $(m_1,m_2)=(1,1)$. The only possibilities are $(m_1,m_2)=(2,1),(1,2)$, but the former fails to satisfy $m_1+1<3$. Therefore, for the MR state the magnetoexcitions result from charge-$e/4$ quasielectrons with $(m_1,m_2)=(1,2)$ \footnote{We remark that quasielectrons resulting from taking $(m_1,m_2)$ or $(m_2,m_1)$ quasiholes in the two layers are identical. What really matters for the construction of magnetoexcitons is where the additional quasihole is placed. This determines which combination out of these two should be taken in this construction. One can equally define the magnetoexciton by reverting the two layers. This will lead to the same final state.}. The wave function of this magnetoexciton (located at the origin) is therefore of the form
\begin{equation}
\label{eq:mext}
    \Psi_{\mathrm{MR};0}^{\mathrm{me}}\!\propto\! \mathcal{A}_{2n+1}\Biggl\{
    \mathrm{Pf}_{2n}\!\left(\frac{2z_i^2z_j^2}{z_i-z_j}\!\right)\!\Psi_L(Z_{2n})^2
    \Biggr\}.
\end{equation}
We remark that it is a generic situation that only quasielectrons of the most elementary charge can merge with a quasihole into a single magnetoexciton due to charge counting.

\subsection{Comparison with local exclusion conditions}
In Ref.~\cite{Yang2019}, an alternative characterization of QH states was introduced, leading to the so-called local exclusion conditions (LEC). In the LEC formalism, concrete predictions for the excitations in FQH fluids were made \cite{Yang2021}. Here, we briefly compare our quasiparticles' root states with the ones resulting from the LEC. We analyze first the case of a Laughlin fluid for which the quasielectron wave function was derived in Ref.~\cite{QP22}. The root pattern for the $\nu=1/3$ Laughlin fluid ($N=3$ particles) reads $100100100$, and adding two quasiholes at the origin modifies the pattern into $001001001$. According to our fusion mechanism, Eq. \eqref{eq:fusionL}, the quasielectron is generated in this case by adding a bare electron on top of the cluster of two quasiholes. The resulting root pattern,  computed by maximizing the quantity $\Delta_J$ [c.f. \eqref{algorithm-root}], is $10100100100$. Comparing this with the pattern for the $N=4$ Laughlin state, $100100100100$, we notice that the quasielectron can be obtained by first adding two zeros from the left (adding two quasiholes), and then flipping the leftmost one into $1$ (adding one electron). This is consistent with our fusion mechanism \cite{QP22}. However, it is different from the excitation proposed in \cite{Yang2021}, where the order of these two operations is opposite, resulting in $11000100100$ (the so-called Type $(5,3)$ LEC-quasielectron \cite{Yang2021}).

Let us extend this analysis to the MR state and its excitations. The root pattern for the $N=4$ MR fluid, determined by the maximization of $\Delta_J$ reads $11001100$, while for the systems with $(2+1)$ and $(1+1)$ quasiholes in the two layers the root patterns are $01010101$ and $01100110$, respectively. For the charge-$e/4$ quasielectron, the corresponding root pattern reads $11010101$ ($N=5$), while for charge-$e/2$ we have $11100110$. We immediately see, by comparing the above patterns to those with quasiholes, that this procedure is again consistent with the fusion mechanism. In other words, first produce a cluster of quasiholes and then add one electron, where the latter corresponds to flipping the leftmost $0$ into $1$. By comparing root patterns, we notice that our charge-$e/2$ quasielectron is identical to the Type $(5,3)$ LEC-QE studied in Ref.~\cite{Yang2021}. For a system with two quasiholes per layer in the MR state, the root pattern is $00110011$ ($N=4$), while for the magnetoexciton given by \eqref{eq:mext} it reads $10110011$, again,  consistent with our fusion mechanism. On the other hand, this excitation is different from the neutral one studied in Ref.~\cite{Yang2021}.  

Finally, we remark that for the (bosonic) Hafnian state $\Psi_{\mathrm{Hf}}^2$ of Eq. \eqref{eq:haf_def}, we expect the quasiparticles and magnetoexcitons to be characterized by the same parameters as in the case of the MR state $\Psi_{\mathrm{Pf}}^2$. This is motivated by the observation that $a+b=4+0=3+1$. 
We did not perform MC simulations to verify these statements; however, under this assumption, one can determine root patterns for all the excitations corresponding to the state $\Psi_{\mathrm{Hf}}^2$. In this case, for the bare Hafnian with $N=4$ the root pattern is $2000200$, while for its charge-$e/4$, charge-$e/2$ quasiparticles and magnetoexciton they read $1200002$, $1200020$, and $1020002$, respectively.

\section{$\!\!\!$Topological quantum~order:
Generalized Composite Operators} 
\label{sec:TQO}

Unveiling the intrinsic topological quantum order inherent in FQH fluids is fundamental for their classification. In this section, we make use of a second quantization representation to derive the composite (generalized Read) operators whose vacuum expectation signals the ODLRO of the fluid's correlations. In Ref. \cite{Mazaheri2015} the second-quantized version of Read's nonlocal (string) order parameter for the Laughlin sequence was explicitly derived, and in \cite{QP22} we expressed the sequence of (bosonic and fermionic) Laughlin second-quantized states in closed form. Here we extend these ideas to non-Abelian fluids. For simplicity, next we focus on the Pfaffian and Hafnian families of states, but we stress that similar ideas can be implemented in principle for other topological fluids. 
 
\subsection{Pfaffian state in second quantization}

We start investigating Pfaffian states in second quantization for both $q$ even and odd, i.e., we consider fermionic and bosonic fluids on an equal footing. We start with a simple fact known from linear algebra: the Pfaffian of a $(2n+2)\times (2n+2)$-matrix $A$ satisfies the recurrence relation,
\begin{equation}
    \mathrm{Pf}_{2n+2}(A)=\sum\limits_{j=2}^{2n+2}(-1)^j A_{1j}\mathrm{Pf}_{2n}(A_{\hat{1}\hat{j}}),
\end{equation}
where $A_{1j}$ is the entry $(1,j)$ of $A$, and $A_{\hat{1}\hat{j}}$ denotes the matrix obtained from $A$ by removing both first and $j$th row and column.

As an immediate consequence (for details see Appendix \ref{app:proofPf}), we get
\begin{equation}
\label{eq:PfF}
    \begin{split}
        &\Psi_{\mathrm{Pf}}^q(Z_{2n+2})=\sum\limits_{j=2}^{2n+2}(-1)^{j(1+q)}(z_1-z_j)^{q-1}\\
        &\times\!\!\!\prod\limits_{\substack{l\neq 1,j}}^{2n+2} \left((z_1-z_l)(z_j-z_l)\right)^q \Psi_{\mathrm{Pf}}^q((Z_{2n+2})_{\hat{1}\hat{j}}) .
    \end{split}
\end{equation}
We remark that for $q$ even the first summation corresponds to the antisymmetrization procedure, while for $q$ odd we have symmetrization. 
The operation $\prod\limits_{\substack{l\neq 1,j}} \left((z_1-z_l)(z_j-z_l)\right)^q$ on top of $\Psi_{\mathrm{Pf}}^q((Z_{2n+2})_{\hat{1}\hat{j}})$ is an insertion of $2q$ quasiholes, $q$ of them at position $z_1$ and the other $q$ at position $z_j$, into a state $\Psi_{\mathrm{Pf}}^q((Z_{2n+2})_{\hat{1}\hat{j}})$ of $2n$ particles, indexed by $2,\ldots, \hat{j}, \ldots, 2n+2$. We would like to express this flux insertion operation in the second quantization language.

In this language, we define the LLL single-particle orbitals in the disk geometry as $\phi_r(z)=\frac{z^r}{\mathcal{N}_r}$, $r\geq 0$, with normalization $\mathcal{N}_r=\sqrt{2\pi 2^r r!}$, so that the integration measure ${\cal D}[z]=d^2z e^{-\frac{1}{2}|z|^2}$ satisfies  $\int \mathcal{D}[z] (z^\ast)^{r'}z^r =\mathcal{N}_r \delta_{r,r'}$. 
We denote creation (respectively annihilation) operators by $a_r^\dagger$ (respectively $a_r$) whenever the formalism is applicable for both fermions and bosons. Otherwise, we use instead $c_r^\dagger$ (respectively $b_r^\dagger$) when we refer exclusively to fermions (respectively bosons).
 In second quantization, the field operator is defined as $\Lambda(z)=\sum\limits_{r\ge 0}\phi_r(z) a_r$ with adjoint $\Lambda^\dagger(z)$. 

To establish the algebraic mechanism of fractionalization, in Ref.~\cite{QP22} we used a compact representation of powers of the quasihole operator $\widehat{U}_N(z)$ in an $N$-particle system in terms of the operators (${\sf j} \in \mathbb{Z}$)
\begin{eqnarray}
    \widehat{S}^{\musSharp{}}_{{\sf j}}=\begin{cases}(-1)^{\sf j}\!\!\!\!\!\!\!\sum\limits_{n_1+\ldots + n_a={\sf j}}\!\!\!\!\!e_{n_1}\ldots e_{n_a}, \quad & {\sf j}\geq 0\\
    0, & \mbox{otherwise}
    \end{cases} ,
\end{eqnarray}
introduced in Refs.~\cite{Chen15, Chen19} to establish a recurrence relation between Laughlin's states ($\nu=\frac{1}{a}$) with different numbers of particles. Here, $e_n$ stands for the elementary symmetric polynomial of degree $n\geq 0$, represented in second quantization  as \cite{Mazaheri2015} 
\begin{eqnarray}
    e_n=\frac{1}{n!}\sum\limits_{r_1,\ldots,r_n} \a_{r_1+1}^\dagger\ldots \a_{r_n+1}^\dagger \a_{r_n}\ldots \a_{r_1}.
\end{eqnarray}
Since the number of particles for Pfaffian fluids is even, $N=2n$, the following follows from \cite[Lemma~1 in Supplementary Notes]{QP22}
\begin{equation}
    \widehat{U}_{2n}(z_1)^q=\sum\limits_{r\ge 0} z_1^r \widehat{S}^{\musSharp{}}_{2nq-r}.
\end{equation}

Using these operators in a mixed (first-and-second quantization) representation we can write Eq. \eqref{eq:PfF} as
\begin{equation}
\label{eq:norm}
\begin{split}
    &\Psi_{\mathrm{Pf}}^q(Z_{2n+2})=(-1)^q\sum\limits_{j=2}^{2n+2}(-1)^{j(1+q)}\\
    &\times \Biggl[\,\sum\limits_{r_1,r_2\ge 0} \sum \limits_{l=0}^{q-1}(-1)^{l+1}\binom{q-1}{l}z_1^{r_1+l}z_j^{q-1+r_2-l}\\
    &\times\widehat{S}^{\musSharp{}}_{2nq-r_1}\widehat{S}^{\musSharp{}}_{2nq-r_2}\Psi_{\mathrm{Pf}}^q((Z_{2n+2})_{\hat{1}\hat{j}})\Biggr],
\end{split}    
\end{equation}
where we have used the binomial expansion
\begin{equation}
    (z_1-z_j)^{q-1}\!=\!\sum\limits_{l=0}^{q-1}\binom{q-1}{l}(-1)^{q-1-l} z_1^l z_j^{q-1-l}.
\end{equation} 
Consequently, in a pure second quantization representation, Eq. \eqref{eq:norm} reads
\begin{equation}
\label{eq:rec_states}
\begin{split}
    &|\Psi_{\mathrm{Pf},2n+2}^q\rangle=\mathsf{N}_{n,q}\!\!\!\sum\limits_{r_1,r_2\ge 0} \sum \limits_{l=0}^{q-1}\Biggl[(-1)^{l+1}\binom{q-1}{l}\\&
    \times \a_{r_1+l}^\dagger\a_{q-1+r_2-l}^\dagger\widehat{S}^{\musSharp{}}_{2nq-r_1}\widehat{S}^{\musSharp{}}_{2nq-r_2}\Biggr]|\Psi_{\mathrm{Pf}, 2n}^q\rangle,
\end{split}    
\end{equation}
where the normalization factor $\mathsf{N}_{n,q}=\frac{(-1)^q}{2(n+1)}$ originates from the (anti-)symmetrization procedure. Indeed, for fermions ($q$ even) the variables $z_j$ with $j=2,\ldots, 2n+2$ are already properly antisymmetrized and it remains to choose $z_1$  out of the set of $2n+2$ variables, hence the factor $\frac{1}{2n+2}$. On the other hand, for the bosonic case ($q$ odd), we have $\mathsf{N}_{n,q}=-\frac{1}{2n+2}$. 

Let us illustrate the above relation with a concrete example, the MR state, i.e., the fermionic Pfaffian state with $q=2$. The four-particle state in second quantization is
\begin{equation}
    \left|\Psi_{\mathrm{Pf},4}^{q=2}\right\rangle=(\c_5^\dagger \c_4^\dagger \c_1^\dagger \c_0^\dagger - 2\c_5^\dagger \c_3^\dagger \c_2^\dagger \c_0^\dagger +10 \c_4^\dagger \c_3^\dagger \c_2^\dagger \c_1^\dagger)|0\rangle,
\end{equation}
while the one with six particles can be decomposed as
\begin{equation}
\begin{split}
    |&\Psi^{q=2}_{\mathrm{Pf},6}\rangle\!=\!\left(\c_9^\dagger\c_8^\dagger\c_5^\dagger\c_4^\dagger\c_1^\dagger\c_0^\dagger-2\c_9^\dagger \c_8^\dagger\c_5^\dagger\c_3^\dagger\c_2^\dagger \c_0^\dagger+10\c_9^\dagger\c_8^\dagger\c_4^\dagger\c_3^\dagger\c_2^\dagger\c_1^\dagger\right.\\
    &\left.-2\c_9^\dagger\c_7^\dagger\c_6^\dagger\c_4^\dagger\c_1^\dagger\c_0^\dagger +4\c_9^\dagger\c_7^\dagger\c_6^\dagger\c_3^\dagger\c_2^\dagger\c_0^\dagger+2\c_9^\dagger\c_7^\dagger\c_5^\dagger\c_4^\dagger\c_2^\dagger\c_0^\dagger
    \right.\\
    &\left. 
    -16\c_9^\dagger\c_7^\dagger\c_5^\dagger\c_3^\dagger\c_2^\dagger\c_1^\dagger-14\c_9^\dagger\c_6^\dagger\c_5^\dagger\c_4^\dagger\c_3^\dagger\c_0^\dagger+28\c_9^\dagger\c_6^\dagger\c_5^\dagger\c_4^\dagger\c_2^\dagger\c_1^\dagger\right.\\
    &\left.
    +10\c_8^\dagger\c_7^\dagger\c_6^\dagger\c_5^\dagger\c_1^\dagger\c_0^\dagger-16\c_8^\dagger\c_7^\dagger\c_6^\dagger\c_4^\dagger\c_2^\dagger\c_0^\dagger +28\c_8^\dagger\c_7^\dagger\c_6^\dagger\c_3^\dagger\c_2^\dagger\c_1^\dagger
    \right.\\
    &\left.
    +28\c_8^\dagger\c_7^\dagger\c_5^\dagger\c_4^\dagger\c_3^\dagger\c_0^\dagger-6\c_8^\dagger\c_7^\dagger\c_5^\dagger\c_4^\dagger\c_2^\dagger\c_1^\dagger -70 \c_8^\dagger\c_6^\dagger\c_5^\dagger\c_4^\dagger\c_3^\dagger\c_1^\dagger
    \right.\\
    &\left.
    +280\c_7^\dagger\c_6^\dagger\c_5^\dagger\c_4^\dagger\c_3^\dagger\c_2^\dagger\right)|0\rangle.
\end{split}    
\end{equation}
The right-hand side of the recurrence relation reduces to
\begin{equation}
\label{ex:RHS}
   \mathsf{N}_{2,2}\!\!\!\!\!\sum\limits_{r_1,r_2\geq 0}(-\c_{r_1}^\dagger \c_{r_2+1}^\dagger + \c_{r_1+1}^\dagger \c_{r_2}^\dagger)\widehat{S}^{\musSharp{}}_{8-r_1}\widehat{S}^{\musSharp{}}_{8-r_2}|\Psi_{\mathrm{Pf},4}^{q=2}\rangle.
\end{equation}
The $\widehat{S}^{\musSharp{}}_{\sf j}$ operators having a nontrivial action on four-particle states are
\begin{equation}
    \begin{split}
         &\widehat{S}^{\musSharp{}}_{8}=e_4^2, \quad \widehat{S}^{\musSharp{}}_{7}=-2e_3e_4, \quad \widehat{S}^{\musSharp{}}_{6}=e_3^2+2e_2e_4, \\
         & \widehat{S}^{\musSharp{}}_{5}=-2(e_1e_4+e_2e_3),\quad \widehat{S}^{\musSharp{}}_{4}=e_2^2+2e_4+2e_1e_3, \\
         &  \widehat{S}^{\musSharp{}}_{2}=2e_2+e_1^2, \quad \widehat{S}^{\musSharp{}}_{1}=-2e_1,\quad
    \widehat{S}^{\musSharp{}}_{0}=1.
    \end{split}
\end{equation}
By a straightforward computation, one can then easily verify that \eqref{ex:RHS} reduces to $\left|\Psi_{\mathrm{Pf},6}^{q=2}\right\rangle$, in complete agreement with  \eqref{eq:rec_states}.

Equation \eqref{eq:rec_states} gives us the recurrence relation between Pfaffian states with different numbers of particles, from which we immediately infer that
\begin{equation}
\label{eq:recPf}
    |\Psi_{\mathrm{Pf},2n}^q\rangle = K_{q,n-1}K_{q,n-2}\ldots K_{q,0}|0\rangle,
\end{equation}
where
\begin{equation}
\label{eq:Koperator}
\begin{split}
    K_{q,n}&=\mathsf{N}_{n,q}\!\!\!\sum\limits_{r_1,r_2\ge 0} \sum \limits_{l=0}^{q-1}\Biggl[(-1)^{l+1}\binom{q-1}{l}\\
    &\a_{r_1+l}^\dagger\a_{q-1+r_2-l}^\dagger\widehat{S}^{\musSharp{}}_{2nq-r_1}\widehat{S}^{\musSharp{}}_{2nq-r_2}\Biggr].
\end{split}    
\end{equation}

Let us now introduce the following operator
\begin{equation}
\label{eq:Composite}
\begin{split}
    \widehat{K}_{q,n}(\eta_1,\eta_2)&=\mathsf{N}_{n,q}(\eta_1-\eta_2)^{q-1}\Lambda^\dagger(\eta_1)\Lambda^\dagger(\eta_2)\\
    &\hspace{1cm}\times \widehat{U}_{2n}(\eta_1)^q \widehat{U}_{2n}(\eta_2)^q,
\end{split}    
\end{equation}
for which we can show (see Appendix \ref{app:proofPf}) that
\begin{equation}
\label{eq:Read}
    \int \mathcal{D}[\eta_1]\mathcal{D}[\eta_2]\widehat{K}_{q,n}(\eta_1,\eta_2)=K_{q,n}.
\end{equation}

In Ref. \cite{Toke}, it was proposed to study correlations in the MR fluid by computation of the correlation function $\langle \Psi_{\mathrm{MR}}| \Lambda^\dagger(z_1')\Lambda^\dagger(z_2')\Lambda(z_1)\Lambda(z_2)|\Psi_{\mathrm{MR}} \rangle$, written in terms of the bare field operators.\footnote{We adjust the notation to be consistent with the conventions we have chosen in the present paper.} The authors used a spherical geometry and placed the primed positions near the north pole of the sphere, and the unprimed ones near the south pole. Their numerical results suggest the vanishing of this parameter for the MR state. We claim (as originally claimed in the context of Laughlin fluids by Read \cite{Read}) that the proper correlation function signaling the ODLRO present in these topological fluids should consider the {\it dressed} field operators $\widehat{K}_{q,n}$ instead of the bare ones, i.e., those responsible for the topological order present in the fluid, when the system is gapped. 
Since bare field operators do not commute with the quasihole operators, this leads to the conclusion that the relative order of operators in the definition of $\widehat{K}_{q,n}$ is important. Indeed, Eq. \eqref{eq:Composite} defines {\it a dressed pair} rather than {\it a pair of dressed operators}. 

By close analogy to Ref.~\cite{QP22}, we consider the flux-number nonconserving quasihole operator for the Pfaffian fluids
\begin{equation}
    \widehat{\mathcal{U}}_{q,q}(z, w)\!=\!\sum\limits_{n\ge 0} \widehat{U}_{2n}(z)^q\widehat{U}_{2n}(w)^q|\Psi_{\mathrm{Pf},2n}^q\rangle \langle\Psi_{\mathrm{Pf},2n}^q|, 
\end{equation}
and define the operator
\begin{eqnarray}
\mathcal{K}_q=\int \mathcal{D}[\eta_1]\mathcal{D}[\eta_2] \ \widehat{\mathcal{K}}_q(\eta_1,\eta_2) \ , \mbox{ where }
\end{eqnarray}
\begin{equation}
       \widehat{\mathcal{K}}_q(\eta_1,\eta_2)\!=\!\mathsf{N}_{n,q} (\eta_1-\eta_2)^{q-1}\Lambda^\dagger(\eta_1)\Lambda^\dagger(\eta_2) \ \widehat{\mathcal{U}}_{q,q}(\eta_1,\eta_2) ,
\end{equation}
and satisfies (see Appendix \ref{app:proofPf})
\begin{equation}
\label{eq:Read2}
    \mathcal{K}_q|\Psi^q_{\mathrm{Pf},2n}\rangle=|\Psi^q_{\mathrm{Pf},2n+2 } \rangle .
\end{equation}

In this way, we can express the Pfaffian state in second quantization 
\begin{equation}
    |\Psi^q_{\mathrm{Pf},2n } \rangle = \mathcal{K}_q^n \ |0\rangle 
\end{equation}
as a nonlocal condensation, where $\ket{0}$ is the state with no particles. 

\subsection{Hafnian state in second quantization}

Given a $(2n+2)\times (2n+2)$ symmetric matrix $B$, the Laplace expansion for its Hafnian leads to the following recurrence formula
\begin{equation}
    \mathrm{Hf}_{2n+2}(B)=\sum\limits_{j=2}^{2n+2} B_{1j}\mathrm{Hf}_{2n}(B_{\hat{1}\hat{j}}).
\end{equation}
Following steps analogous to those considered for the Pfaffian case, we end up with
\begin{equation}
\begin{split}
        &\Psi_{\mathrm{Hf}}^q(Z_{2n+2})=\sum\limits_{j=1}^{2n+2}\Biggl[(-1)^{jq}(z_1-z_j)^{q-1}\\
        &\times \!\!\!\prod\limits_{k\neq 1,j}^{2n+2}\!\!\left((z_1-z_k)(z_j-z_k)\right)^q \Psi_{\mathrm{Hf}}^q((Z_{2n})_{\hat{1}\hat{j}})\Biggr].
\end{split}
\end{equation}
In contrast to Pfaffian states, $q$ odd corresponds now to fermions, while  $q$ even to bosons. This is in complete agreement with our previous discussion about generating Pfaffian- and Hafnian-like states from Halperin bilayer systems. 

Therefore, in the second quantization formalism, we have
\begin{equation}
\begin{split}
    &|\Psi_{\mathrm{Hf},2n+2}^q\rangle=\mathsf{N}_{n,q}\!\!\!\sum\limits_{r_1,r_2\ge 0} \sum \limits_{l=0}^{q-1}\Biggl[(-1)^{l+1}\binom{q-1}{l}\\
    &\times\a_{r_1+l}^\dagger\a_{q-1+r_2-l}^\dagger\widehat{S}^{\musSharp{}}_{2nq-r_1}\widehat{S}^{\musSharp{}}_{2nq-r_2}\Biggr]|\Psi_{\mathrm{Hf}, 2n}^q\rangle,
\end{split}
\end{equation}
with the normalization factor $\mathsf{N}_{n,q}=\frac{(-1)^q}{2(n+1)}$ resulting from the (anti-)symmetrization procedure. As an immediate consequence, we deduce that 
\begin{equation}
\label{eq:recHf}
    |\Psi_{\mathrm{Hf},2n}^q\rangle = K_{q,n-1}K_{q,n-2}\ldots K_{q,0}|0\rangle,
\end{equation}
with exactly the same composite (generalized Read) operator $K_{q,n}$ defined in Eq. \eqref{eq:Koperator}. Note that $K_{1,0}$ is the zero operator in the fermionic case $\a^\dagger_r=\c^\dagger_r$. At first sight, the fact that Eqs. \eqref{eq:recPf} and \eqref{eq:recHf} look identical might seem  puzzling but note that the choice of $\a^\dagger_r$ operators is now opposite to the one for Pfaffians. Indeed, in the Hafnian case, for $q$ odd we take $\a^\dagger_r$ to be a fermionic operator $\c^\dagger_r$, while for $q$ even we have $\a^\dagger_r=\b^\dagger_r$. This is yet another manifestation of the fact that both Pfaffian and Hafnian families of states are of the same nature. They are the result of symmetrization or antisymmetrization of Halperin bilayer systems $\Psi_{q+s\, q+s\,q-s}$ and they differ essentially only by the parity of the parameter $s$.

Consequently, also for the Hafnian state the operator $\widehat{K}_{q,n}(\eta_1,\eta_2)$ is given by Eq. \eqref{eq:Composite}, satisfying \eqref{eq:Read}, and the Hafnian state 
\begin{equation}
    |\Psi^q_{\mathrm{Hf},2n } \rangle = \mathcal{K}_q^n|0\rangle
\end{equation}
can be expressed as a nonlocal condensation. Notice that a simple change of representation can uncover a deep physical result. In particular, the nature of correlations in Pfaffian and Hafnian families is manifest from the {\it dressed paired} form of the composite operator, which is quite different from the Laughlin fluid case \cite{Read, Mazaheri2015}. 

\section{Conclusions and outlook}
\label{sec:Conclusions}

In this work, we developed a scheme, the symmetrized multicluster construction (SMC), to generate, LLL bosonic and fermionic, FQH fluids of arbitrary filling factor. The main idea of the construction relies on separating the set of particle coordinates into clusters (that we call layers because it was Halperin who first proposed a bilayer extension of Laughlin fluids) that correlate among themselves by means of specific pairing or higher-order correlation terms. A symmetrization or antisymmetrization procedure is applied next to all coordinate labels, depending on the character of the fluid. This process can be implemented in a hierarchical fashion. It turns out that this simple construction generates translationally and rotationally invariant topological fluid states which host Abelian and non-Abelian excitations. Remarkably, the same construction extends seamlessly to quasiholes, quasiparticles (quasielectrons) and magnetoexcitons. In particular, we extended previous work 
\cite{QP22} to non-Abelian fluids and showed that a similar, albeit more subtle, fusion mechanism emerges for quasiparticles (quasielectrons) that display the correct quantum numbers. This is a relevant achievement considering that it may help establish the correct fusion category associated with a given topological fluid. We argue that the non-Abelian character of the fluid state that results from our SMC is congruous with the multiclustering and nonuniform nature of the multilayer system. For instance, the  Ising anyon excitations of the Pfaffian family can be read off from the elementary quasiholes and their quasielectrons, while the excitations in the $\nu=3/5$ state discussed in Sec.~\ref{sec:multi} are related to Fibonacci anyons \cite{Hormozi09}.

The idea of implementing a symmetrization process over Halperin multilayer systems goes back to the work by Cappelli {\it et al.} \cite{Cappelli01}, where only the multilayer analog of Halperin $331$ state was considered [and referred to as {\it generalized $(331)$ Abelian theory}] (as well as the possibility of placing $m$ quasiholes homogeneously among the layers, under the assumption that $m$ is a multiplicity of the number of layers $n_l$. These additional conditions were motivated by conformal field theory). Reference \cite{Regnault08} generalized this idea to arbitrary homogeneous Halperin multilayer systems. The latter scheme is equivalent to our SMC under the additional homogeneity assumption. Our hierarchical construction is more general than the multilayer scheme, and includes, as particular examples, the constructions above. The most significant difference with other proposals is the way quasiparticles (quasielectrons) are constructed. In contrast to approaches based on conformal field theory \cite{Hermanns10} or composite fermions \cite{Milovanovic10, Wojs11, Sreejith2011a, Sreejith2013, Rodriguez2012} that use the idea of (anti-)symmetrization of Halperin multilayer systems, in our scheme quasiparticles satisfy a remarkable fusion mechanism.

First and second quantization representations of these systems highlight different aspects of the quantum correlations present in those fluids. For example, while the fusion mechanism is evident in first quantization the {\it dressed paired} nature of a Pfaffian fluid, for instance, is manifest in second quantization. Composite (generalized Read) operators, which signal the intrinsic topological order (or ODLRO) of the fluid, is another case where second quantization is king. We illustrated the methodology by deriving generalized composite operators for the non-Abelian Pfaffian and Hafnian families of states.

A few outstanding open problems remain. We envisage the resolution of some of these in future publications. One of these problems is a proof of completeness (or overcompletness) of the set of (anti-)symmetric, translationally and rotationally (i.e., homogeneous) invariant, holomorphic polynomials generated by our hierarchical  multicluster scheme. Another is the topological classification of that same set. Can one associate a knot invariant to an arbitrary element in that set? In the current paper, we conjectured a {\it K-complexity} criterium for incompressibility. What are the constraints in the construction of translationally and rotationally invariant parent Hamiltonians \cite{Ahari22, Greiterbook, Simon07, Simon12}, stabilizing those {\it K}-incompressible fluids, leading to {\it gap}-incompressibility? Since by construction parent Hamiltonians for FQH fluids are typically positive-semidefinite \cite{Ahari22}, their highest density zero modes are $K$-incompressible because of the ground state monotonicity theorem \cite{Ahari22}. An interesting question is if for an arbitrary $K$-incompressible state one can always find a parent Hamiltonian for which it is its highest density zero mode and is gap-incompressible. Yet another problem we intend to study in the future is  the braiding and topological spin structure of the quasiparticles introduced in the present paper. Finally, new numerical algorithms for efficient simulation of SMC states are desirable. One potential bypass is to apply ideas from Ref.~\cite{Balram2019}, where total antisymmetrization also led to inefficient simulations, and replace the original state by a parton-like state suitably tuned to reproduce its universal properties.

{\it Note added:} While preparing this manuscript, we became aware of a work in parallel by F.~Zhang, M.~Schossler, A.~Seidel, and L.~Chen \cite{Zhang23} which also contains a second-quantized presentation of the MR states equivalent to ours.

\acknowledgements
The work of AB was partially funded by the Deutsche Forschungsgemeinschaft (DFG, German Research Foundation) under Germany’s Excellence Strategy – EXC-2111 – 390814868. This research was undertaken thanks in part to funding from the Canada First Research Excellence Fund. We acknowledge insightful discussions with  L. Mazza, A. Nardin and A.~Seidel.

\appendix

\section{Proof of Eq. \eqref{eq:HafPROOF}} 
\label{app:proof1}

We will prove this statement by induction. First, notice that it holds for $n=1$. For the inductive step, we use the recurrence formula for Hafnian,
\begin{equation}
   \mathrm{Hf}_{2n+2}(A)=\sum_{j=2}^{2n+2} A_{1j} \, \mathrm{Hf}_{2n}(A_{\hat 1 \hat j}),
\end{equation}
where $A_{\hat 1 \hat j}$ is the matrix obtained from a $(2n+2)\times (2n+2)$ matrix $A$ by crossing out the first and $j$th columns and rows. Using this relation, the right-hand side $\mathrm{RHS}_{n+1}$ of Eq.~\eqref{eq:HafPROOF} for $2(n+1)$ particles takes the form
\begin{equation}
    \begin{split}
        &\mathrm{RHS}_{n+1}\!=\!\frac{2(n+1)^2}{2(n+1)(2n+1)}\frac{\Psi_{L}(Z_{2n+2})^{2}}{\Psi_{L}(Z_{2n+2})^{2}_{\hat{1}\hat{j}}(z_1-z_j)^2}\\
        \times\! \!\!&\sum\limits_{j =2}^{2n+2}\!\!\left\{\!\!\frac{2^n (n!)^2}{(2n)!}\mathrm{Hf}_{2n}\!\Biggl[\!\!\left(\!\!\frac{1}{(z_k-z_l)^2}\!\!\right)_{\!\!\hat{1}\hat{j}}\!\Biggr]\!\Psi_{L}(Z_{2n+2})^{2}_{\hat{1}\hat{j}}\!\!\right\}\!.
    \end{split}
\end{equation}
Since
\begin{equation}
    \begin{split}
        &\frac{\Psi_{L}(Z_{2n+2})^{2}}{\Psi_{L}(Z_{2n+2})^{2}_{\hat{1}\hat{j}}(z_1-z_j)^2}\!\!=\!\!\prod\limits_{l\neq 1,j}^{2n+2}\!((z_1-z_l)(z_l-z_j))^2,
    \end{split}
\end{equation}
and using the inductive hypothesis, we can rewrite the above equation as
\begin{equation}
    \mathrm{RHS}_{n+1}\!\!=\frac{n+1}{2n+1}\!\!\sum\limits_{j=2}^{2n+2}\!\bigl[\mathcal{S}_{2n}^{\hat{1}\hat{j}}\bigr]\!\!\prod\limits_{l\neq 1,j}^{2n+2}((z_1-z_l)(z_l-z_j))^2,   
\end{equation}
where $\bigl[\mathcal{S}_{2n}^{\hat{1}\hat{j}}\bigr]$ denotes the left-hand side of Eq.~\eqref{eq:HafPROOF} but applied to the set $Z_{2n+2}\setminus\{z_1,z_j\}$. Using basic properties of the symmetrization, we therefore get 
\begin{equation}
\begin{aligned}
   \frac{\mathrm{RHS}_{n+1}}{n+1}&\!= \mathcal{S}_{2n+2}\Biggl\{\!\!\Biggl(\prod\limits_{i<j}^{n}(z_i-z_j)^4(z_{n+1+i}-z_{n+1+j})^4\Biggr)\\
    \times\!\!\!\!\!\!\!\!\! &\prod\limits_{l\neq n+1,2n+2}^{2n+2}\!\!\!\!\!\!((z_l-z_{n+1})(z_l-z_{2n+2}))^2\Biggr\}.
\end{aligned}   
\end{equation}
The goal is to show that $\mathrm{RHS}_{n+1}$ is the same as
\begin{equation}
   \mathcal{S}_{2n+2}\!\left(\prod\limits_{i<j}^{n+1}(z_i-z_j)^4(z_{n+1+i}-z_{n+1+j})^4\right)\!.
\end{equation}
We will achieve this using (adjusted to our situation) techniques from Ref.~\cite[Appendix~C]{Jeong}. First, notice that the above claim is true for $n=1$. To simplify the notation, we denote in this Appendix $[2n+2]=Z_{2n+2}$ and $[2n]=Z_{2n+2}\setminus\{z_{n+1},z_{2n+2}\}$, and consider two polynomials:
\begin{equation}
\label{eq:defP}
    P_n([2n]) =\mathcal{S}_{2n}\!\left(\prod\limits_{i<j}^{n}(z_i-z_j)^4(z_{n+i}-z_{n+j})^4\right),
\end{equation}
\begin{equation}
    \begin{split}Q_{n}([2n])&=n\mathcal{S}_{2n}\Biggl\{\left(\prod\limits_{i<j}^{n-1}(z_i-z_j)^4(z_{n+i}-z_{n+j})^4\right)\\
    &\hspace{1cm}\times\!\!\!\prod\limits_{l\neq n,2n}^{2n}((z_l-z_{n})(z_l-z_{2n}))^2\Biggr\}. 
    \end{split}
\end{equation}
We show that $P_{n}=Q_{n}$. This trivially holds for $n=1$. Next, denoting by $[\hat{Z}]$ the set $\{(z_1,\ldots,z_{2n+2})\in Z_{2n+2}\, : \, z_{n+1}=z_{2n+2}=z\}$, we demonstrate that
\begin{equation}
         P_{n+1}([2n+2])|_{[\hat{Z}]}=c_n\!\!\!\!\!\prod\limits_{z_k\in [2n]}\!\!\!(z_k-z)^4 P_n([2n]),
    \label{coef:c}
\end{equation}
and
\begin{equation}
  \label{coef:d}
      Q_{n+1}([2n+2])|_{[\hat{Z}]} =d_n\!\!\!\!\!\prod\limits_{z_k\in [2n]}\!\!\!(z_k-z)^4 P_n([2n]).
\end{equation}
with some combinatorial factors $c_n$ and $d_n$.

We start with the proof of \eqref{coef:c}. First, notice that for every permutation $\sigma\in S_{2n+2}$, after applying the condition $z_{n+1}=z_{2n+2}=z$, the corresponding term in the definition of $P$ [see Eq. \eqref{eq:defP}] gives either zero or produces the $S_{2n}$-symmetric factor $\prod\limits_{z_k\in [2n]}\!\!\!(z_k-z)^4$ times $\tau\left(\prod\limits_{i<j}^{n}(z_i-z_j)^4(z_{n+i}-z_{n+j})^4\right)$, with certain $\tau\in S_{2n}$. Moreover, in this way, we are getting all the permutations from $S_{2n}$ an equal number of times. Therefore, \eqref{coef:c} indeed holds and $c_n=\frac{\Delta_n}{(2n+2)!}$, where $\Delta_n$ is the number of permutations from $S_{2n+2}$ that lead to nonzero contributions. The only permutations that gives zero after putting $z_{n+1}=z_{2n+2}=z$ are of the following form:
\begin{itemize}
    \item $\sigma(z_{n+1})=z_{n+1+i}$ and $\sigma(z_{2n+2})=z_{2n+2}$ with $i\leq n$,
    \item $\sigma(z_{n+1})=z_{n+1}$ and $\sigma(z_{2n+2})=z_i$ with $i\leq n$,
    \item $\sigma(z_{2n+2})=z_{n+1}$ and $\sigma(z_{n+1})=z_i$ with $i\leq n$,
    \item $\sigma(z_{n+1})=z_{2n+2}$ and $\sigma(z_{2n+2})=z_{n+1+i}$ with $i\leq n$,
    \item $\sigma(z_{n+1})=z_i$ and $z_{2n+2}=z_j$ with $i,j\leq n$,
    \item $\sigma(z_{n+1})=z_{n+1+i}$ and $\sigma(z_{2n+2})=z_{n+1+j}$ with $i,j\leq n$.
\end{itemize}
The first four of them are essentially of the same type and each of them appears $n(2n)!$ times, while the last two have multiplicities $n(n-1)(2n)!$. As a result,
\begin{equation}
    \Delta_n=(2n+2)!-\left[4n(2n)!+2n(n-1)(2n)!\right],
\end{equation}
and therefore $ c_n=\frac{n+1}{2n+1}$.

For the proof of \eqref{coef:d} we observe that the terms following from applying the symmetrization ${\cal S}_{2n+2}$ in the definition of $Q_{n+1}$, after imposing the condition $z_{n+1}=z_{2n+2}=z$, are either zero, or produce the factor $\prod\limits_{z_k\in [2n]}(z_k-z)^4$ times either $P_n$ or $Q_n$. Taking into account all the permutations from the group $S_{2n+2}$ leads to
\begin{equation}
  \begin{split}
      &Q_{n+1}([2n+2])|_{[\hat{Z}]}\\
      &\!=\!\frac{n+1}{(2n+2)!}\!\left(\alpha_n P_n([2n])+\frac{\gamma_n}{n}Q_n([2n])\right)\!\!\! \!\!\prod\limits_{z_k\in [2n]}\!\!\!\!(z_k-z)^4.
  \end{split}
\end{equation}
Here $\alpha_n$ and $\gamma_n$ are the numbers of permutations from $S_{2n+2}$ that produces $P_n$ and $Q_n$, respectively. The only permutations that lead to $P_n$ are those acting as $S_2$ on $\{z_{n+1},z_{2n+2}\}$. Therefore $\alpha_n=2(2n)!$. On the other hand, the ones that produce terms $Q_n$ are of either of the following two types:
\begin{itemize}
    \item $\sigma(z_{n+1})=z_{i}$ with $i\leq n$ and $\sigma(z_{2n+2})=z_{n+1+j}$ with $j\leq n$,
    \item $\sigma(z_{n+1})=z_{n+1+i}$ with $i\leq n$ and $\sigma(z_{2n+2})=z_{j}$ with $j\leq n$.
\end{itemize}
Therefore $\gamma_n=2n^2(2n)!$. By inductive hypothesis, we know that $P_n=Q_n$. Therefore
\begin{equation}
    \begin{split}
      &Q_{n+1}([2n+2])|_{[\hat{Z}]}\\
      &\!=\!\frac{n+1}{(2n+2)!}\!\left[2(2n)! +\frac{2n^2(2n)!}{n}\right]\! P_n([2n])\!\!\!\!\! \prod\limits_{z_k\in [2n]}\!\!\!\!(z_k-z)^4,
  \end{split}
\end{equation}
so that $d_n=\frac{n+1}{2n+1}=c_n$. Therefore, comparing \eqref{coef:c} and \eqref{coef:d} we get
$P_{n+1}([2n+2])|_{[\hat{Z}]}=Q_{n+1}([2n+2])|_{[\hat{Z}]}$,
so that
\begin{equation}
    \begin{split}
        &P_{n+1}([2n+2])-Q_{n+1}([2n+2])\\
        &=(z_{n+1}-z_{2n+2})^s R([2n+2]),
    \end{split}
\end{equation}
with $s\geq 1$ and a polynomial $R$ such that it does not vanish on $[\hat{Z}]$. Since 
$\frac{\partial P_{n+1}([2n+2])|_{[\hat{Z}]}}{\partial z} = \frac{\partial Q_{n+1}([2n+2])|_{[\hat{Z}]}}{\partial z},$
we have $s(z_{n+1}-z_{2n+2})^{s-1}R |_{[\hat{Z}]}=0$, and, as a result, $s\geq 2$. Therefore there exists a polynomial $T_1$ such that
\begin{equation}
\label{eq:poly1}
    \begin{split}
        &P_{n+1}([2n+2])-Q_{n+1}([2n+2])\\
        &=\prod\limits_{i<j}^{2n+2} (z_i-z_j)^2 T_1([2n+2]).
    \end{split}
\end{equation}
Expanding the right-hand side of the above equation in the powers of $z_1$, the leading term is an expression of the form $z_1^{\mu_1} T_2(z_2,\ldots,z_{2n+2})$ with some polynomial $T_2$. Notice that by the presence of $\prod\limits_{i<j}^{2n} (z_i-z_j)^2$ the exponent $\mu_1$ has to be at least $2(n+1)(2n+1)$. On the other hand, the monomial $z_1^{\nu_1}$ in the expansion of the difference $P_{n+1}([2n+2])-Q_{n+1}([2n+2])$ has $\nu_1\leq 4n(n+1)$. Therefore the polynomial $T_1$ has to be identically zero. This finishes the proof.

\begin{widetext}
\section{Proof of Eq.\eqref{eq:PfQhP}}
\label{app:profPfQh} 

Similarly to the case of the ground state, for $q$ even, we have ($m_1,m_2$ are non-negative integers)
\begin{equation}
\mathcal{A}_{2n}\Psi_{q+1\, q+1\, q-1;\, \eta_1,\eta_2}^{2\mathrm{qh}}(Z_{2n})=\varepsilon_n \mathcal{A}_{2n}\Biggl[\prod\limits_{i=1}^n(z_i-\eta_1)^{m_1}\prod\limits_{k=1}^n(z_{n+k}-\eta_2)^{m_2}{\rm det}_{n}\!\left(\frac{1}{z_i-z_{n+k}}\right)\Biggr] \Psi_L(Z_{2n})^q,
\end{equation}
while for $q$ odd the following is true
\begin{equation}
\begin{split}
    &\mathcal{S}_{2n}\Psi_{q+1\, q+1\, q-1;\, \eta_1,\eta_2}^{2\mathrm{qh}}(Z_{2n})\\
    &=\varepsilon_n\mathcal{S}_{2n}\Biggl[\prod\limits_{i=1}^n(z_i-\eta_1)^{m_1}\prod\limits_{k=1}^n(z_{n+k}-\eta_2)^{m_2}{\rm det}_{n}\!\left(\frac{1}{z_i-z_{n+k}}\right) \Psi_L(Z_{2n})\Biggr]\Psi_L(Z_{2n})^{q-1}\\
     &=\varepsilon_n\mathcal{A}_{2n}\Biggl[\prod\limits_{i=1}^n(z_i-\eta_1)^{m_1}\prod\limits_{k=1}^n(z_{n+k}-\eta_2)^{m_2}{\rm det}_{n}\!\left(\frac{1}{z_i-z_{n+k}}\right)\Biggr] \Psi_L(Z_{2n})\Psi_L(Z_{2n})^{q-1}\\
     &=\varepsilon_n \mathcal{A}_{2n}\Biggl[\prod\limits_{i=1}^n(z_i-\eta_1)^{m_1}\prod\limits_{k=1}^n(z_{n+k}-\eta_2)^{m_2}{\rm det}_{n}\!\left(\frac{1}{z_i-z_{n+k}}\right)\Biggr] \Psi_L(Z_{2n})^q.
\end{split}    
\end{equation}

To finish the proof notice that
    \begin{equation}
    \begin{split}
        &\mathcal{A}_{2n}\left[\prod\limits_{i=1}^n(z_i-\eta_1)^{m_1}\prod\limits_{k=1}^n(z_{n+k}-\eta_2)^{m_2}{\rm det}_{n}\!\left(\frac{1}{z_i-z_{n+j}}\right)\right]\\
        &=\sum\limits_{\sigma\in S_n}\mathrm{sgn}(\sigma)\mathcal{A}_{2n}\Biggl(\prod\limits_{i=1}^n(z_i-\eta_1)^{m_1}\prod\limits_{k=1}^n(z_{n+k}-\eta_2)^{m_2}\prod\limits_{j=1}^n\frac{1}{z_j-z_{n+\sigma(j)}}\Biggr)\\
        &=n!\, \mathcal{A}_{2n}\Biggl(\prod\limits_{i=1}^n(z_i-\eta_1)^{m_1}\prod\limits_{k=1}^n(z_{n+k}-\eta_2)^{m_2}\prod\limits_{j=1}^n\frac{1}{z_j-z_{n+j}}\Biggr)\\
        &=n! \, \mathcal{A}_{2n}\prod\limits_{j=1}^n\frac{(z_j-\eta_1)^{m_1}(z_{n+j}-\eta_2)^{m_2}}{z_j-z_{n+j}}=-\frac{(n!)^2}{(2n)!} \, \mathrm{Pf}_{2n}\left(\frac{(z_i-\eta_1)^{m_1}(z_j-\eta_2)^{m_2}+(i\leftrightarrow j)}{z_i-z_j}\right).
    \end{split}
\end{equation}
\end{widetext}

\section{Proof of Eq. \eqref{eq:hafQH}} \label{app:proof2}

In order to prove the claim \eqref{eq:hafQH}, we follow exactly the same steps as for the Hafnian state without quasiholes. For $n=1$, the formula is true. Next, the right-hand side $\mathrm{RHS}_{n+1}$ for $2(n+1)$ particles takes a form

\begin{widetext}
\begin{equation}
\begin{split}
    &\mathrm{RHS}_{n+1}\!=\!\frac{(n+1)^2}{(2n+2)(2n+1)}\sum\limits_{j=2}^{2n+2}\Biggl\{\frac{(n!)^2}{(2n)!}\mathrm{Hf}_{2n}\Biggl[\!\left(\frac{(z_k-\eta_1)^{m_1}(z_l-\eta_2)^{m_2}+(k\leftrightarrow l)}{(z_k-z_l)^2}\right)_{\!\!\hat{1}\hat{j}}\!\Biggr]\Psi_{L}(Z_{2n+2})^2_{\hat{1}\hat{j}}\Biggr\}\\
    &\hspace{5cm}\times\frac{\Psi_L(Z_{2n+2})^2\left[(z_1-\eta_1)^{m_1}(z_j-\eta_2)^{m_2}+(1\leftrightarrow j)\right]}{\Psi_{L}(Z_{2n+2})^2_{\hat{1}\hat{j}}(z_1-z_j)^2}\\
    &=\frac{n+1}{2(2n+1)}\sum\limits_{j=2}^{2n+2}\bigl[\mathcal{S}_{2n}^{\hat{1}\hat{j}}\bigr]\left(\prod\limits_{l\neq 1,j}^{2n+2}(z_l-z_1)^2(z_l-z_j)^2\right)\left[(z_1-\eta_1)^{m_1}(z_j-\eta_2)^{m_2}+(1\leftrightarrow j)\right]\\
    & =\frac{n+1}{2}\mathcal{S}_{2n+2} \Biggl\{
        \prod\limits_{i<j}^n (z_i-\eta_1)^{m_1}(z_{n+1+i}-\eta_2)^{m_2}\prod\limits_{i<j}^n (z_i-z_j)^4 (z_{n+1+i}-z_{n+1+j})^4\\
        & \times \left[(z_{n+1}-\eta_1)^{m_1}(z_{2n+2}-\eta_2)^{m_2}+(z_{2n+2}-\eta_1)^{m_1}(z_{n+1}-\eta_2)^{m_2}\right] \!\!\!\prod\limits_{l\neq n+1,2n+2}^{2n+2} (z_l-z_{n+1})^2(z_{l}-z_{2n+2})^2
        \Biggr\},
\end{split}
\end{equation}
\end{widetext}
where in the first line the inductive hypothesis was used, and $\bigl[\mathcal{S}_{2n}^{\hat{1}\hat{j}}\bigr]$ denotes the left-hand side of Eq.~\eqref{eq:hafQH} but applied to the set $Z_{2n+2}\setminus\{z_1,z_j\}$.

Using the notation introduced in the previous Appendix, let us denote the polynomial from the last line by $Q^\eta_{n+1}([2n+2])$. Our goal is to show that it is equal to  
\begin{equation}
\begin{split}
   &P^\eta_{n+1}([2n+2])=\mathcal{S}_{2n}\Biggl[\prod\limits_{i=1}^n(z_i-\eta_1)^{m_1}(z_{n+i}-\eta_2)^{m_2}\\
   &\hspace{3cm}\times\prod\limits_{i<j}^n(z_i-z_j)^4(z_{n+i}-z_{n+j})^4\Biggr].
\end{split}   
\end{equation}
Following the same steps as for the Hafnian fluid we show that
\begin{equation}
\begin{split}
&Q^\eta_{n+1}([2n+2])|_{[\hat{Z}]}\\
&=c_n\!\!\!\!\! \prod\limits_{z_k\in [2n]}\!\!\!(z_k-z)^4(z-\eta_1)^{m_1}(z-\eta_2)^{m_2} P^\eta_n([2n]),
\end{split}
\end{equation}
and
\begin{equation}
    \begin{split}
        &P^\eta_{n+1}([2n+2])|_{[\hat{Z}]}\\
        &=c_n\!\!\!\!\!\prod\limits_{z_k\in [2n]}\!\!\!(z_k-z)^4(z-\eta_1)^{m_1}(z-\eta_2)^{m_2} P^\eta_n([2n]).
    \end{split}
\end{equation}
Therefore 
\begin{equation}
    \begin{split}
          &P^\eta_{n+1}([2n+2])-Q^\eta_{n+1}([2n+2])\\
          &=(z_{n+1}-z_{2n+2})^2 R^\eta([2n+2])
    \end{split}
\end{equation}
with some polynomial $R^\eta$ that does not vanish on $[\hat{Z}]$. Since the derivatives of the polynomials $Q^\eta_{n+1}([2n+2])|_{[\hat{Z}]}$ and $P^\eta_{n+1}([2n+2])|_{[\hat{Z}]}$ are equal, we deduce the existence of a polynomial $S^\eta$ such that
\begin{equation}
    \begin{split}
        &P^\eta_{n+1}([2n+2])-Q^\eta_{n+1}([2n+2])\\
        &=\prod\limits_{i<j}^{2n+2}(z_i-z_j)^2 S^\eta([2n+2]).
    \end{split}
\end{equation}
By power counting, we infer that the last polynomial has to be identically zero, and this finishes the proof.

\section{Efficient simulation of Pfaffian states} 
\label{app:Pfaffiansimul}

In our Monte Carlo simulations, we need to update Pfaffians as well as quasiholes in Pfaffian-like states. We next briefly describe an efficient procedure. For a $2n\times 2n$ antisymmetric matrix $A$ with entries $A_{ij}=\frac{1}{z_i-z_j}$, let $B$ denote this matrix after a one-particle update, say at position $z_{i_0}$. From Cayley's formula \cite{Bajdich08, Cayley}, we obtain
\begin{equation}
\label{eq:cayley}
    \frac{\mathrm{Pf}_{2n}(B)}{\mathrm{Pf}_{2n}(A)}=\sum\limits_{j=1}^{2n}B_{i_0j} \left(A^{-1}\right)_{ji_0}.
\end{equation}
This requires $O(2n)$ operations in addition to the computation of the inverse $A^{-1}$. This can be implemented, e.g., by using Gauss's method with a pivoting strategy involved (to avoid numerical instabilities).    

To effectively simulate systems of quasiholes with one additional electron we make use of the following identity
\begin{equation}
\begin{split}
    &N\mathcal{A}_N\left(e^{\frac{z_{N}\eta^\ast}{2}}f(Z_{N-1})\right)\\
    &=f(Z_{N-1})e^{\frac{z_{N}\eta^\ast}{2}} - \sum\limits_{i=1}^{N-1} e^{\frac{z_{i}\eta^\ast}{2}} f(Z_{[i]}^{N}),
\end{split}    
\end{equation}
where $Z_{[i]}^{N}=(z_1,\ldots, \overset{i}{z_{N}}, \ldots, z_{N-1})$, $N$ is odd, and $f$ is either
\begin{equation}
    f(Z_{N-1})=\mathrm{Pf}_{N-1}\left(\frac{z_i+z_j}{z_i-z_j}\right)\Psi_L(Z_{N-1})^q
\end{equation}
or
\begin{equation}
    f(Z_{N-1})=\mathrm{Pf}_{N-1}\left(\frac{2z_iz_j}{z_i-z_j}\right)\Psi_L(Z_{N-1})^q,
\end{equation}
with $q$ even. In this case, for the computation of the Pfaffian, we use the method presented in Ref.~\cite{numeric} based on complex Householder transformations.

\begin{widetext}
\section{Pfaffian states in second quantization - proofs}
\label{app:proofPf}

First, we show that Eq.~\eqref{eq:PfF} is indeed true. We have the following series of equalities:
\begin{equation}
\begin{split}
    \Psi_{\mathrm{Pf}}^q(Z_{2n+2})\!&=\!\sum\limits_{j=2}^{2n+2}\!\frac{(-1)^j}{z_1-z_j}\!\Biggl[\mathrm{Pf}_{2n}\!\left(\frac{1}{z_a-z_b}\right)_{\!\!\hat{1}\hat{j}}\!\prod\limits_{\substack{k<l\\ k\neq 1,j\\l\neq j}}^{2n+2}\!\!(z_k-z_l)^q\!\Biggr] (z_1-z_j)^q\!\prod\limits_{\substack{l=2\\ l\neq j}}^{2n+2}\!\!(z_1-z_l)^{q}\!\prod\limits_{k=2}^{j-1}(z_k-z_j)^q\!\!\prod\limits_{l=j+1}^{2n+2}\!\!(z_j-z_l)^q\\
    &=\sum\limits_{j=2}^{2n+2}(-1)^j(z_1-z_j)^{q-1}\!\prod\limits_{\substack{l=2\\ l\neq j}}^{2n+2}\!\!(z_1-z_l)^{q}\!\!\prod\limits_{l=j+1}^{2n+2}\!\!(z_j-z_l)^q  \!\prod\limits_{l=2}^{j-1}\!\left((-1)^q(z_j-z_l)^q\right) \Psi_{\mathrm{Pf}}^q((Z_{2n+2})_{\hat{1}\hat{j}})\\
    &=\sum\limits_{j=2}^{2n+2}(-1)^{j(1+q)}(z_1-z_j)^{q-1}\!\prod\limits_{\substack{l=2\\ l\neq j}}^{2n+2}\!\!(z_1-z_l)^q\!\prod\limits_{\substack{l=2\\ l\neq j}}^{2n+2}\!\!(z_j-z_l)^q\Psi_{\mathrm{Pf}}^q((Z_{2n+2})_{\hat{1}\hat{j}})\\
    &=\sum\limits_{j=2}^{2n+2}(-1)^{j(1+q)}(z_1-z_j)^{q-1}\!\prod\limits_{\substack{l\neq 1,j}}^{2n+2}\!\! \left((z_1-z_l)(z_j-z_l)\right)^q \Psi_{\mathrm{Pf}}^q((Z_{2n+2})_{\hat{1}\hat{j}}).
\end{split}
\label{eq:rec0}
\end{equation}

Next, we present the proof of Eq.~\eqref{eq:Read}. We have
\begin{equation}
\begin{split}
    &\int \mathcal{D}[\eta_1]\mathcal{D}[\eta_2]\widehat{K}_{q,n}(\eta_1,\eta_2)\\
    =&\mathsf{N}_{n,q} \int \mathcal{D}[\eta_1]\mathcal{D}[\eta_2](\eta_1-\eta_2)^{q-1}\!\!\!\sum\limits_{r_1,r_2\ge 0}\phi_{r_1}^\ast(\eta_1)\phi_{r_2}^\ast(\eta_2)a^\dagger_{r_1}a^\dagger_{r_2}\!\!\sum\limits_{r_1',r_2'\ge 0} \eta_1^{r_1'}\eta_2^{r_2'}\widehat{S}^{\musSharp{}}_{2nq-r_1'} \widehat{S}^{\musSharp{}}_{2nq-r_2'}\\
    =&\mathsf{N}_{n,q}\sum\limits_{\substack{r_1,r_2\\ r_1',r_2'}}\int \mathcal{D}[\eta_1]\mathcal{D}[\eta_2](\eta_1-\eta_2)^{q-1}\frac{(\eta_1^\ast)^{r_1}(\eta_2^\ast)^{r_2}\eta_1^{r_1'}\eta_2^{r_2'} }{\mathcal{N}_{r_1}\mathcal{N}_{r_2}}a^\dagger_{r_1} a^\dagger_{r_2}\widehat{S}^{\musSharp{}}_{2nq-r_1'} \widehat{S}^{\musSharp{}}_{2nq-r_2'}\\
    =&\mathsf{N}_{n,q}\sum\limits_{\substack{r_1,r_2\\ r_1',r_2'}}\int \mathcal{D}[\eta_1]\mathcal{D}[\eta_2]\sum\limits_{l=0}^{q-1}\binom{q-1}{l}(-1)^{q-1-l} \eta_1^l\eta_2^{q-1-l}\frac{(\eta_1^\ast)^{r_1}(\eta_2^\ast)^{r_2}\eta_1^{r_1'}\eta_2^{r_2'} }{\mathcal{N}_{r_1}\mathcal{N}_{r_2}}a^\dagger_{r_1}a^\dagger_{r_2}\widehat{S}^{\musSharp{}}_{2nq-r_1'} \widehat{S}^{\musSharp{}}_{2nq-r_2'}\\
    =&\mathsf{N}_{n,q} \sum\limits_{\substack{r_1,r_2\\ r_1',r_2'}} \sum\limits_{l=0}^{q-1}\mathcal{N}_{r_1}\mathcal{N}_{r_2}\delta_{r_1,r_1'+l}\delta_{r_2,r_2'+q-1-l}\binom{q-1}{l}(-1)^{q-1-l} a_{r_1}^\dagger a_{r_2}^\dagger \widehat{S}^{\musSharp{}}_{2nq-r_1'} \widehat{S}^{\musSharp{}}_{2nq-r_2'}\\
    =&\mathsf{N}_{n,q}\!\!\!\sum\limits_{r_1,r_2\ge 0}\sum\limits_{l=0}^{q-1}\binom{q-1}{l}(-1)^{l+1}\a^\dagger_{r_1+l}\a^\dagger_{r_2+q-1-l}\widehat{S}^{\musSharp{}}_{2nq-r_1} \widehat{S}^{\musSharp{}}_{2nq-r_2}=K_{q,n}.
\end{split}    
\end{equation}
Finally, we verify \eqref{eq:Read2}. It is a result of the following series of equalities:
\begin{equation}
    \begin{split}
        \mathcal{K}_q|\Psi^q_{\mathrm{Pf},2n}\rangle&=\int \mathcal{D}[\eta_1]\mathcal{D}[\eta_2]\widehat{\mathcal{K}}_q(\eta_1,\eta_2)|\Psi^q_{\mathrm{Pf},2n}\rangle\\
        &=\int \mathcal{D}[\eta_1]\mathcal{D}[\eta_2]\mathsf{N}_{n,q} (\eta_1-\eta_2)^{q-1}\Lambda^\dagger(\eta_1)\Lambda^\dagger(\eta_2)\widehat{\mathcal{U}}_{q,q}(\eta_1,\eta_2)|\Psi^q_{\mathrm{Pf},2n}\rangle\\
        &=\int \mathcal{D}[\eta_1]\mathcal{D}[\eta_2] \mathsf{N}_{n,q}(\eta_1-\eta_2)^{q-1}\Lambda^\dagger (\eta_1) \Lambda^\dagger(\eta_2)\widehat{U}_{2n}(\eta_1)^q\widehat{U}_{2n}(\eta_2)^q |\Psi^q_{\mathrm{Pf},2n}\rangle\\
        &=\int \mathcal{D}[\eta_1]\mathcal{D}[\eta_2]\widehat{K}_{q,n}(\eta_1,\eta_2) |\Psi^q_{\mathrm{Pf},2n}\rangle = K_{q,n} |\Psi^q_{\mathrm{Pf},2n} \rangle = |\Psi^q_{\mathrm{Pf},2n+2 } \rangle.
    \end{split}
\end{equation}
\end{widetext}


\bibliography{Paper}

\providecommand{\noopsort}[1]{}\providecommand{\singleletter}[1]{#1}%
\begin{thebibliography}{51}%
\makeatletter
\providecommand \@ifxundefined [1]{%
 \@ifx{#1\undefined}
}%
\providecommand \@ifnum [1]{%
 \ifnum #1\expandafter \@firstoftwo
 \else \expandafter \@secondoftwo
 \fi
}%
\providecommand \@ifx [1]{%
 \ifx #1\expandafter \@firstoftwo
 \else \expandafter \@secondoftwo
 \fi
}%
\providecommand \natexlab [1]{#1}%
\providecommand \enquote  [1]{``#1''}%
\providecommand \bibnamefont  [1]{#1}%
\providecommand \bibfnamefont [1]{#1}%
\providecommand \citenamefont [1]{#1}%
\providecommand \href@noop [0]{\@secondoftwo}%
\providecommand \href [0]{\begingroup \@sanitize@url \@href}%
\providecommand \@href[1]{\@@startlink{#1}\@@href}%
\providecommand \@@href[1]{\endgroup#1\@@endlink}%
\providecommand \@sanitize@url [0]{\catcode `\\12\catcode `\$12\catcode
  `\&12\catcode `\#12\catcode `\^12\catcode `\_12\catcode `\%12\relax}%
\providecommand \@@startlink[1]{}%
\providecommand \@@endlink[0]{}%
\providecommand \url  [0]{\begingroup\@sanitize@url \@url }%
\providecommand \@url [1]{\endgroup\@href {#1}{\urlprefix }}%
\providecommand \urlprefix  [0]{URL }%
\providecommand \Eprint [0]{\href }%
\providecommand \doibase [0]{https://doi.org/}%
\providecommand \selectlanguage [0]{\@gobble}%
\providecommand \bibinfo  [0]{\@secondoftwo}%
\providecommand \bibfield  [0]{\@secondoftwo}%
\providecommand \translation [1]{[#1]}%
\providecommand \BibitemOpen [0]{}%
\providecommand \bibitemStop [0]{}%
\providecommand \bibitemNoStop [0]{.\EOS\space}%
\providecommand \EOS [0]{\spacefactor3000\relax}%
\providecommand \BibitemShut  [1]{\csname bibitem#1\endcsname}%
\let\auto@bib@innerbib\@empty
\bibitem [{\citenamefont {Jain}(2007)}]{Jainbook}%
  \BibitemOpen
  \bibfield  {author} {\bibinfo {author} {\bibfnamefont {J.~K.}\ \bibnamefont
  {Jain}},\ }\href {https://doi.org/10.1017/CBO9780511607561} {\emph {\bibinfo
  {title} {Composite Fermions}}}\ (\bibinfo  {publisher} {Cambridge University
  Press},\ \bibinfo {year} {2007})\BibitemShut {NoStop}%
\bibitem [{\citenamefont {Ahari}\ \emph {et~al.}(2023)\citenamefont {Ahari},
  \citenamefont {Bandyopadhyay}, \citenamefont {Nussinov}, \citenamefont
  {Seidel},\ and\ \citenamefont {Ortiz}}]{Ahari22}%
  \BibitemOpen
  \bibfield  {author} {\bibinfo {author} {\bibfnamefont {M.~T.}\ \bibnamefont
  {Ahari}}, \bibinfo {author} {\bibfnamefont {S.}~\bibnamefont
  {Bandyopadhyay}}, \bibinfo {author} {\bibfnamefont {Z.}~\bibnamefont
  {Nussinov}}, \bibinfo {author} {\bibfnamefont {A.}~\bibnamefont {Seidel}},\
  and\ \bibinfo {author} {\bibfnamefont {G.}~\bibnamefont {Ortiz}},\ }\href
  {https://doi.org/10.21468/SciPostPhys.15.2.043} {\bibfield  {journal}
  {\bibinfo  {journal} {SciPost Phys.}\ }\textbf {\bibinfo {volume} {15}},\
  \bibinfo {pages} {043} (\bibinfo {year} {2023})}\BibitemShut {NoStop}%
\bibitem [{\citenamefont {Bochniak}\ \emph {et~al.}(2022)\citenamefont
  {Bochniak}, \citenamefont {Nussinov}, \citenamefont {Seidel},\ and\
  \citenamefont {Ortiz}}]{QP22}%
  \BibitemOpen
  \bibfield  {author} {\bibinfo {author} {\bibfnamefont {A.}~\bibnamefont
  {Bochniak}}, \bibinfo {author} {\bibfnamefont {Z.}~\bibnamefont {Nussinov}},
  \bibinfo {author} {\bibfnamefont {A.}~\bibnamefont {Seidel}},\ and\ \bibinfo
  {author} {\bibfnamefont {G.}~\bibnamefont {Ortiz}},\ }\href
  {https://doi.org/10.1038/s42005-022-00946-8} {\bibfield  {journal} {\bibinfo
  {journal} {Commun Phys}\ }\textbf {\bibinfo {volume} {5}},\ \bibinfo {pages}
  {171} (\bibinfo {year} {2022})}\BibitemShut {NoStop}%
\bibitem [{\citenamefont {Halperin}(1983)}]{Halperin83}%
  \BibitemOpen
  \bibfield  {author} {\bibinfo {author} {\bibfnamefont {B.~I.}\ \bibnamefont
  {Halperin}},\ }\href@noop {} {\bibfield  {journal} {\bibinfo  {journal}
  {Helv. Phys. Acta}\ }\textbf {\bibinfo {volume} {56}},\ \bibinfo {pages} {75}
  (\bibinfo {year} {1983})}\BibitemShut {NoStop}%
\bibitem [{\citenamefont {Greiter}\ \emph
  {et~al.}(1992{\natexlab{a}})\citenamefont {Greiter}, \citenamefont {Wen},\
  and\ \citenamefont {Wilczek}}]{Greiter92a}%
  \BibitemOpen
  \bibfield  {author} {\bibinfo {author} {\bibfnamefont {M.}~\bibnamefont
  {Greiter}}, \bibinfo {author} {\bibfnamefont {X.~G.}\ \bibnamefont {Wen}},\
  and\ \bibinfo {author} {\bibfnamefont {F.}~\bibnamefont {Wilczek}},\ }\href
  {https://doi.org/10.1103/PhysRevB.46.9586} {\bibfield  {journal} {\bibinfo
  {journal} {Phys. Rev. B}\ }\textbf {\bibinfo {volume} {46}},\ \bibinfo
  {pages} {9586} (\bibinfo {year} {1992}{\natexlab{a}})}\BibitemShut {NoStop}%
\bibitem [{\citenamefont {Jeong}\ \emph {et~al.}(2017)\citenamefont {Jeong},
  \citenamefont {Lu}, \citenamefont {Lee}, \citenamefont {Hashimoto},
  \citenamefont {Chung},\ and\ \citenamefont {Park}}]{Jeong}%
  \BibitemOpen
  \bibfield  {author} {\bibinfo {author} {\bibfnamefont {J.-S.}\ \bibnamefont
  {Jeong}}, \bibinfo {author} {\bibfnamefont {H.}~\bibnamefont {Lu}}, \bibinfo
  {author} {\bibfnamefont {K.~H.}\ \bibnamefont {Lee}}, \bibinfo {author}
  {\bibfnamefont {K.}~\bibnamefont {Hashimoto}}, \bibinfo {author}
  {\bibfnamefont {S.~B.}\ \bibnamefont {Chung}},\ and\ \bibinfo {author}
  {\bibfnamefont {K.}~\bibnamefont {Park}},\ }\href
  {https://doi.org/10.1103/PhysRevB.96.125148} {\bibfield  {journal} {\bibinfo
  {journal} {Phys. Rev. B}\ }\textbf {\bibinfo {volume} {96}},\ \bibinfo
  {pages} {125148} (\bibinfo {year} {2017})}\BibitemShut {NoStop}%
\bibitem [{\citenamefont {Kuo}(2020)}]{Kuo}%
  \BibitemOpen
  \bibfield  {author} {\bibinfo {author} {\bibfnamefont {E.-J.}\ \bibnamefont
  {Kuo}},\ }\href@noop {} {\bibinfo {title} {Halperin states can produce any
  filling fractions}} (\bibinfo {year} {2020}),\ \Eprint
  {https://arxiv.org/abs/2004.09035} {arXiv:2004.09035 [math-ph]} \BibitemShut
  {NoStop}%
\bibitem [{\citenamefont {Cappelli}\ \emph {et~al.}(2001)\citenamefont
  {Cappelli}, \citenamefont {Georgiev},\ and\ \citenamefont
  {Todorov}}]{Cappelli01}%
  \BibitemOpen
  \bibfield  {author} {\bibinfo {author} {\bibfnamefont {A.}~\bibnamefont
  {Cappelli}}, \bibinfo {author} {\bibfnamefont {L.~S.}\ \bibnamefont
  {Georgiev}},\ and\ \bibinfo {author} {\bibfnamefont {I.~T.}\ \bibnamefont
  {Todorov}},\ }\href
  {https://doi.org/https://doi.org/10.1016/S0550-3213(00)00774-4} {\bibfield
  {journal} {\bibinfo  {journal} {Nuclear Physics B}\ }\textbf {\bibinfo
  {volume} {599}},\ \bibinfo {pages} {499} (\bibinfo {year}
  {2001})}\BibitemShut {NoStop}%
\bibitem [{\citenamefont {Regnault}\ \emph {et~al.}(2008)\citenamefont
  {Regnault}, \citenamefont {Goerbig},\ and\ \citenamefont
  {Jolicoeur}}]{Regnault08}%
  \BibitemOpen
  \bibfield  {author} {\bibinfo {author} {\bibfnamefont {N.}~\bibnamefont
  {Regnault}}, \bibinfo {author} {\bibfnamefont {M.~O.}\ \bibnamefont
  {Goerbig}},\ and\ \bibinfo {author} {\bibfnamefont {T.}~\bibnamefont
  {Jolicoeur}},\ }\href {https://doi.org/10.1103/PhysRevLett.101.066803}
  {\bibfield  {journal} {\bibinfo  {journal} {Phys. Rev. Lett.}\ }\textbf
  {\bibinfo {volume} {101}},\ \bibinfo {pages} {066803} (\bibinfo {year}
  {2008})}\BibitemShut {NoStop}%
\bibitem [{\citenamefont {Hermanns}(2010)}]{Hermanns10}%
  \BibitemOpen
  \bibfield  {author} {\bibinfo {author} {\bibfnamefont {M.}~\bibnamefont
  {Hermanns}},\ }\href {https://doi.org/10.1103/PhysRevLett.104.056803}
  {\bibfield  {journal} {\bibinfo  {journal} {Phys. Rev. Lett.}\ }\textbf
  {\bibinfo {volume} {104}},\ \bibinfo {pages} {056803} (\bibinfo {year}
  {2010})}\BibitemShut {NoStop}%
\bibitem [{\citenamefont {Hansson}\ \emph {et~al.}(2007)\citenamefont
  {Hansson}, \citenamefont {Chang}, \citenamefont {Jain},\ and\ \citenamefont
  {Viefers}}]{Hansson07}%
  \BibitemOpen
  \bibfield  {author} {\bibinfo {author} {\bibfnamefont {T.~H.}\ \bibnamefont
  {Hansson}}, \bibinfo {author} {\bibfnamefont {C.-C.}\ \bibnamefont {Chang}},
  \bibinfo {author} {\bibfnamefont {J.~K.}\ \bibnamefont {Jain}},\ and\
  \bibinfo {author} {\bibfnamefont {S.}~\bibnamefont {Viefers}},\ }\href
  {https://doi.org/10.1103/PhysRevB.76.075347} {\bibfield  {journal} {\bibinfo
  {journal} {Phys. Rev. B}\ }\textbf {\bibinfo {volume} {76}},\ \bibinfo
  {pages} {075347} (\bibinfo {year} {2007})}\BibitemShut {NoStop}%
\bibitem [{\citenamefont {Hansson}\ \emph {et~al.}(2009)\citenamefont
  {Hansson}, \citenamefont {Hermanns},\ and\ \citenamefont
  {Viefers}}]{Hansson09}%
  \BibitemOpen
  \bibfield  {author} {\bibinfo {author} {\bibfnamefont {T.~H.}\ \bibnamefont
  {Hansson}}, \bibinfo {author} {\bibfnamefont {M.}~\bibnamefont {Hermanns}},\
  and\ \bibinfo {author} {\bibfnamefont {S.}~\bibnamefont {Viefers}},\ }\href
  {https://doi.org/10.1103/PhysRevB.80.165330} {\bibfield  {journal} {\bibinfo
  {journal} {Phys. Rev. B}\ }\textbf {\bibinfo {volume} {80}},\ \bibinfo
  {pages} {165330} (\bibinfo {year} {2009})}\BibitemShut {NoStop}%
\bibitem [{\citenamefont {Hansson}\ \emph {et~al.}(2017)\citenamefont
  {Hansson}, \citenamefont {Hermanns}, \citenamefont {Simon},\ and\
  \citenamefont {Viefers}}]{Hansson17}%
  \BibitemOpen
  \bibfield  {author} {\bibinfo {author} {\bibfnamefont {T.~H.}\ \bibnamefont
  {Hansson}}, \bibinfo {author} {\bibfnamefont {M.}~\bibnamefont {Hermanns}},
  \bibinfo {author} {\bibfnamefont {S.~H.}\ \bibnamefont {Simon}},\ and\
  \bibinfo {author} {\bibfnamefont {S.~F.}\ \bibnamefont {Viefers}},\ }\href
  {https://doi.org/10.1103/RevModPhys.89.025005} {\bibfield  {journal}
  {\bibinfo  {journal} {Rev. Mod. Phys.}\ }\textbf {\bibinfo {volume} {89}},\
  \bibinfo {pages} {025005} (\bibinfo {year} {2017})}\BibitemShut {NoStop}%
\bibitem [{\citenamefont {Moore}\ and\ \citenamefont
  {Read}(1991)}]{MooreRead91}%
  \BibitemOpen
  \bibfield  {author} {\bibinfo {author} {\bibfnamefont {G.}~\bibnamefont
  {Moore}}\ and\ \bibinfo {author} {\bibfnamefont {N.}~\bibnamefont {Read}},\
  }\href {https://doi.org/https://doi.org/10.1016/0550-3213(91)90407-O}
  {\bibfield  {journal} {\bibinfo  {journal} {Nuclear Physics B}\ }\textbf
  {\bibinfo {volume} {360}},\ \bibinfo {pages} {362} (\bibinfo {year}
  {1991})}\BibitemShut {NoStop}%
\bibitem [{\citenamefont {Greiter}\ \emph {et~al.}(1991)\citenamefont
  {Greiter}, \citenamefont {Wen},\ and\ \citenamefont {Wilczek}}]{Greiter91}%
  \BibitemOpen
  \bibfield  {author} {\bibinfo {author} {\bibfnamefont {M.}~\bibnamefont
  {Greiter}}, \bibinfo {author} {\bibfnamefont {X.-G.}\ \bibnamefont {Wen}},\
  and\ \bibinfo {author} {\bibfnamefont {F.}~\bibnamefont {Wilczek}},\ }\href
  {https://doi.org/10.1103/PhysRevLett.66.3205} {\bibfield  {journal} {\bibinfo
   {journal} {Phys. Rev. Lett.}\ }\textbf {\bibinfo {volume} {66}},\ \bibinfo
  {pages} {3205} (\bibinfo {year} {1991})}\BibitemShut {NoStop}%
\bibitem [{\citenamefont {Read}\ and\ \citenamefont {Rezayi}(1999)}]{RR99}%
  \BibitemOpen
  \bibfield  {author} {\bibinfo {author} {\bibfnamefont {N.}~\bibnamefont
  {Read}}\ and\ \bibinfo {author} {\bibfnamefont {E.}~\bibnamefont {Rezayi}},\
  }\href {https://doi.org/10.1103/PhysRevB.59.8084} {\bibfield  {journal}
  {\bibinfo  {journal} {Phys. Rev. B}\ }\textbf {\bibinfo {volume} {59}},\
  \bibinfo {pages} {8084} (\bibinfo {year} {1999})}\BibitemShut {NoStop}%
\bibitem [{\citenamefont {Girvin}\ and\ \citenamefont
  {MacDonald}(1987)}]{Girvin87}%
  \BibitemOpen
  \bibfield  {author} {\bibinfo {author} {\bibfnamefont {S.~M.}\ \bibnamefont
  {Girvin}}\ and\ \bibinfo {author} {\bibfnamefont {A.~H.}\ \bibnamefont
  {MacDonald}},\ }\href {https://doi.org/10.1103/PhysRevLett.58.1252}
  {\bibfield  {journal} {\bibinfo  {journal} {Phys. Rev. Lett.}\ }\textbf
  {\bibinfo {volume} {58}},\ \bibinfo {pages} {1252} (\bibinfo {year}
  {1987})}\BibitemShut {NoStop}%
\bibitem [{\citenamefont {Tong}(2016)}]{Tong}%
  \BibitemOpen
  \bibfield  {author} {\bibinfo {author} {\bibfnamefont {D.}~\bibnamefont
  {Tong}},\ }\href@noop {} {\bibinfo {title} {Lectures on the quantum hall
  effect}} (\bibinfo {year} {2016}),\ \Eprint
  {https://arxiv.org/abs/1606.06687} {arXiv:1606.06687 [hep-th]} \BibitemShut
  {NoStop}%
\bibitem [{\citenamefont {Ho}(1995)}]{Ho95}%
  \BibitemOpen
  \bibfield  {author} {\bibinfo {author} {\bibfnamefont {T.-L.}\ \bibnamefont
  {Ho}},\ }\href {https://doi.org/10.1103/PhysRevLett.75.1186} {\bibfield
  {journal} {\bibinfo  {journal} {Phys. Rev. Lett.}\ }\textbf {\bibinfo
  {volume} {75}},\ \bibinfo {pages} {1186} (\bibinfo {year}
  {1995})}\BibitemShut {NoStop}%
\bibitem [{\citenamefont {Barkeshli}\ and\ \citenamefont
  {Wen}(2010)}]{Barkeshli10}%
  \BibitemOpen
  \bibfield  {author} {\bibinfo {author} {\bibfnamefont {M.}~\bibnamefont
  {Barkeshli}}\ and\ \bibinfo {author} {\bibfnamefont {X.-G.}\ \bibnamefont
  {Wen}},\ }\href {https://doi.org/10.1103/PhysRevB.81.045323} {\bibfield
  {journal} {\bibinfo  {journal} {Phys. Rev. B}\ }\textbf {\bibinfo {volume}
  {81}},\ \bibinfo {pages} {045323} (\bibinfo {year} {2010})}\BibitemShut
  {NoStop}%
\bibitem [{\citenamefont {Greiter}\ \emph
  {et~al.}(1992{\natexlab{b}})\citenamefont {Greiter}, \citenamefont {Wen},\
  and\ \citenamefont {Wilczek}}]{Greiter92}%
  \BibitemOpen
  \bibfield  {author} {\bibinfo {author} {\bibfnamefont {M.}~\bibnamefont
  {Greiter}}, \bibinfo {author} {\bibfnamefont {X.}~\bibnamefont {Wen}},\ and\
  \bibinfo {author} {\bibfnamefont {F.}~\bibnamefont {Wilczek}},\ }\href
  {https://doi.org/https://doi.org/10.1016/0550-3213(92)90401-V} {\bibfield
  {journal} {\bibinfo  {journal} {Nuclear Physics B}\ }\textbf {\bibinfo
  {volume} {374}},\ \bibinfo {pages} {567} (\bibinfo {year}
  {1992}{\natexlab{b}})}\BibitemShut {NoStop}%
\bibitem [{\citenamefont {Wan}\ \emph {et~al.}(2008)\citenamefont {Wan},
  \citenamefont {Hu}, \citenamefont {Rezayi},\ and\ \citenamefont
  {Yang}}]{Wan08}%
  \BibitemOpen
  \bibfield  {author} {\bibinfo {author} {\bibfnamefont {X.}~\bibnamefont
  {Wan}}, \bibinfo {author} {\bibfnamefont {Z.-X.}\ \bibnamefont {Hu}},
  \bibinfo {author} {\bibfnamefont {E.~H.}\ \bibnamefont {Rezayi}},\ and\
  \bibinfo {author} {\bibfnamefont {K.}~\bibnamefont {Yang}},\ }\href
  {https://doi.org/10.1103/PhysRevB.77.165316} {\bibfield  {journal} {\bibinfo
  {journal} {Phys. Rev. B}\ }\textbf {\bibinfo {volume} {77}},\ \bibinfo
  {pages} {165316} (\bibinfo {year} {2008})}\BibitemShut {NoStop}%
\bibitem [{\citenamefont {Ortiz}\ \emph {et~al.}(2013)\citenamefont {Ortiz},
  \citenamefont {Nussinov}, \citenamefont {Dukelsky},\ and\ \citenamefont
  {Seidel}}]{Ortiz2013}%
  \BibitemOpen
  \bibfield  {author} {\bibinfo {author} {\bibfnamefont {G.}~\bibnamefont
  {Ortiz}}, \bibinfo {author} {\bibfnamefont {Z.}~\bibnamefont {Nussinov}},
  \bibinfo {author} {\bibfnamefont {J.}~\bibnamefont {Dukelsky}},\ and\
  \bibinfo {author} {\bibfnamefont {A.}~\bibnamefont {Seidel}},\ }\href
  {https://doi.org/10.1103/PhysRevB.88.165303} {\bibfield  {journal} {\bibinfo
  {journal} {Phys. Rev. B}\ }\textbf {\bibinfo {volume} {88}},\ \bibinfo
  {pages} {165303} (\bibinfo {year} {2013})}\BibitemShut {NoStop}%
\bibitem [{\citenamefont {Seidel}\ and\ \citenamefont {Yang}(2008)}]{Seidel08}%
  \BibitemOpen
  \bibfield  {author} {\bibinfo {author} {\bibfnamefont {A.}~\bibnamefont
  {Seidel}}\ and\ \bibinfo {author} {\bibfnamefont {K.}~\bibnamefont {Yang}},\
  }\href {https://doi.org/10.1103/PhysRevLett.101.036804} {\bibfield  {journal}
  {\bibinfo  {journal} {Phys. Rev. Lett.}\ }\textbf {\bibinfo {volume} {101}},\
  \bibinfo {pages} {036804} (\bibinfo {year} {2008})}\BibitemShut {NoStop}%
\bibitem [{\citenamefont {Cauchy}(1841)}]{Cauchy41}%
  \BibitemOpen
  \bibfield  {author} {\bibinfo {author} {\bibfnamefont {A.~L.}\ \bibnamefont
  {Cauchy}},\ }\href@noop {} {\emph {\bibinfo {title} {Exercices D'analyse et
  de Physique Mathématique}}},\ Vol.~\bibinfo {volume} {2}\ (\bibinfo
  {publisher} {Bachelier},\ \bibinfo {year} {1841})\BibitemShut {NoStop}%
\bibitem [{\citenamefont {Tőke}\ \emph {et~al.}(2007)\citenamefont {Tőke},
  \citenamefont {Regnault},\ and\ \citenamefont {Jain}}]{Toke}%
  \BibitemOpen
  \bibfield  {author} {\bibinfo {author} {\bibfnamefont {C.}~\bibnamefont
  {Tőke}}, \bibinfo {author} {\bibfnamefont {N.}~\bibnamefont {Regnault}},\
  and\ \bibinfo {author} {\bibfnamefont {J.~K.}\ \bibnamefont {Jain}},\ }\href
  {https://doi.org/https://doi.org/10.1016/j.ssc.2007.03.061} {\bibfield
  {journal} {\bibinfo  {journal} {Solid State Communications}\ }\textbf
  {\bibinfo {volume} {144}},\ \bibinfo {pages} {504} (\bibinfo {year}
  {2007})}\BibitemShut {NoStop}%
\bibitem [{\citenamefont {Nayak}\ and\ \citenamefont {Wilczek}(1996)}]{Nayak}%
  \BibitemOpen
  \bibfield  {author} {\bibinfo {author} {\bibfnamefont {C.}~\bibnamefont
  {Nayak}}\ and\ \bibinfo {author} {\bibfnamefont {F.}~\bibnamefont
  {Wilczek}},\ }\href
  {https://doi.org/https://doi.org/10.1016/0550-3213(96)00430-0} {\bibfield
  {journal} {\bibinfo  {journal} {Nuclear Physics B}\ }\textbf {\bibinfo
  {volume} {479}},\ \bibinfo {pages} {529} (\bibinfo {year}
  {1996})}\BibitemShut {NoStop}%
\bibitem [{\citenamefont {Kivelson}\ and\ \citenamefont
  {Schrieffer}(1982)}]{Kivelson}%
  \BibitemOpen
  \bibfield  {author} {\bibinfo {author} {\bibfnamefont {S.}~\bibnamefont
  {Kivelson}}\ and\ \bibinfo {author} {\bibfnamefont {J.~R.}\ \bibnamefont
  {Schrieffer}},\ }\href {https://doi.org/10.1103/PhysRevB.25.6447} {\bibfield
  {journal} {\bibinfo  {journal} {Phys. Rev. B}\ }\textbf {\bibinfo {volume}
  {25}},\ \bibinfo {pages} {6447} (\bibinfo {year} {1982})}\BibitemShut
  {NoStop}%
\bibitem [{\citenamefont {Comparin}\ \emph {et~al.}(2022)\citenamefont
  {Comparin}, \citenamefont {Opler}, \citenamefont {Macaluso}, \citenamefont
  {Biella}, \citenamefont {Polychronakos},\ and\ \citenamefont
  {Mazza}}]{Comparin2022}%
  \BibitemOpen
  \bibfield  {author} {\bibinfo {author} {\bibfnamefont {T.}~\bibnamefont
  {Comparin}}, \bibinfo {author} {\bibfnamefont {A.}~\bibnamefont {Opler}},
  \bibinfo {author} {\bibfnamefont {E.}~\bibnamefont {Macaluso}}, \bibinfo
  {author} {\bibfnamefont {A.}~\bibnamefont {Biella}}, \bibinfo {author}
  {\bibfnamefont {A.~P.}\ \bibnamefont {Polychronakos}},\ and\ \bibinfo
  {author} {\bibfnamefont {L.}~\bibnamefont {Mazza}},\ }\href
  {https://doi.org/10.1103/PhysRevB.105.085125} {\bibfield  {journal} {\bibinfo
   {journal} {Phys. Rev. B}\ }\textbf {\bibinfo {volume} {105}},\ \bibinfo
  {pages} {085125} (\bibinfo {year} {2022})}\BibitemShut {NoStop}%
\bibitem [{\citenamefont {Nardin}\ \emph {et~al.}(2023)\citenamefont {Nardin},
  \citenamefont {Ardonne},\ and\ \citenamefont {Mazza}}]{Nardin2023}%
  \BibitemOpen
  \bibfield  {author} {\bibinfo {author} {\bibfnamefont {A.}~\bibnamefont
  {Nardin}}, \bibinfo {author} {\bibfnamefont {E.}~\bibnamefont {Ardonne}},\
  and\ \bibinfo {author} {\bibfnamefont {L.}~\bibnamefont {Mazza}},\ }\href
  {https://doi.org/10.1103/PhysRevB.108.L041105} {\bibfield  {journal}
  {\bibinfo  {journal} {Phys. Rev. B}\ }\textbf {\bibinfo {volume} {108}},\
  \bibinfo {pages} {L041105} (\bibinfo {year} {2023})}\BibitemShut {NoStop}%
\bibitem [{\citenamefont {Cayley}(1849)}]{Cayley}%
  \BibitemOpen
  \bibfield  {author} {\bibinfo {author} {\bibfnamefont {A.}~\bibnamefont
  {Cayley}},\ }\href {https://doi.org/doi:10.1515/crll.1849.38.93} {\bibfield
  {journal} {\bibinfo  {journal} {Journal f{\"u}r die reine angewandte
  Mathematik}\ }\textbf {\bibinfo {volume} {38}},\ \bibinfo {pages} {93}
  (\bibinfo {year} {1849})}\BibitemShut {NoStop}%
\bibitem [{\citenamefont {Ma}\ \emph {et~al.}(2024)\citenamefont {Ma},
  \citenamefont {Peterson}, \citenamefont {Scarola},\ and\ \citenamefont
  {Yang}}]{Ma22}%
  \BibitemOpen
  \bibfield  {author} {\bibinfo {author} {\bibfnamefont {K.~K.}\ \bibnamefont
  {Ma}}, \bibinfo {author} {\bibfnamefont {M.~R.}\ \bibnamefont {Peterson}},
  \bibinfo {author} {\bibfnamefont {V.}~\bibnamefont {Scarola}},\ and\ \bibinfo
  {author} {\bibfnamefont {K.}~\bibnamefont {Yang}},\ }in\ \href
  {https://doi.org/https://doi.org/10.1016/B978-0-323-90800-9.00135-9} {\emph
  {\bibinfo {booktitle} {Encyclopedia of Condensed Matter Physics (Second
  Edition)}}},\ \bibinfo {editor} {edited by\ \bibinfo {editor} {\bibfnamefont
  {T.}~\bibnamefont {Chakraborty}}}\ (\bibinfo  {publisher} {Academic Press},\
  \bibinfo {address} {Oxford},\ \bibinfo {year} {2024})\ \bibinfo {edition}
  {second edition}\ ed.,\ pp.\ \bibinfo {pages} {324--365}\BibitemShut
  {NoStop}%
\bibitem [{\citenamefont {Yang}(2019)}]{Yang2019}%
  \BibitemOpen
  \bibfield  {author} {\bibinfo {author} {\bibfnamefont {B.}~\bibnamefont
  {Yang}},\ }\href {https://doi.org/10.1103/PhysRevB.100.241302} {\bibfield
  {journal} {\bibinfo  {journal} {Phys. Rev. B}\ }\textbf {\bibinfo {volume}
  {100}},\ \bibinfo {pages} {241302} (\bibinfo {year} {2019})}\BibitemShut
  {NoStop}%
\bibitem [{\citenamefont {Yang}\ and\ \citenamefont {Balram}(2021)}]{Yang2021}%
  \BibitemOpen
  \bibfield  {author} {\bibinfo {author} {\bibfnamefont {B.}~\bibnamefont
  {Yang}}\ and\ \bibinfo {author} {\bibfnamefont {A.~C.}\ \bibnamefont
  {Balram}},\ }\href {https://doi.org/10.1088/1367-2630/abd49d} {\bibfield
  {journal} {\bibinfo  {journal} {New Journal of Physics}\ }\textbf {\bibinfo
  {volume} {23}},\ \bibinfo {pages} {013001} (\bibinfo {year}
  {2021})}\BibitemShut {NoStop}%
\bibitem [{\citenamefont {Mazaheri}\ \emph {et~al.}(2015)\citenamefont
  {Mazaheri}, \citenamefont {Ortiz}, \citenamefont {Nussinov},\ and\
  \citenamefont {Seidel}}]{Mazaheri2015}%
  \BibitemOpen
  \bibfield  {author} {\bibinfo {author} {\bibfnamefont {T.}~\bibnamefont
  {Mazaheri}}, \bibinfo {author} {\bibfnamefont {G.}~\bibnamefont {Ortiz}},
  \bibinfo {author} {\bibfnamefont {Z.}~\bibnamefont {Nussinov}},\ and\
  \bibinfo {author} {\bibfnamefont {A.}~\bibnamefont {Seidel}},\ }\href
  {https://doi.org/10.1103/PhysRevB.91.085115} {\bibfield  {journal} {\bibinfo
  {journal} {Phys. Rev. B}\ }\textbf {\bibinfo {volume} {91}},\ \bibinfo
  {pages} {085115} (\bibinfo {year} {2015})}\BibitemShut {NoStop}%
\bibitem [{\citenamefont {Chen}\ and\ \citenamefont {Seidel}(2015)}]{Chen15}%
  \BibitemOpen
  \bibfield  {author} {\bibinfo {author} {\bibfnamefont {L.}~\bibnamefont
  {Chen}}\ and\ \bibinfo {author} {\bibfnamefont {A.}~\bibnamefont {Seidel}},\
  }\href {https://doi.org/10.1103/PhysRevB.91.085103} {\bibfield  {journal}
  {\bibinfo  {journal} {Phys. Rev. B}\ }\textbf {\bibinfo {volume} {91}},\
  \bibinfo {pages} {085103} (\bibinfo {year} {2015})}\BibitemShut {NoStop}%
\bibitem [{\citenamefont {Chen}\ \emph {et~al.}(2019)\citenamefont {Chen},
  \citenamefont {Bandyopadhyay}, \citenamefont {Yang},\ and\ \citenamefont
  {Seidel}}]{Chen19}%
  \BibitemOpen
  \bibfield  {author} {\bibinfo {author} {\bibfnamefont {L.}~\bibnamefont
  {Chen}}, \bibinfo {author} {\bibfnamefont {S.}~\bibnamefont {Bandyopadhyay}},
  \bibinfo {author} {\bibfnamefont {K.}~\bibnamefont {Yang}},\ and\ \bibinfo
  {author} {\bibfnamefont {A.}~\bibnamefont {Seidel}},\ }\href
  {https://doi.org/10.1103/PhysRevB.100.045136} {\bibfield  {journal} {\bibinfo
   {journal} {Phys. Rev. B}\ }\textbf {\bibinfo {volume} {100}},\ \bibinfo
  {pages} {045136} (\bibinfo {year} {2019})}\BibitemShut {NoStop}%
\bibitem [{\citenamefont {Read}(1989)}]{Read}%
  \BibitemOpen
  \bibfield  {author} {\bibinfo {author} {\bibfnamefont {N.}~\bibnamefont
  {Read}},\ }\href {https://doi.org/10.1103/PhysRevLett.62.86} {\bibfield
  {journal} {\bibinfo  {journal} {Phys. Rev. Lett.}\ }\textbf {\bibinfo
  {volume} {62}},\ \bibinfo {pages} {86} (\bibinfo {year} {1989})}\BibitemShut
  {NoStop}%
\bibitem [{\citenamefont {Hormozi}\ \emph {et~al.}(2009)\citenamefont
  {Hormozi}, \citenamefont {Bonesteel},\ and\ \citenamefont
  {Simon}}]{Hormozi09}%
  \BibitemOpen
  \bibfield  {author} {\bibinfo {author} {\bibfnamefont {L.}~\bibnamefont
  {Hormozi}}, \bibinfo {author} {\bibfnamefont {N.~E.}\ \bibnamefont
  {Bonesteel}},\ and\ \bibinfo {author} {\bibfnamefont {S.~H.}\ \bibnamefont
  {Simon}},\ }\href {https://doi.org/10.1103/PhysRevLett.103.160501} {\bibfield
   {journal} {\bibinfo  {journal} {Phys. Rev. Lett.}\ }\textbf {\bibinfo
  {volume} {103}},\ \bibinfo {pages} {160501} (\bibinfo {year}
  {2009})}\BibitemShut {NoStop}%
\bibitem [{\citenamefont {Milovanovic}\ and\ \citenamefont
  {Jolicoeur}(2010)}]{Milovanovic10}%
  \BibitemOpen
  \bibfield  {author} {\bibinfo {author} {\bibfnamefont {M.~V.}\ \bibnamefont
  {Milovanovic}}\ and\ \bibinfo {author} {\bibfnamefont {T.}~\bibnamefont
  {Jolicoeur}},\ }\href {https://doi.org/10.1142/S0217979210055081} {\bibfield
  {journal} {\bibinfo  {journal} {International Journal of Modern Physics B}\
  }\textbf {\bibinfo {volume} {24}},\ \bibinfo {pages} {549} (\bibinfo {year}
  {2010})}\BibitemShut {NoStop}%
\bibitem [{\citenamefont {Sreejith}\ \emph
  {et~al.}(2011{\natexlab{a}})\citenamefont {Sreejith}, \citenamefont
  {T\ifmmode~\mbox{\H{o}}\else \H{o}\fi{}ke}, \citenamefont {W\'ojs},\ and\
  \citenamefont {Jain}}]{Wojs11}%
  \BibitemOpen
  \bibfield  {author} {\bibinfo {author} {\bibfnamefont {G.~J.}\ \bibnamefont
  {Sreejith}}, \bibinfo {author} {\bibfnamefont {C.}~\bibnamefont
  {T\ifmmode~\mbox{\H{o}}\else \H{o}\fi{}ke}}, \bibinfo {author} {\bibfnamefont
  {A.}~\bibnamefont {W\'ojs}},\ and\ \bibinfo {author} {\bibfnamefont {J.~K.}\
  \bibnamefont {Jain}},\ }\href
  {https://doi.org/10.1103/PhysRevLett.107.086806} {\bibfield  {journal}
  {\bibinfo  {journal} {Phys. Rev. Lett.}\ }\textbf {\bibinfo {volume} {107}},\
  \bibinfo {pages} {086806} (\bibinfo {year} {2011}{\natexlab{a}})}\BibitemShut
  {NoStop}%
\bibitem [{\citenamefont {Sreejith}\ \emph
  {et~al.}(2011{\natexlab{b}})\citenamefont {Sreejith}, \citenamefont
  {W\'ojs},\ and\ \citenamefont {Jain}}]{Sreejith2011a}%
  \BibitemOpen
  \bibfield  {author} {\bibinfo {author} {\bibfnamefont {G.~J.}\ \bibnamefont
  {Sreejith}}, \bibinfo {author} {\bibfnamefont {A.}~\bibnamefont {W\'ojs}},\
  and\ \bibinfo {author} {\bibfnamefont {J.~K.}\ \bibnamefont {Jain}},\ }\href
  {https://doi.org/10.1103/PhysRevLett.107.136802} {\bibfield  {journal}
  {\bibinfo  {journal} {Phys. Rev. Lett.}\ }\textbf {\bibinfo {volume} {107}},\
  \bibinfo {pages} {136802} (\bibinfo {year} {2011}{\natexlab{b}})}\BibitemShut
  {NoStop}%
\bibitem [{\citenamefont {Sreejith}\ \emph {et~al.}(2013)\citenamefont
  {Sreejith}, \citenamefont {Wu}, \citenamefont {W\'ojs},\ and\ \citenamefont
  {Jain}}]{Sreejith2013}%
  \BibitemOpen
  \bibfield  {author} {\bibinfo {author} {\bibfnamefont {G.~J.}\ \bibnamefont
  {Sreejith}}, \bibinfo {author} {\bibfnamefont {Y.-H.}\ \bibnamefont {Wu}},
  \bibinfo {author} {\bibfnamefont {A.}~\bibnamefont {W\'ojs}},\ and\ \bibinfo
  {author} {\bibfnamefont {J.~K.}\ \bibnamefont {Jain}},\ }\href
  {https://doi.org/10.1103/PhysRevB.87.245125} {\bibfield  {journal} {\bibinfo
  {journal} {Phys. Rev. B}\ }\textbf {\bibinfo {volume} {87}},\ \bibinfo
  {pages} {245125} (\bibinfo {year} {2013})}\BibitemShut {NoStop}%
\bibitem [{\citenamefont {Rodriguez}\ \emph {et~al.}(2012)\citenamefont
  {Rodriguez}, \citenamefont {Sterdyniak}, \citenamefont {Hermanns},
  \citenamefont {Slingerland},\ and\ \citenamefont {Regnault}}]{Rodriguez2012}%
  \BibitemOpen
  \bibfield  {author} {\bibinfo {author} {\bibfnamefont {I.~D.}\ \bibnamefont
  {Rodriguez}}, \bibinfo {author} {\bibfnamefont {A.}~\bibnamefont
  {Sterdyniak}}, \bibinfo {author} {\bibfnamefont {M.}~\bibnamefont
  {Hermanns}}, \bibinfo {author} {\bibfnamefont {J.~K.}\ \bibnamefont
  {Slingerland}},\ and\ \bibinfo {author} {\bibfnamefont {N.}~\bibnamefont
  {Regnault}},\ }\href {https://doi.org/10.1103/PhysRevB.85.035128} {\bibfield
  {journal} {\bibinfo  {journal} {Phys. Rev. B}\ }\textbf {\bibinfo {volume}
  {85}},\ \bibinfo {pages} {035128} (\bibinfo {year} {2012})}\BibitemShut
  {NoStop}%
\bibitem [{\citenamefont {Greiter}(2011)}]{Greiterbook}%
  \BibitemOpen
  \bibfield  {author} {\bibinfo {author} {\bibfnamefont {M.}~\bibnamefont
  {Greiter}},\ }\href {https://doi.org/10.1017/CBO9780511607561} {\emph
  {\bibinfo {title} {Mapping of Parent Hamiltonians}}}\ (\bibinfo  {publisher}
  {Springer Berlin},\ \bibinfo {year} {2011})\BibitemShut {NoStop}%
\bibitem [{\citenamefont {Simon}\ \emph {et~al.}(2007)\citenamefont {Simon},
  \citenamefont {Rezayi},\ and\ \citenamefont {Cooper}}]{Simon07}%
  \BibitemOpen
  \bibfield  {author} {\bibinfo {author} {\bibfnamefont {S.~H.}\ \bibnamefont
  {Simon}}, \bibinfo {author} {\bibfnamefont {E.~H.}\ \bibnamefont {Rezayi}},\
  and\ \bibinfo {author} {\bibfnamefont {N.~R.}\ \bibnamefont {Cooper}},\
  }\href {https://doi.org/10.1103/PhysRevB.75.195306} {\bibfield  {journal}
  {\bibinfo  {journal} {Phys. Rev. B}\ }\textbf {\bibinfo {volume} {75}},\
  \bibinfo {pages} {195306} (\bibinfo {year} {2007})}\BibitemShut {NoStop}%
\bibitem [{\citenamefont {Davenport}\ and\ \citenamefont
  {Simon}(2012)}]{Simon12}%
  \BibitemOpen
  \bibfield  {author} {\bibinfo {author} {\bibfnamefont {S.~C.}\ \bibnamefont
  {Davenport}}\ and\ \bibinfo {author} {\bibfnamefont {S.~H.}\ \bibnamefont
  {Simon}},\ }\href {https://doi.org/10.1103/PhysRevB.85.075430} {\bibfield
  {journal} {\bibinfo  {journal} {Phys. Rev. B}\ }\textbf {\bibinfo {volume}
  {85}},\ \bibinfo {pages} {075430} (\bibinfo {year} {2012})}\BibitemShut
  {NoStop}%
\bibitem [{\citenamefont {Balram}\ \emph {et~al.}(2019)\citenamefont {Balram},
  \citenamefont {Barkeshli},\ and\ \citenamefont {Rudner}}]{Balram2019}%
  \BibitemOpen
  \bibfield  {author} {\bibinfo {author} {\bibfnamefont {A.~C.}\ \bibnamefont
  {Balram}}, \bibinfo {author} {\bibfnamefont {M.}~\bibnamefont {Barkeshli}},\
  and\ \bibinfo {author} {\bibfnamefont {M.~S.}\ \bibnamefont {Rudner}},\
  }\href {https://doi.org/10.1103/PhysRevB.99.241108} {\bibfield  {journal}
  {\bibinfo  {journal} {Phys. Rev. B}\ }\textbf {\bibinfo {volume} {99}},\
  \bibinfo {pages} {241108} (\bibinfo {year} {2019})}\BibitemShut {NoStop}%
\bibitem [{\citenamefont {Zhang}\ \emph {et~al.}(2023)\citenamefont {Zhang},
  \citenamefont {Schossler}, \citenamefont {Seidel},\ and\ \citenamefont
  {Chen}}]{Zhang23}%
  \BibitemOpen
  \bibfield  {author} {\bibinfo {author} {\bibfnamefont {F.}~\bibnamefont
  {Zhang}}, \bibinfo {author} {\bibfnamefont {M.}~\bibnamefont {Schossler}},
  \bibinfo {author} {\bibfnamefont {A.}~\bibnamefont {Seidel}},\ and\ \bibinfo
  {author} {\bibfnamefont {L.}~\bibnamefont {Chen}},\ }\href
  {https://doi.org/10.1103/PhysRevB.108.075142} {\bibfield  {journal} {\bibinfo
   {journal} {Phys. Rev. B}\ }\textbf {\bibinfo {volume} {108}},\ \bibinfo
  {pages} {075142} (\bibinfo {year} {2023})}\BibitemShut {NoStop}%
\bibitem [{\citenamefont {Bajdich}\ \emph {et~al.}(2008)\citenamefont
  {Bajdich}, \citenamefont {Mitas}, \citenamefont {Wagner},\ and\ \citenamefont
  {Schmidt}}]{Bajdich08}%
  \BibitemOpen
  \bibfield  {author} {\bibinfo {author} {\bibfnamefont {M.}~\bibnamefont
  {Bajdich}}, \bibinfo {author} {\bibfnamefont {L.}~\bibnamefont {Mitas}},
  \bibinfo {author} {\bibfnamefont {L.~K.}\ \bibnamefont {Wagner}},\ and\
  \bibinfo {author} {\bibfnamefont {K.~E.}\ \bibnamefont {Schmidt}},\ }\href
  {https://doi.org/10.1103/PhysRevB.77.115112} {\bibfield  {journal} {\bibinfo
  {journal} {Phys. Rev. B}\ }\textbf {\bibinfo {volume} {77}},\ \bibinfo
  {pages} {115112} (\bibinfo {year} {2008})}\BibitemShut {NoStop}%
\bibitem [{\citenamefont {González-Ballestero}\ \emph
  {et~al.}(2011)\citenamefont {González-Ballestero}, \citenamefont {Robledo},\
  and\ \citenamefont {Bertsch}}]{numeric}%
  \BibitemOpen
  \bibfield  {author} {\bibinfo {author} {\bibfnamefont {C.}~\bibnamefont
  {González-Ballestero}}, \bibinfo {author} {\bibfnamefont {L.}~\bibnamefont
  {Robledo}},\ and\ \bibinfo {author} {\bibfnamefont {G.}~\bibnamefont
  {Bertsch}},\ }\href
  {https://doi.org/https://doi.org/10.1016/j.cpc.2011.04.025} {\bibfield
  {journal} {\bibinfo  {journal} {Computer Physics Communications}\ }\textbf
  {\bibinfo {volume} {182}},\ \bibinfo {pages} {2213} (\bibinfo {year}
  {2011})}\BibitemShut {NoStop}%
\end{thebibliography}%

\end{document}